\documentclass[aps,twocolumn,superscriptaddress,floatfix,letterpaper]{revtex4-1}

\newcommand{\tAw}[3]{{#1}_{e_{#2#3} }^{v_{#2}v_{#3}}}

\newcommand{\tC}[5]{{#1}_{v_{#2}v_{#3}v_{#4}v_{#5};
\phi_{#2#3#5} \phi_{#3#4#5}
}^{
e_{#2#3} e_{#2#4} e_{#2#5} e_{#3#4} e_{#3#5} e_{#4#5};
\phi_{#2#3#4} \phi_{#2#4#5}
}}

\usepackage{tikz}
\usepackage{tikz-cd}
\usetikzlibrary{arrows}
\usetikzlibrary{intersections}
\usetikzlibrary{shapes.geometric}

\newcommand{\id}{\mathop{\mathrm{id}}\nolimits}
\newcommand{\ot}{\otimes}

\usepackage[all]{xy}
\newcommand{\g}[1]{\vcenter{\xybox{\xygraph{!{/r2pc/:0}!~:{@{|}|>/-2.5pt/@{(}}
#1}}}}

\newcommand{\gdir}[1]{\vcenter{\xybox{\xygraph{!{/r2pc/:0}!~:{@{-}|-@{>}}#1}}}}

\usepackage{amsthm,amsfonts,amsmath,amssymb}

\usepackage{euscript}
\usepackage{bbold,bbm}

\usepackage{enumerate}
\usepackage{hyperref}
\usepackage{graphics}
\usepackage{xypic}

\usepackage{graphicx}

\newcommand\void[1] {}

\theoremstyle{definition}

\numberwithin{thm}{section}

\newcommand\CC           {\mathcal{C}}
\newcommand\DC           {\mathcal{D}}

\newcommand\CM          {\mathcal{M}}

\newcommand\Irr {\mathrm{Irr}}

\newcommand{\pf}{\begin{proof}}
\newcommand{\epf}{\end{proof}}

\begin{document}

\title{
Gapped domain walls between 2+1D topologically ordered states
}

\author{Tian Lan} 
\affiliation{Institute for Quantum Computing,
  University of Waterloo, Waterloo, Ontario N2L 3G1, Canada}

\author{Xueda Wen}
\affiliation{Department of Physics, Massachusetts Institute of
Technology, Cambridge, Massachusetts 02139, USA}

\author{Liang Kong} 
\affiliation{Shenzhen Institute for Quantum Science and Engineering, and Department of Physics, Southern University of Science and Technology, Shenzhen 518055, China}

\author{Xiao-Gang Wen}
\affiliation{Department of Physics, Massachusetts Institute of
Technology, Cambridge, Massachusetts 02139, USA}

\begin{abstract}

The 2+1D topological order can be characterized by the mapping-class-group
representations for Riemann surfaces of genus-1, genus-2, etc.  In this paper,
we use those representations to determine the possible gapped boundaries of a
2+1D topological order, as well as the domain walls between two topological
orders.  We find that mapping-class-group representations for both genus-1 and
genus-2 surfaces are needed to determine the gapped domain walls and
boundaries.  Our systematic theory is based on the fixed-point partition
functions for the walls (or the boundaries), which completely characterize the
gapped domain walls (or the boundaries).  The mapping-class-group
representations give rise to conditions that must be satisfied by the
fixed-point partition functions, which leads to a systematic theory.  Such
conditions can be viewed as bulk topological order determining the
(non-invertible) gravitational anomaly at the domain wall, and our theory can
be viewed as finding all types of the gapped domain wall given a
(non-invertible) gravitational anomaly.  We also developed a systematic theory
of gapped domain walls (boundaries) based on the structure coefficients of
condensable algebras.

\end{abstract}

\maketitle

{\small \setcounter{tocdepth}{2} \tableofcontents }

\section{Introduction}

Topological order is a new kind of order in gapped quantum states of matter
beyond Landau symmetry breaking theory.\cite{Wtop,WNtop} In \Ref{Wrig,KW9327},
it was conjectured that \emph{the non-Abelian geometric phases\cite{WZ8411}
(both the $U(1)$ part and the non-Abelian part) of degenerate ground states
generated by the automorphism of Riemann surfaces can completely characterize
different topological orders}.\cite{Wrig} 

The non-Abelian geometric phases
contain an universal non-Abelian part\cite{Wrig,KW9327} and a path dependent
$U(1)$ part\cite{Wrig,W1221}. 
The non-Abelian part carries information about the  projective representation
of mapping class group (MCG) of the space manifold.  For torus, the MCG is
$\Ga_1=SL(2,\Z)$, which is generated by a 90$^\circ$ rotation and a Dehn twist.
The associated non-Abelian geometric phases for such two generators are denoted
by $S$ and $T$, which are unitary matrices.  $S$ and $T$ generate a projective
representation of MCG $SL(2,\Z)$ for torus.  

The Abelian part of the non-Abelian geometric phases are also important: it is
related to the gravitational Chern-Simons term\cite{KF9732,HLP1242,KW1458} in
the partition function and carries information about the chiral central charge
$c$ for the gapless edge excitations.\cite{Wedgerev,Wtoprev}  The data
$(S,T,c)$ is a quite complete description of 2+1D topological orders.  However,
to obtain a full description of 2+1D topological orders, the modular data for
genus-1 surface is not enough,\cite{MS170802796,BW180505736,KS180603158} we
must also use the non-Abelian geometric phases (\ie the mapping class group
representations) for genus-2 surfaces.\cite{WW190810381}


In this paper, we will study the boundary of topological orders, or more
generally, the domain wall between two topological orders.  We will see how the
boundary properties are determined by the bulk topological orders.  We would
like to consider the following issues: \emph{What is the data that allow us to
characterize different gapped domain walls between topologically ordered
states? How to classify gapped domain walls?}
\Ref{KS10080654,WW1263,L13017355,K13064254} studied those issues for the case
of 2+1D Abelian topological orders, using condensable set of bosonic
topological excitations.
\Ref{BH08124596,BW10065479,KK11045047,FV12034568,FV13073632,K13078244,HW14080014}
considered this problem for the more general case of 2+1D non-Abelian
topological orders, using boson condensation\cite{BMc0602115,BS08080627} and/or
condensable algebra\cite{K13078244}.  Some discussions for the boundaries of
topological orders beyond 2+1D can be found in
\Ref{KW1458,FV14095723,KW170200673}.  In particular, it was pointed out that
the boundary effective theory of a bulk topological order has a gravitational
anomaly.\cite{W1313,LW1384,K1467,KW1458} In fact, the bulk topological order
gives an one-to-one classification of gravitational anomaly in one low
dimension (realized by the boundary).\cite{KW1458} Thus in some sense
\emph{types of bosonic topological order = types of bosonic gravitational
anomaly in one lower dimension}.  (Note that bosonic gravitational anomaly is a
property of an effective theory that cannot be realized as a local bosonic
system with finite cut-off.  Two effective theories are said to have the same
type of gravitational anomaly if one effective theory can change into another
effective theory, possibly via phase transitions in the same
dimension.\cite{KW1458})

\begin{figure}[tb]
\centering
\includegraphics[scale=0.4]{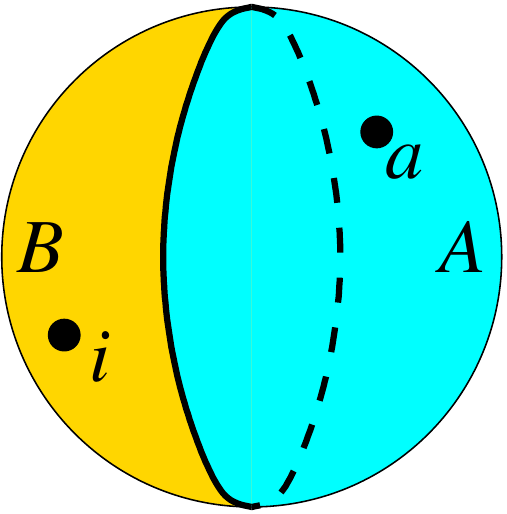}
\caption{
 Two topologically ordered phases $\cA$ and $\cB$, each occupies a hemisphere
of a 2D space $S^2$.  There is an excitation of type $a$ in the phase $\cA$ and
an excitation of type $i$ in the phase $\cB$.  } \label{S2ABai} \end{figure}

In this paper, we will rederive the simple results based on the $(S,T,c)$, for
the boundaries of general  2+1D non-Abelian topological orders introduced in
\Ref{LWW1414}.  Our new approach also allows us to generalize the approach in
\Ref{LWW1414} to high genus Riemann surface which may lead to a complete
description of the gaped boundary of 2+1D topological orders.  
%
%
We try to address the following question: \emph{given a 2+1D topological order
described by $(S,T,c)$ and the data from higher genus, how to describe and
classify different gapped 1+1D boundary phases?}  If we find that there is
no gapped 1+1D boundary for a type of 2+1D  topological order, then such a type
of topological order must have a gapless boundary.  It can also be stated in
the following way: given a type of gravitational anomaly determined, how to
describe different gapped 1+1D phases?  If we find that there is no gapped
1+1D phase for a type of gravitational anomaly, then such a type of
gravitational anomaly will require (or ensure) the 1+1D phases to be gapless.

\section{Characterization of gapped domain walls}
\label{walldata}

\subsection{Dimensions of fusion spaces}

To introduce data that can characterize different gapped domain walls, let us
consider a 2D space $S^2$. Half of $S^2$ is occupied by the phase $\cA$ and the
other half by the phase $\cB$.  Let us assume the domain wall between the
topological phases is gapped.  We put a type-$a$ topological excitation in the
phase $\cA$ and a type-$i$ topological excitation in the phase $\cB$.  We denote such a
configuration as $S^2_{\cB\cA;ia}$ (see Fig. \ref{S2ABai}).  The ground state space
for such a configuration is a fusion space.  The dimension of the fusion space
(\ie the ground state degeneracy), denoted as  $M_{\cB\cA}^{ia}\in \N$, is the data
that characterizes the gapped domain wall between $\cA$ and $\cB$ phases.  We may also
view $M_{\cB\cA}^{ia}$ as the fusion coefficients for the fusion of type-$i$
particle, type-$a$ particle, and the domain wall $BA$.

\subsection{Weighted wave function overlap}

To obtain more data to characterize the domain walls, let us consider the
degenerate ground states of the topologically ordered phase $\cA$, described
by normalized wave function $|\psi^{\cA}_{I_{\cA}}\>$, where the index
$I_{\cA}$ label different ground states on a closed Riemann surface of genus
$g$.  Similarly, we have the degenerate ground states of the topologically
ordered phase $\cB$: $|\psi^{\cB}_{I_{\cB}}\>$.  

\begin{figure}[tb] 
\centerline{ \includegraphics[scale=0.4]{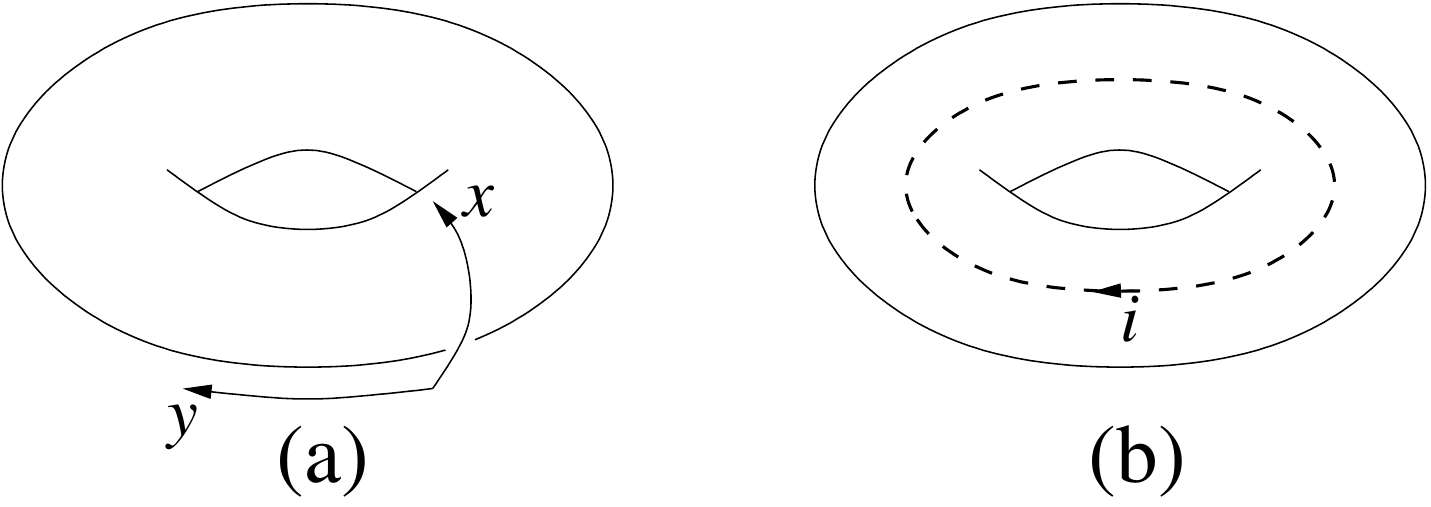} } 
\caption{
(a): The ground state $|\one\>$ on a torus that corresponds to the trivial
particle $\one$ can be represented by an empty solid torus.  (b): The other
ground state $|i\>$ that corresponds to a type-$i$ topological excitation can be
represented by an solid torus with a world-line of the type-$i$ in the center.  }
\label{sttT} 
\end{figure}

The degenerate ground states on a closed surface can be obtained as path
integral on the ``solid surface'' (a 3 dimensional manifold whose boundary is
the surface) with a world-line of a topological excitation (see Fig.
\ref{sttT}). Thus we can label different ground states using the label of
those topological excitations.  Also, the above construction of degenerate
ground states  using world-lines give rise to a natural basis for the
degenerate ground states. We will refer such a basis as the \emph{excitation
basis}.

For example, on a genus-1 surface, the degenerate ground states are labeled by
$I_{\cB}=i$ for the phase $\cB$ and $I_{\cA}=a$ for the phase $\cA$.  Here
$i=1,\cdots,N_{\cB}$ label the types of the topological excitations in the
phase $\cB$ and $a=1,\cdots,N_{\cA}$ label the types of the topological
excitations in the phase $\cA$.  For genus-1 surface, those excitation-labeled
ground states happen to be orthogonal.  We see that the ground state
degeneracies on genus-1 surface are given by $N_{\cA}$ for phase $\cA$ and
$N_{\cB}$ for phase $\cB$.

\begin{figure}[tb] 
\centerline{ \includegraphics[scale=0.5]{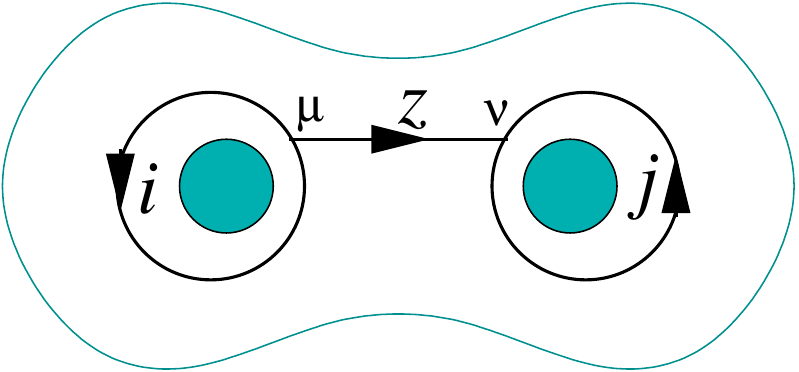} } 
\caption{
A filled genus-2 Riemaniann surface (denoted as $\Si_2^\text{fill}$), which has
three world-lines $i$, $j$, and $z$ in the interior.
} 
\label{O-O} 
\end{figure}

As another example, on a genus-2 surface, the degenerate ground states are
labeled by $I_{\cB}=(i,j,z,\mu,\nu)$ where $\mu=1,\cdots,N^{i\bar i}_{B;z}$ and
$\nu=1,\cdots,N^{j\bar j}_{B;\bar z}$ for the phase $\cB$.  The degenerate ground
states are labeled by $I_{\cA}=(a,b,c,\al,\bt)$ where $\al=1,\cdots,N^{a\bar
a}_{A;c}$ and $\bt=1,\cdots,N^{b\bar b}_{A;\bar c}$ for the phase $\cA$ (see
Fig.\ref{O-O}).  Here $i,j,z$ label the types of the topological excitations in
the phase $\cB$ and $a,b,c$ label the types of the topological excitations in the
phase $\cA$.  $N^{ij}_{B,z}$ are the fusion coefficients of the topological
excitations in the phase $\cA$ and $N^{ab}_{A,c}$ are the fusion coefficients of
the topological excitations in the phase $\cB$.  We see that the ground state
degeneracies on genus-2 surface are given by $\sum_{z=1}^{N_{\cB}} N^{i\bar
i}_{B;z}N^{j\bar j}_{B;\bar z}$ for phase $\cB$ and $\sum_{c=1}^{N_{\cA}} N^{a\bar
a}_{A;c}N^{b\bar b}_{A;\bar c}$ for phase $\cA$.

Motivated by the wave function overlap\cite{MW1418,HMW1457,MW1427,MW160609639}
that can characterize different topological orders, here we will use the
\emph{weighted wave function overlap} to characterize different domain walls. We
conjecture that\cite{KW1458,WW180109938}\\
\emph{
the weighted
overlap of the degenerate ground states on a closed genus-$g$ surface $\Si_g$
for topologically ordered phases $\cA$ and $\cB$ have the following form 
\begin{align}
\label{olap}
\<\psi^{\cB}_{I_{\cB}}|\ee^{- H_W}|\psi^{\cA}_{I_{\cA}}\> = \ee^{-\si_{\cB\cA} A_{\Si_g}+o(\frac{1}{A_{\Si_g}})}
W_{\cB\cA,g}^{I_{\cB}I_{\cA}} 
\end{align} 
where $H_W$ is local hermitian operator like a Hamiltonian of a quantum
system, $A_{\Si_g}$ is the area of the surface $\Si_g$ and
$W_{\cB\cA,g}^{I_{\cB}I_{\cA}}$ is a topological invariant that characterize
the domain between the phases $\cA$ and $\cB$.  }

In general, $W_{\cB\cA,g}^{I_{\cB}I_{\cA}}$ depends on the choices of $H_W$
which correspond to different  choices of domain walls.  A concrete
calculation of the wave function overlap, $W_{\cB\cA,g}^{I_{\cB}I_{\cA}}$, for
a simple topological order is presented in Section \ref{Z2example}.  We will show
in next section that the wave function overlap can also be viewed as partition
function of the domain wall.

When $\cB$ is a trivial product state, the index $I_{\cB}$ is always fixed to
be $1$, since $\cB$ has no ground state degeneracy.  In this case we
simplify
\begin{align}
\label{W1}
 W_{\cB\cA,g}^{1,I_{\cA}}  =  W_{\cA,g}^{I_{\cA}}
\end{align}
by dropping index for trivial phase $\cB$ and $I_{\cB}$.
Similarly, we simplify
\begin{align}
\label{M1}
 M_{\cB\cA}^{1a}  =  M_{\cA}^{a}
\end{align}
Those are data that describe a gapped boundary of topological order $\cA$.

We note that $W_{\cB\cA,g}^{I_{\cB}I_{\cA}}$ are in general complex numbers,
whose phase can be changed by a change of phases for the ground states
\begin{align}
 |\psi^{\cA}_{I_{\cA}}\> \to \ee^{\ii \th_{I_{\cA}}} |\psi^{\cA}_{I_{\cA}}\>, \ \ \ \
 |\psi^{\cB}_{I_{\cB}}\> \to \ee^{\ii \phi_{I_{\cB}}} |\psi^{\cB}_{I_{\cB}}\>.
\end{align}
Also $W_{\cB\cA,g}^{I_{\cB}I_{\cA}}$ may depend on some choices of basis that cannot be
fixed.  So $W_{\cB\cA,g}^{I_{\cB}I_{\cA}}$ by themselves are not a topological invariant,
and they are not even physical.  So we need to find a way mode out those phase
and basis dependence.  When we said $W_{\cB\cA,g}^{I_{\cB}I_{\cA}}$ are ``topological
invariant'', we mean that they are topological invariant up to those
phase and basis choices.  

For genus-$1$ surface $\Si_1$, we may choose the
phases of $|\psi^{\cA}_\one\>$ and $|\psi^{\cB}_\one\>$ to make $W_{\cB\cA,1}^{\one\one}$
real and positive. (Here $\one$ correspond to the trivial particle.) We then
choose the phases of $|\psi^{\cA}_a\>$ to make $W_{\cB\cA,1}^{\one a}$ real and
positive.  Similarly, we choose the phases of $|\psi^{\cB}_i\>$ to make
$W_{\cB\cA,1}^{i \one}$ real and positive.  With such a choice, we find that
$W_{\cB\cA,1}^{I_{\cB}I_{\cA}}=M_{\cB\cA}^{ia}$, which is the dimension of the fusion space
defined in the last subsection (see Section \ref{MW} for an explanation).  For
higher genus $g>1$, $W_{\cB\cA,g}^{I_{\cB}I_{\cA}}$, after moding out some gauge
redundancy, are new topological invariants.  
We hope $W_{\cB\cA,g}^{I_{\cB}I_{\cA}}$ carry enough information
to fully characterize a gapped domain wall.



\section{The conditions on the domain-wall data}

In this section, we are going to derive some conditions on the data that
characterize the gapped domain walls.  For example we like to show that the
dimension of the fusion state $M_{\cB\cA}^{ia}$ (which are non-negative integers)
and the weighted wave function overlap for torus $W_{\cB\cA,1}^{ia}$ (which can be
complex numbers) are actually equal to each other $M_{\cB\cA}^{ia} =
W_{\cB\cA,1}^{ia}$.  This is quite an amazing relation. 

To derive those conditions, we need to introduce topological path integral, \ie
the path integral for triangulated space-time whose value is re-triangulation
invariant. This is done in Appendix \ref{toppath}.  We also need to introduce
an algebraic (\ie a categorical) approach to evaluate those  topological path
integrals, which is done in Appendix \ref{catcal}.  Using those results, we can
derive the conditions on the domain-wall data.

\subsection{Why $M^{ia}_{\cB\cA}= W^{ia}_{\cB\cA,1}$?}
\label{MW}

First, we like to show $M^{ia}_{\cB\cA}= W^{ia}_{\cB\cA,1}$.  Let us consider
the topological path integral on space-time $D^2_{i}\times S^1$, where $D^2_i$
is a disk with a puncture.  Such a puncture corresponds to a world-line of a
type-$i$ topological excitation wrapping around $S^1$.  The topological path
integral on $D^2_{i}\times S^1$ only sum over the degrees of freedom in the
bulk, and leave the degrees of freedom boundary fixed.  Thus
$Z^\text{top}(D^2_{i}\times S^1)$ corresponds to a wave function $|\psi_i\>$ on
$S^1\times S^1 = \prt( D^2_{i}\times S^1)$ (for details, see
\Ref{WW180109938}).  First, we like to show that such a wave function
$|\psi_i\>$ is automatically normalized.  According to \Ref{WW180109938},
$\<\psi_i|\psi_i\>$ is given by $Z^\text{top}(S^2_{i,\bar i}\times S^1)$ where
$S^2_{i,\bar i}\times S^1$ is obtained by gluing two $D^2_{i}\times S^1$
together along its boundary.  $S^2_{i,\bar i}\times S^1$ contains two
world-lines of type-$i$ and type-$\bar i$.  

Let us evaluate the above partition function on $S^2_{i,\bar i}\times S^1$,
which turns out to be $Z^\text{top}(S^2_{i,\bar i}\times S^1)=1$:
\begin{align}
\label{Zii}
Z^\text{top} &\bpm \includegraphics[scale=.40]{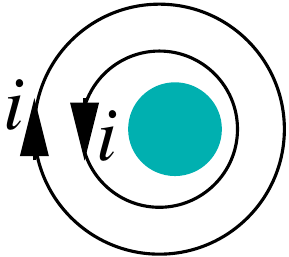} \epm 
= \sum_{\al\bt,k=\one}  Y^{\bar i i}_{\bar k,\al\bt} 
Z^\text{top} \bpm \includegraphics[scale=.40]{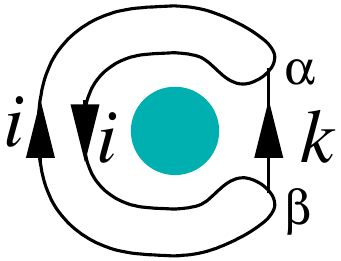} \epm 
\nonumber \\
&= \sum_{\al\bt,k=\one}  Y^{\bar i i}_{\bar k,\al\bt}  O^{\bar i i,\al\bt}_k
Z^\text{top} \bpm \includegraphics[scale=.40]{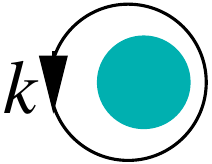} \epm  
\nonumber\\
&=
\sum_{\al\bt}Y^{\bar i i}_{\one,\al\bt}  O^{\bar i i,\al\bt}_\one
Z^\text{top}(S^2\times S^1)=1,
\end{align}
where we have used $N^{i\bar i}_\one=N^{\bar i i}_\one=1$, so $\al=\bt=1$ and
the $\sum_{\al\bt}$ only contain one term.  $k$ has to be the trivial particle
$\one$, since $k$ lives on a sphere $S^2$ alone. We have also used $Y^{\bar i
i}_{\one,11} O^{\bar i i,11}_\one =1$ (see appendix \ref{catcal} for the definition
of $Y$-move and $O$-move) and $Z^\text{top}(S^2\times S^1)=1$.  So
the topological path integral on $D^2_{i}\times S^1$ gives rise to normalized
wave function $|\psi_i\>$.

Now, instead of a path integral on $S^2_{i,\bar i}\times S^1$ which gives us
$Z^\text{top}(S^2_{i,\bar i}\times S^1)=1$, let us consider a path integral on
$S^2_{\cB\cA;ia}\times S^1$, where $S^2_{\cB\cA;ia}$ is described by Fig.
\ref{S2ABai}.  In space-time, the particle $a$ and $i$ correspond to
world-lines wrapping around $S^1$ which is viewed as a space direction.  The
boundary between the two hemisphere in Fig.  \ref{S2ABai} is viewed as another
space direction.  The topological partition function for space-time
$S^2_{\cB\cA;ia}\times S^1$ corresponds to the weighted wave function overlap
$\<\psi_i^{\cB}|\ee^{-H_W}|\psi_a^{\cA}\>$ with a fine tune choice of $H_W$, which has
a form
\begin{align}
 \<\psi_i^{\cB}|\ee^{-H_W}|\psi_a^{\cA}\>=Z^\text{top}(S^2_{\cB\cA;ia}\times S^1)
=
\ee^{ -\si_{\cB\cA} A_{\Si_1}} 
W_{\cB\cA,1}^{ia} .
\nonumber
\end{align}
Here $\si_{\cB\cA}$ is the energy density of the domain wall between the $\cA$ and $\cB$
phases, and $A_{\Si_1}$ is the space-time area occupied by the domain wall.
For our topological path integral, the energy density of the domain wall
$\si_{\cB\cA}$ is fine tuned to zero. Thus we actually have
\begin{align}\label{WF_overlap}
 \<\psi_i^{\cB}|\ee^{-H_W}|\psi_a^{\cA}\>=Z^\text{top}(S^2_{\cB\cA;ia}\times S^1)
=
W_{\cB\cA,1}^{ia} .
\end{align}

In the above, we have regarded the $S^1$ in $S^2_{\cB\cA;ia}\times S^1$ as a space
direction. Now we regard the $S^1$ in $S^2_{\cB\cA;ia}\times S^1$ as the time
direction, and $S^2_{\cB\cA;ia}$ in $S^2_{\cB\cA;ia}\times S^1$ as the space.  In this
case, the area independent part of the partition function has a different
meaning: it becomes the ground state degeneracy on the space $S^2_{\cB\cA;ia}$
which is denoted by $ M_{\cB\cA}^{ia}$.  Thus $M^{ia}_{\cB\cA}= W^{ia}_{\cB\cA,1}$.  It is
interesting to see that the weighted wave function overlaps for torus
$W^{ia}_{\cB\cA,1}$ (after fine-tuned into topological ones), are always given by
non-negative integers in the excitation basis.

When viewed as partition function,  $W^{ia}_{\cB\cA,1}$ is a multi-component
partition function (labeled by $i$ and $a$) on a torus.  Such a multi-component
partition function describe a gapped theory on the torus (\ie the domain wall)
that has a non-invertible gravitational anomaly.\cite{JW190513279} Thus the
conditions on $W^{ia}_{\cB\cA,1}$ (see next subsection) are a special case of the
modular covariance condition discussed in \Ref{JW190513279}.

\subsection{Invariance under the MCG action}

\begin{figure}[tb]
\centering
\includegraphics[scale=0.5]{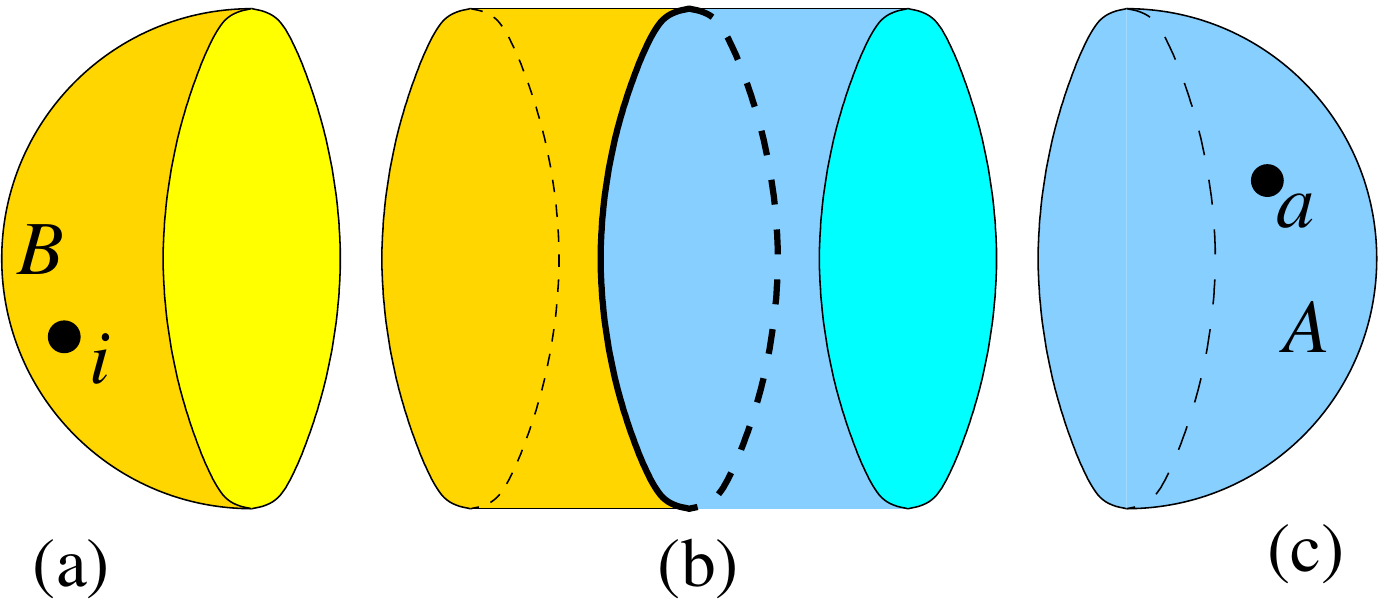}
\caption{
Divide $S^2$ in Fig. \ref{S2ABai} into three pieces.
} 
\label{DCD} 
\end{figure}

Next, we divide the space-time $S^2_{\cB\cA;ia}\times S^1$ into three pieces
$D^2_{A;a}\times S^1$ (see Fig. \ref{DCD}a), $D^2_{B;i}\times S^1$ (see Fig.
\ref{DCD}c), and $S^1_{\cB\cA}\times S^1\times S^1$ (see Fig. \ref{DCD}b), where
$D^2_{A;a}$ is a disk occupied by the phase $\cA$ and the type-$a$ particle,
$D^2_{B;i}$ is a disk occupied by the phase $\cB$ and the type-$i$ particle, and
$S^1_{\cB\cA}\times S^1$ is a cylinder occupied by the $\cA$ and $\cB$ phases and the
domain wall.  We can use the three pieces, $D^2_{A;a}\times S^1$,
$D^2_{B;i}\times S^1$, and $S^1_{\cB\cA}\times S^1\times S^1$, to asamble the same
closed space-time in two ways: (1) we glue  $D^2_{A;a}\times S^1$ and
$S^1_{\cB\cA}\times S^1\times S^1$ directly along $S^1\times S^1$ boundary, and
glue $S^1_{\cB\cA}\times S^1\times S^1$ and $D^2_{B;i}\times S^1$ with a twist
$\hat U \in $ $\Ga_1=SL(2,\Z)$ for the boundary $S^1\times S^1$; (2)  we glue
$D^2_{A;a}\times S^1$ and $S^1_{\cB\cA}\times S^1\times S^1$ with a twist $\hat U
\in $ $\Ga_1=SL(2,\Z)$ and glue $S^1_{\cB\cA}\times S^1\times S^1$ and
$D^2_{B;i}\times S^1$  directly.

The two ways to assemble the closed space-time \emph{only differ by a
re-triangulation}, and they must produce the same topological partition
function (see Appendix \ref{pathtop} and \ref{pathtopB}). This leads to a matrix relation
\begin{align}
\label{MNNM1}
 R^U_{\cB} W_{\cB\cA,1}=W_{\cB\cA,1} R^U_{\cA}.
\end{align}
We notice that the path integral on $D^2_{A;a}\times S^1$ produce the basis
state $|a;A\>$ in the excitation basis for the phase $\cA$ on the surface $\prt
(D^2_{A;a}\times S^1)$.  The action of the MCG
transformation $\hat U$ on $|a;A\>$ is described by the unitary representation
$R^U_{\cA}$ of $\hat U$ for the phase $\cA$.  Similarly, the path integral on
$D^2_{B;i}\times S^1$ produce the basis state $|i;B\>$ in the quasiparticle
basis for the phase $\cB$.  The action of the MCG transformation $\hat U$ on
$|i;B\>$ is described by the unitary representation $R^U_{\cB}$ of $\hat U$ for the
phase $\cB$.  This is how do we obtain \eqn{MNNM1}.

\subsection{A consistent condition between the fusions in the bulk and on the
wall}
\label{Sec: ConsistentCondition}

In this section, we like to obtain some additional conditions.  Let us consider
the following fusion process: we bring a type-$a$ excitation in phase $\cA$ and a
type-$i$ excitation in phase $\cB$ to the domain wall $w_{\cB\cA}$ between the two
phase and fuse them into an excitation of type $s$ on the domain wall. The
corresponding fusion algebra is given by
\begin{align}
 i\otimes_{w_{\cB\cA}} a = M^{ia}_{\cB\cA;s} s, \ \ \ \  M^{ia}_{\cB\cA;s} \in \N. 
\end{align} 
It turns out that 
\begin{align}
\label{MN}
 M_{\cB\cA}^{ia}= M^{ia}_{\cB\cA;1}
\end{align}
where $s=1$ represents the trivial type of the excitations on the domain wall.

Now let us compute the dimension of the fusion space of $S^2_{\cB\cA}$ with
excitations of types $a$ and $b$ in the phase $\cA$ and excitations of types $i$
and $j$ in the phase $\cB$.  There are two ways to compute the dimension of the
fusion space: (1) we may fuse $a$ and $b$ into $c$ and fuse $i$ and $j$ into
$k$, or (2) we may first fuse $a$ and $i$ into $s$ on the domain wall and fuse
$i$ and $j$ into $\bar s$  on the domain, and then fuse $s$ and $\bar s$ into
the trivial excitation on the domain wall.  The two fusion path should
produce the same result, and we obtain
\begin{align}
\label{NNNMM}
\sum_{c,k} N^{ab}_{A;c}
 N^{ij}_{B;k} M_{\cB\cA}^{kc} =\sum_{s}
M^{ia}_{\cB\cA;s}
M^{jb}_{\cB\cA;\bar s}
\end{align}
The two equations \eqn{MN} and \eqn{NNNMM} imply the condition \eqn{stable}.

\begin{figure}[tb]
  \centering
  \includegraphics[scale=0.5]{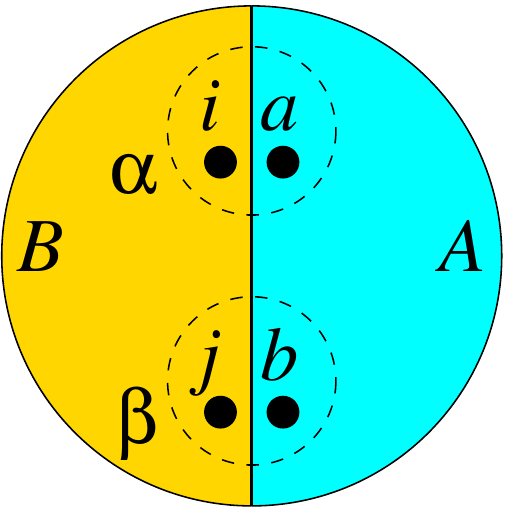}
  \caption{A 2D space $S^2$ is occupied by two topologically ordered phases,
the phase $\cA$ and the phase $\cB$.  Fusing an excitation of type $a$ in the phase $\cA$
and an excitation of type $i$ in the phase $\cB$ produce an excitation of type $s$ on the domain wall between the phase $\cA$ and the phase $\cB$.  } 
\label{S2ABabij} 
\end{figure}

\subsection{The conditions on genus-1 data for a gapped domain wall}

Consider two topological orders, phases $\cA$
and $\cB$, described by $(S^{\cA}, T^{\cA},c^{\cA})$ and $(S^{\cB},T^{\cB},c^{\cB})$.  Suppose there are
$N^{\cA}$ and $N^{\cB}$ types of topological excitations in the phase $\cA$ and the phase
B, then the ranks of their modular matrices are $N^{\cA}$ and $N^{\cB}$ respectively.
We find the following necessry conditions for the phase $\cA$ and the phase $\cB$  to
be connected by a gapped domain wall:\cite{LWW1414}
\begin{align}
c^{\cA} &= c^{\cB}.
\label{c=c}
\end{align}
there exist non-zero 
\begin{align}
 \label{Winteger}
M_{\cB\cA}^{ia}\in\N,  
\end{align}
such that
\begin{align}
  \sum_j S^{\cB}_{ij} M_{\cB\cA}^{ja} &=  \sum_b M_{\cB\cA}^{ib} S^{\cA}_{ba},
\nonumber\\
  \sum_j T^{\cB}_{ij} M_{\cB\cA}^{ja} &=  \sum_b M_{\cB\cA}^{ib} T^{\cA}_{ba}.
  \label{commute}\\
  M_{\cB\cA}^{ia}M_{\cB\cA}^{jb} &\leq\sum_{kc} N^{ij}_{B;k} M_{\cB\cA}^{kc} N^{ab}_{A;c}\,.
  \label{stable}
\end{align}
Here $\N$ denotes the set of non-negative integers.  $a,b,c,\dots$ and
$i,j,k,\dots$ are indices for the particle types in phases $\cA$ and $\cB$.
$N^{ab}_{A;c}$ and $N^{ij}_{B;k}$ are fusion coefficient of phases $\cA$ and $\cB$.
We may call Eq.\eqref{commute} the commuting condition, and Eq.\eqref{stable}
the stable condition.\cite{LWW1414}

In fact the matrix $M_{\cB\cA}$ label a gapped domain wall between phases $\cA$ and $\cB$.
Eqns.  \eq{c=c}, \eq{Winteger}, \eq{commute}, and \eq{stable} is a set of
necessary conditions a gapped domain wall $M_{\cB\cA}$ must satisfy, i.e., if there
is a gapped domain wall, we will have a non-zero $M_{\cB\cA}$ that satisfies
those conditions.  This implies that if there is no non-zero solution of
$M_{\cB\cA}$, the domain wall must be gapless.  However, it is not clear if those
condition are sufficient for a gapped domain wall to exist.
In some simple examples, the solutions $M_{\cB\cA}$ are in one-to-one
correspondence with gapped domain walls.  However, for some complicated
examples~\cite{D13127466}, 
a $M_{\cB\cA}$ may correspond to more than one type of
gapped domain wall.  This indicates that some additional data are needed to
completely characterize gapped domain walls.

\subsection{Wave function overlap on genus-2 Riemaniann surfaces and
topological path integral}\label{sec.wfg2}

In the above, we have developed a theory on the domain wall based on the data
$W^{ia}_{\cB\cA;1}$, the wave function overlap on torus.  However, such a data is
insufficient to fully characterize the domain wall, \ie there are known
different domain walls to produce the same $W^{ia}_{\cB\cA;1}$.  So in this section,
we consider the overlap of normalized ground state wave functions,
$|\psi^{\cA}_{I_{\cA}}\>$ and $|\psi^{\cB}_{I_{\cB}}\>$, on genus-2 surfaces, which gives rise
to data $W^{I_{\cB}I_{\cA}}_{\cB\cA,2}$ (see \eqn{olap}).

To understand the structure of the  data $W^{I_{\cB}I_{\cA}}_{\cB\cA,2}$, let us generate
the degenerate ground states on a genus-2 surface $\Si_2$ via a path integral.
We first consider a filled genus-2 surface denoted as $\Si_2^\text{fill}$:
$\Si_2 = \prt \Si_2^\text{fill} $.  The path integral on $\Si_2^\text{fill} $
will generate a ground state on $\Si_2$. To generate other ground states, we
need to add three world-lines $i,j,z$ in the interior of the filled genus-2
surface (see Fig. \ref{O-O}).  Thus the linearly independent states on $\Si_2$
are labeled by $i,j,z,\mu,\nu$, where $\mu=1,\cdots, N^{i\bar i}_z$ and
$\nu=1,\cdots, N^{j\bar j}_z$.  Thus, for topological phase $\cA$, the label
$I_{\cA}$ correspond to $I_{\cA} \sim (i_{\cA},j_{\cA},z_{\cA},\mu_{\cA},\nu_{\cA})$. For topological phase
$\cB$, the label $I_{\cB}$ correspond to $I_{\cB} \sim (i_{\cB},j_{\cB},z_{\cB},\mu_{\cB},\nu_{\cB})$.  So
the wave function overlap on genus-2 surfaces is given by
\begin{align}
 W^{I_{\cB}I_{\cA}}_{\cB\cA,2}=W^{(i_{\cB},j_{\cB},z_{\cB},\mu_{\cB},\nu_{\cB}),(i_{\cA},j_{\cA},z_{\cA},\mu_{\cA},\nu_{\cA})}_{\cB\cA,2}
\end{align}

\begin{figure}[tb] 
\centerline{ \includegraphics[scale=0.5]{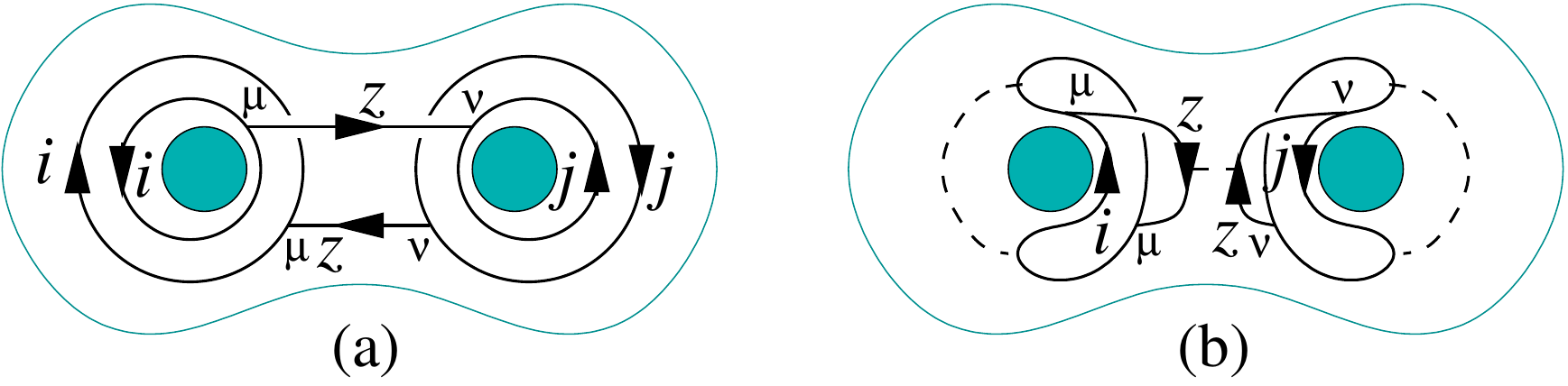} } 
\caption{
A path integral on a closed 3-dimensional space-time
$\Si_2^\text{fill} \sqcup \Si_2^\text{fill}$.
} 
\label{O-O2} 
\end{figure}

The wave functions, $|\Psi_{i,j,z,\mu,\nu}\>$, given by the path integral on
$\Si_2^\text{fill}$ may not be normalized.  In the following we will consider
their normalization.  First, we express the normalization via path integral
\begin{align}
&\ \ \ \ 
 \< \Psi_{i,j,z,\mu,\nu} | \Psi_{i,j,z,\mu,\nu}\>
\nonumber\\
&=Z^\text{top}[\Si_2^\text{fill}(i,j,z,\mu,\nu) \sqcup \Si_2^\text{fill}(i,j,z,\mu,\nu)]
\end{align}
where $\Si_2^\text{fill}(i,j,z,\mu,\nu) \sqcup \Si_2^\text{fill}(i,j,z,\mu,\nu)$ is the closed 3-dimensional
space-time obtained by gluing two $\Si_2^\text{fill}(i,j,z,\mu,\nu)$'s along their $\Si_2$
boundary (see Fig. \ref{O-O2}).  
Also $\Si_2^\text{fill}(i,j,z,\mu,\nu)$ contain the worldlines
$i,j,z$ ( Fig. \ref{O-O2}).
We next deform the world-lines in Fig.
\ref{O-O2}a to those in Fig. \ref{O-O2}b using \eqn{PhiY}.  This gives rise to
a factor $Y^{i\bar i}_{\one;11} Y^{j\bar j}_{\one;11} Y^{z\bar z}_{\one;11}$.
Note that dash lines in Fig. \ref{O-O2}b all go through $S^2$ and hence the
dash lines must correspond to world lines of the trivial particles (\ie
type-$\one$).  We then drop the three dash lines in Fig. \ref{O-O2}b.  Next, we
deform the two cluster of the world lines into two $\Th$ graphs, which give
rise to a phase factor since the world lines may get twisted (see Fig.
\ref{twist}).  Last we use \eqn{PhiO} to reduce the two $\Th$ graphs into two
loops of the $z$ world lines, which gives rise to a factor $ O^{i\bar
i;\mu\mu}_z O^{j\bar j;\nu\nu}_z$ (here the repeated indices $\mu\mu$ and
$\nu\nu$ are not summed).  The two $z$ loops produces $d_z^2$.

Collecting all those factors, we find that $\< \Psi_{i,j,z,\mu,\nu} |
\Psi_{i,j,z,\mu,\nu}\>$ is given by (where $\mu,\nu$ are not summed), up to a
phase factor,
\begin{align}\label{eq.norm}
&\ \ \ \
 Y^{i\bar i}_{\one;11}
 Y^{j\bar j}_{\one;11}
 Y^{z\bar z}_{\one;11} 
O^{i\bar i;\mu\mu}_z O^{j\bar j;\nu\nu}_z 
d_z^2 Z^\text{top}(\Si_2^\text{fill} \sqcup \Si_2^\text{fill})
\\
&=
\sqrt{\frac{d_1}{d_id_{\bar i}}}
\sqrt{\frac{d_1}{d_jd_{\bar j}}}
\sqrt{\frac{d_1}{d_zd_{\bar z}}}
\sqrt{\frac{d_id_{\bar i}}{d_z}}
\sqrt{\frac{d_jd_{\bar j}}{d_z}}
d_z^2 Z^\text{top}(\Si_2^\text{fill} \sqcup \Si_2^\text{fill})
\nonumber\\
&=Z^\text{top}(\Si_2^\text{fill} \sqcup \Si_2^\text{fill}),
\nonumber 
\end{align}
where $Z^\text{top}(\Si_2^\text{fill} \sqcup \Si_2^\text{fill})$ is the topological partition function on
$\Si_2^\text{fill} \sqcup \Si_2^\text{fill}$ without worldlines.

Since $\< \Psi_{i,j,z,\mu,\nu} | \Psi_{i,j,z,\mu,\nu}\>$ is positive,
the ambiguous phase factor must be $1$.
We see that the normalized ground state wave function is given by the
topological path integral on $\Si_2^\text{fill}$ with the world lines $i,j,z$:
\begin{align}\label{WF_genus2}
  | \psi_{i,j,z,\mu,\nu}\> = \frac{\cZ^\text{top}[\Si_2^\text{fill}(i,j,z,\mu,\nu)]}{\sqrt{Z^\text{top}(\Si_2^\text{fill} \sqcup \Si_2^\text{fill})}}.
\end{align}

This result allows us to express the
wave function overlap in terms of path integral on
$\Si_2^\text{fill} \sqcup \Si_2^\text{fill}$:
\begin{align}\label{WFoverlap_genus2}
&
W^{(i_{\cB},j_{\cB},z_{\cB},\mu_{\cB},\nu_{\cB}),(i_{\cA},j_{\cA},z_{\cA},\mu_{\cA},\nu_{\cA})}_{\cB\cA,2}=
\\
& \frac{Z^\text{top} [\Si_{2}^\text{fill}(\cA;i_{\cA},j_{\cA},z_{\cA},\mu_{\cA},\nu_{\cA}) \sqcup \Si_{2}^\text{fill}(\cB;i_{\cB},j_{\cB},z_{\cB},\mu_{\cB},\nu_{\cB})]}
{\sqrt{Z^\text{top}_{\Si_2;\cA}Z^\text{top}_{\Si_2;\cB}} }. 
\nonumber 
\end{align}
But now $\Si_{2}^\text{fill}(\cA;i_{\cA},j_{\cA},z_{\cA},\mu_{\cA},\nu_{\cA})$ is occupied by a topological phase $\cA$ and
the worldlines of $i_{\cA}$, $j_{\cA}$, $z_{\cA}$.  $\Si_{2}^\text{fill}(\cB;i_{\cB},j_{\cB},z_{\cB},\mu_{\cB},\nu_{\cB})$
is occupied by a topological phase $\cB$ and the worldlines of $i_{\cB}$,
$j_{\cB}$, $z_{\cB}$.  Also, $Z^\text{top}_{\Si_2;\cA}\equiv Z^\text{top}[\Si_{2}^\text{fill}(\cA) \sqcup \Si_{2}^\text{fill}(\cA)]$ is
the topological partition function on space-time $\Si^\text{fill}_{2}(\cA) \sqcup \Si^\text{fill}_{2}(\cA) $ filed with phase $\cA$ without worldlines.
Similarly, $Z^\text{top}_{\Si_2;\cB}\equiv Z^\text{top}[\Si_{2}(\cB)^\text{fill} \sqcup \Si_{2}^\text{fill}(\cB)]$ is
the topological partition function for phase $\cB$.

Using the same reasoning as in \eqn{MNNM1}, we find that the rectangular matrix
$W_{\cB\cA,2}$ satisfies
\begin{align}
\label{MNNM2}
 R^U_{\cB} W_{\cB\cA,2}=W_{\cB\cA,2} R^U_{\cA},
\end{align}
where $R^U_{\cA,\cB}$ are the representations of genus-$2$ MCG $\Ga_2$ for the
phase $\cA$ and $\cB$.

\begin{figure}[tb]
\centering
\includegraphics[scale=0.4]{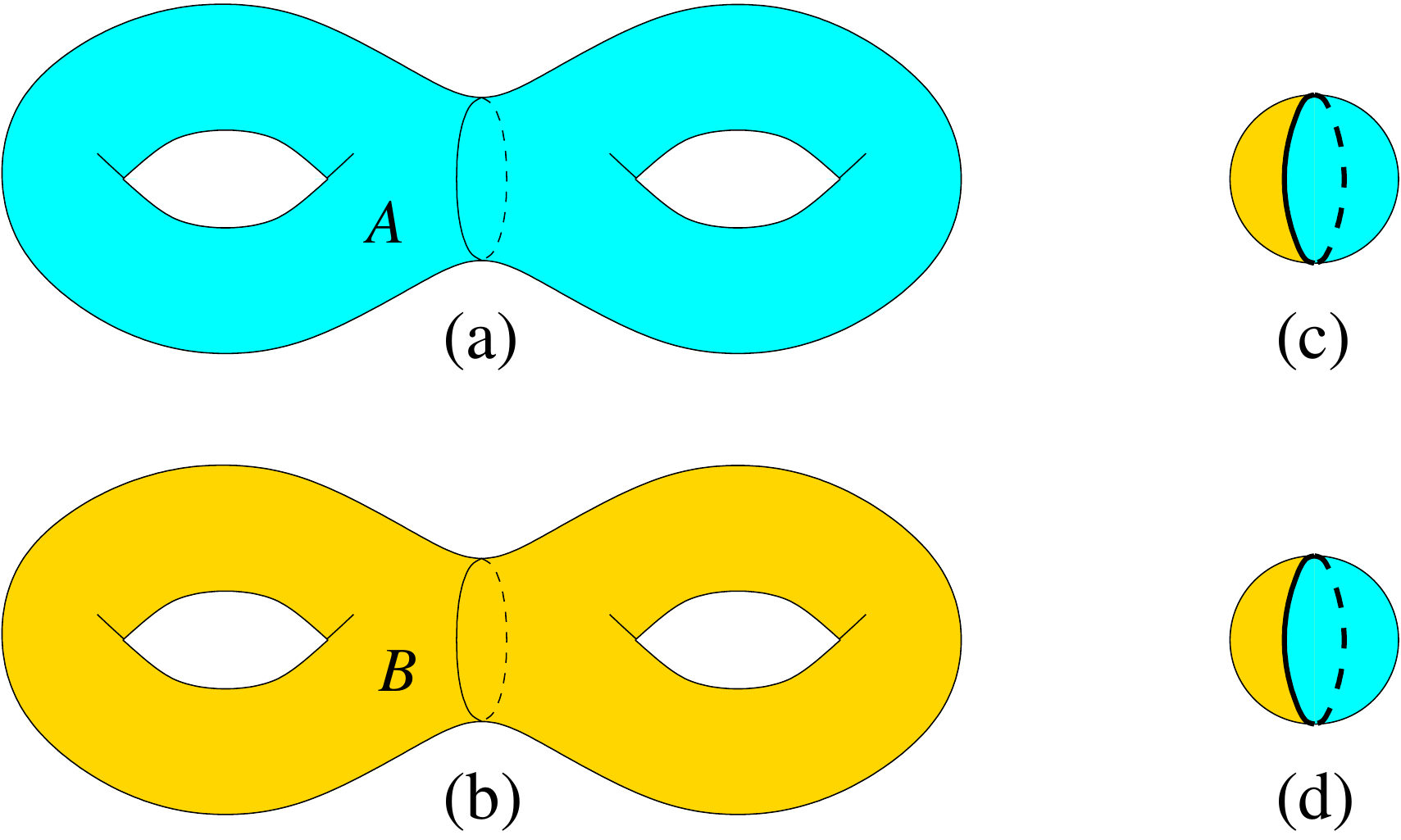}
\caption{
(a) and (b) are two 3-dimensional manifolds with topological order $\cA$ and $\cB$,
whose surfaces are genus-2 Riemann surface.  (c) and (d) are two 3-dimensional
balls.
} 
\label{g2s2} 
\end{figure}

\begin{figure}[tb]
\centering
\includegraphics[scale=0.4]{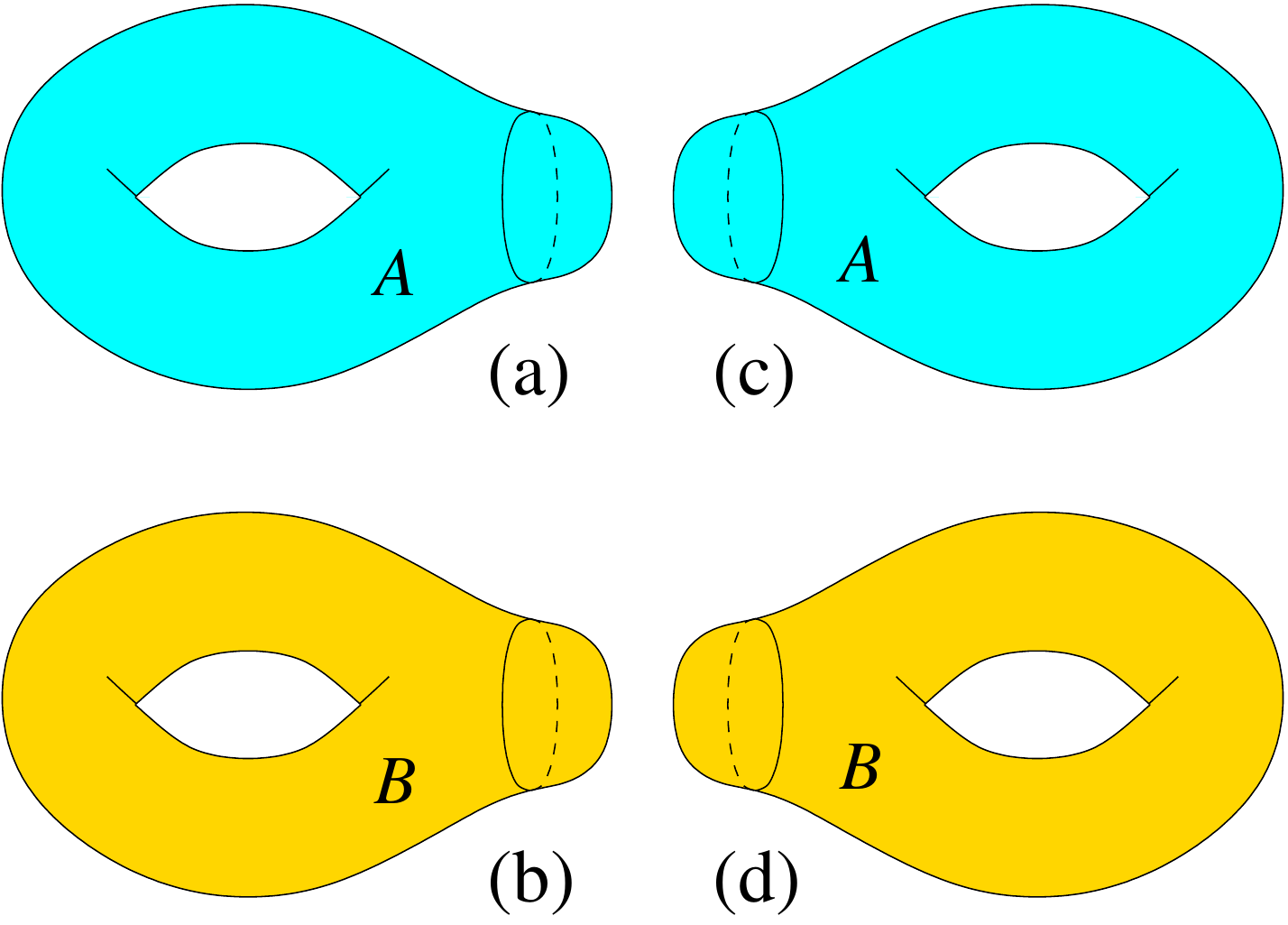}
\caption{
(a), (b), (c), and (d) are four 
3-dimensional solid tori  whose surfaces are genus-1 Riemann surface.}
\label{g1g1} 
\end{figure}

The wave function overlap on genus-2 surface $W_{\cB\cA,2}$ and the wave function
overlap on genus-1 surface $W_{\cB\cA,1}$ are related:
\begin{align}
\label{Genus2_Genus1a}
&\ \ \ \
 W^{(i_{\cB},j_{\cB},\one,1,1),(i_{\cA},j_{\cA},\one,1,1)}_{\cB\cA,2}
\sqrt{Z^\text{top}_{\Si_2;\cA}Z^\text{top}_{\Si_2;\cB}}\times 
\nonumber\\
&\ \ \ \ \ \ \ \ \ \ \
 Z^\text{top}[B^3(\cA)\sqcup B^3(\cB)]
\nonumber\\
&= W^{i_{\cB}i_{\cA}}_{\cB\cA,1} W^{j_{\cB}j_{\cA}}_{\cB\cA,1}
\end{align}
Here
$W^{(i_{\cB},j_{\cB},\one,1,1),(i_{\cA},j_{\cA},\one,1,1)}_{\cB\cA,2}\sqrt{Z^\text{top}_{\Si_2;\cA}Z^\text{top}_{\Si_2;\cB}}$
is the topological partition function on the space-time obtained by gluing Fig.
\ref{g2s2}a and Fig. \ref{g2s2}b together, and  $ Z^\text{top}[B^3(\cA)\sqcup
B^3(\cB)]$ is the topological partition function on the space-time obtained by
gluing Fig.  \ref{g2s2}c and Fig. \ref{g2s2}d together (which gives rise to a
$S^3$ occupied by phase $\cA$ and $\cB$).  Also, $W^{i_{\cB}i_{\cA}}_{\cB\cA,1}$
is the topological partition function on the space-time obtained by gluing Fig.
\ref{g1g1}a and Fig. \ref{g1g1}b together.  It is the topological partition
function on $S^3$ where half of $S^3$ is occupied by phase $\cA$, and the other
half by phase $\cB$.  $W^{j_{\cB}j_{\cA}}_{\cB\cA,1}$ is the topological
partition function on the space-time obtained by gluing Fig. \ref{g1g1}c and
Fig. \ref{g1g1}d together.

The equality of \eqn{Genus2_Genus1a} is obtained by noticing that the space-time
manifolds in Fig. \ref{g2s2} correspond to the space-time manifolds in Fig.
\ref{g1g1}, after some cutting and gluing.  To see this, we first cut Fig.
\ref{g2s2}a and Fig.  \ref{g2s2}b into left and right pieces.  Then gluing Fig.
\ref{g2s2}a(left), Fig. \ref{g2s2}b(left), and Fig. \ref{g2s2}c together along
their surfaces, we obtain a space-time obtained by gluing Fig. \ref{g1g1}a and
Fig. \ref{g1g1}b together along their surfaces.  Gluing Fig.
\ref{g2s2}a(right), Fig.  \ref{g2s2}b(right), and Fig.  \ref{g2s2}d together, we
obtain a space-time obtained by gluing Fig.  \ref{g1g1}c and Fig.  \ref{g1g1}d
together.

For convenience, we like to introduce a normalized wave function overlap via
\begin{align}
	\t{W}^{I_{\cB}J_{\cA}}_{\cB\cA,g}
	\equiv N_{\cB\cA,g} W^{I_{\cB}J_{\cA}}_{\cB\cA,g}
\end{align}
where we choose the normalization $N_{\cB\cA,g}$ such that $\hat
W^{I_{\cB}J_{\cA}}_{\cB\cA,g}=1$ when $I_{\cB},J_{\cA}$ correspond to trivial
wordlines, \ie for $g=1,2$ we have
\begin{align}
	\t{W}^{\one\one}_{\cB\cA,1}&=1,
	\nonumber\\
	\t{W}^{(\one,\one,\one,1,1),(\one,\one,\one,1,1)}_{\cB\cA,2}&=1.
\end{align}
Thus we have
\begin{align}
&\ \ \ \
	\t{W}^{i_{\cB}j_{\cA}}_{\cB\cA,1} = W^{i_{\cB}j_{\cA}}_{\cB\cA,1},
\end{align}
and
\begin{align}
&\ \ \ \
\t{W}^{(i_{\cB},j_{\cB},z_{\cB},\mu_{\cB},\nu_{\cB}),(i_{\cA},j_{\cA},z_{\cA},\mu_{\cA},\nu_{\cA})}_{\cB\cA,2}
\nonumber\\
&=
\sqrt{Z^\text{top}_{\Si_2;\cA}Z^\text{top}_{\Si_2;\cB}}
Z^\text{top}[B^3(\cA)\sqcup B^3(\cB)]
\nonumber\\
&\ \ \ \
W^{(i_{\cB},j_{\cB},z_{\cB},\mu_{\cB},\nu_{\cB}),(i_{\cA},j_{\cA},z_{\cA},\mu_{\cA},\nu_{\cA})}_{\cB\cA,2}
\end{align}
The normalized wave function overlap for genus-1 and genus-2 have a simpler
relation
\begin{align}
\label{Genus2_Genus1}
\t{W}^{(i_{\cB},j_{\cB},\one,1,1),(i_{\cA},j_{\cA},\one,1,1)}_{\cB\cA,2}
= \t{W}^{i_{\cB}i_{\cA}}_{\cB\cA,1} \t{W}^{j_{\cB}j_{\cA}}_{\cB\cA,1}
\end{align}

%

\section{The wave function overlap in a lattice Hamiltonian model}
\label{Z2example}

In this section we try to compute the wave function overlap in a concrete
lattice Hamiltonian model. Above, in the path integral formulation in Euclidean
spacetime, we learned the lesson that a gapped domain wall $W$ between phases
$\cA,\cB$ (a defect along space direction), after a proper Wick-rotation,
becomes an operator $\ee^{-H_W}$ (a defect along the time direction) that
determines the weighted wave function overlap.  However, in the Hamiltonian
formulation there is no natural equivalence between space and time. To apply
the previous results, we first try to analyze the physical picture in a lattice
model.

Assume that there is a gapped domain wall at $x=0$, phase $\cA$ in the region
$x<0$ a and phase $\cB$ in the region $x>0$. Clearly the Hamiltonian near $x=0$
can not be the same as those far from $x=0$. For simplicity, we may assume that
for some small positive number $\varepsilon$, the Hamiltonian is uniform  in
the regions $x<-\varepsilon$ and $x>\varepsilon$. On the other hand, in the
region $-\varepsilon<x<\varepsilon$, the Hamiltonian is uniform along $y$ and
$t$ directions, but changes with $x$. The Wick-rotated version of this picture,
is that there are uniform phases $\cA$ at time $t<-\varepsilon$ and $\cB$ at
time $t>\varepsilon$, and during time $-\varepsilon<t<\varepsilon$, the
Hamiltonian evolves from that of $\cA$ to that of $\cB$. Thus $\ee^{-H_W}$
should correspond to the accumulated evolution operator during
$-\varepsilon<t<\varepsilon$.

For the domain wall Hamiltonian in the region $-\varepsilon<x<\varepsilon$, the
requirement is that it should keep the spectrum of the whole system gapped.
However, we have no idea how this requirement is translated for $\ee^{-H_W}$.
There is further another subtlety, that phases $\cA,\cB$ may not be defined on
the same microscopic lattice, in particular, any generalized local unitary
transformations can be inserted to deform $\cA$ or $\cB$. In the followings, we
try to propose some reasonable assumptions that allows us to calculate the
quantity $W^{I_{\cB} I_{\cA}}_{\cB\cA,g}$.
\begin{itemize}
  \item First, to remove the ambiguity of generalized local unitary
    transformations, we assume that proper generalized local unitary
    transformations have been applied such that $\cA,\cB$ are already on the same
    microscopic lattice and their Hamiltonians satisfy the following relation;
  \item We focus on the gapped domain walls that are directly induced by anyon
    condensations. Suppose $H^\text{ac}$ is the Hermitian operator that forces anyon
    condensation in $\cA$. (If only Abelian anyons are condensed, $H^\text{ac}$ is
      simply the sum of hopping string operators of the condensed anyons, that
      forces a total zero-momentum state of the condensed anyons. We will give
    more detail later in the toric code example.) We require the Hamiltonian $H_{\cB}$ of
    phase $\cB$ to be:
    \[ H_{\cB}=H_{\cA}+ h H^\text{ac},\quad h\to +\infty.\]
    where $H_{\cA}$ is the Hamiltonian of phase $\cA$.
  \item Then we assume that $\ee^{-H_W}$ acts on the ground state subspace of
    $H_{\cA}$
    (or $H_{\cB}$) by multiplying a constant factor that may depend on the system
    size.
\end{itemize}

Now we look at the example of toric code model\cite{K032} which realize a $Z_2$
topological order:\cite{RS9173,W9164} 
\begin{align}
  H_{\cA}=-\sum_v \prod_{i\in star(v)}\sigma_i^z -\sum_p \prod_{i\in \partial
  p}\sigma_i^x.
\end{align}
The spins are on the links of the lattice. The first term sums over all
vertices and $star(v)$ denotes the legs of the vertex $v$. The second term sums
over all plaquettes and $\partial p$ denotes the boundary edges of the
plaquette $p$. There are three types of nontrivial anyons $e,m,f$: $e$ is
created by string operators of the form $\prod \sigma_i^x$ along links, $m$ is
created by string operators of the form $\prod \sigma_i^z$ along links on the
dual lattice, and $f$ is the fusion of $e$ and $m$. Both $e$ and $m$ can be
condensed. The term that forces the condensation of $e$ is
$H^\text{ac}_{e}=-\sum_i \sigma_i^x$, and that for $m$ is
$H^\text{ac}_{m}=-\sum_i \sigma_i^z$.  According to the above assumptions, we
next calculate the overlap between the ground states of $H_{\cA}$ and $H_{\cB}=H_{\cA} + h
H^\text{ac}_{e/m}$. It is easy to see that, under the $h\to+\infty$ limit, the
ground state of $H_{\cB}$ is simply the spin polarized state $ \otimes_i \frac
{|0\>_i +|1\>_i}{\sqrt2} = \otimes_i |+\>_i $ or $\otimes_i |0\>_i$.

\begin{figure}[tb]
  \centering
  \includegraphics{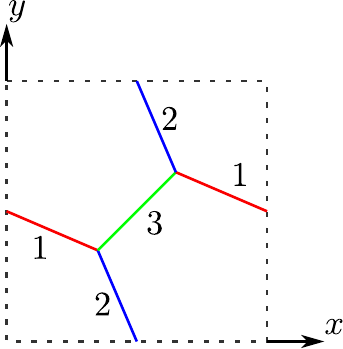}
  \caption{The simplest triangulated torus with three links and two vertices.}
  \label{fig:tc3bit}
\end{figure}

Now we write down the ground states of $\cA$ on a torus with the simplest lattice
of only three links (see Fig.~\ref{fig:tc3bit})
According to the anyon flux along $y$ direction (measured by string operators
winding around $x$ direction), the four ground states are
\begin{align}
  |\psi_\one\>&=\frac 1{\sqrt{2}} \left(|000\>+|101\>\right),\nonumber\\
  |\psi_e\>   &=\frac 1{\sqrt{2}} \left(|110\>+|011\>\right),\nonumber\\
  |\psi_m\>   &=\frac 1{\sqrt{2}} \left(|000\>-|101\>\right),\nonumber\\
  |\psi_f\>   &=\frac 1{\sqrt{2}} \left(|110\>-|011\>\right).
\end{align}
The corresponding ground states of $\cB$ are
\begin{align}
  e \text{ condenses:  } &|\psi^\text{ac}_{e}\>= |{+}{+}{+}\>,\nonumber\\
  m \text{ condenses:  } &|\psi^\text{ac}_{m}\>= |000\>.
\end{align}
We obtain:
\begin{align}
  \<\psi^\text{ac}_{e}|\psi_a\>&=\frac 12 (1,1,0,0),\nonumber\\
  \<\psi^\text{ac}_{m}|\psi_a\>&=\frac 1{\sqrt{2}} (1,0,1,0),\quad  a=\one,e,m,f.
\end{align}
which are indeed solutions to \eqref{commute} and \eqref{stable} after dropping
the prefactors. One reason for the different prefactors of $e,m$ condensations
is that the number of excitations the lattice can host is different. $e$ is
hosted by vertices while $m$ is hosted by plaquettes and there are two vertices
and only one plaquette in Fig.~\ref{fig:tc3bit}.

Let us generalize the above calculation to arbitrary system size and arbitrary
lattice.  We note that the four ground states on a torus for the toric-code
model $H_{\cA}$ are sum of all closed strings formed by $|1\>$'s on the links, with
fixed even-odd winding numbers in the two directions of torus, which are given
by $|ee\>$, $|eo\>$, $|oe\>$, and $|oo\>$.  The four ground states in the
quasiparticle basis are given by
\begin{align}
  |\psi_\one\>&=\frac 1{\sqrt{2}} \left(|ee\>+|oe\>\right),\nonumber\\
  |\psi_e\>   &=\frac 1{\sqrt{2}} \left(|eo\>+|oo\>\right),\nonumber\\
  |\psi_m\>   &=\frac 1{\sqrt{2}} \left(|ee\>-|oe\>\right),\nonumber\\
  |\psi_f\>   &=\frac 1{\sqrt{2}} \left(|eo\>-|oo\>\right).
\end{align}
The ground states for the two possible $\cB$ phases are
\begin{align}
 |\psi^\text{ac}_{e}\> = \bigotimes_i |+\>_i,\ \ \ \ \
 |\psi^\text{ac}_{m}\> = \bigotimes_i |0\>_i.
\end{align}
To compute the overlap between $|\psi^\text{ac}_{e,m}\>$ with
$|\psi_{1,e,m,f}\>$, let $N_l$ be the number of links, $N_v$ the number
vertices and $N_p$ the number of plaquettes of the triangulated torus (see Fig.
\ref{fig:tc3bit}). On the torus whose genus $g=1$, they satisfy
\begin{align}
  N_l=N_v+N_p.
\end{align}
We note that each of $|ee\>$, $|eo\>$, $|oe\>$, and
$|oo\>$ is an equal-weight  superposition of $2^{N_p-1}=2^{N_l-N_v-1}$ closed string
configurations. 

Since $|\psi^\text{ac}_{m}\>$ correspond to a single no-string configuration,
it only overlaps with $|ee\>$ (which contains a no-string configuration). We
find
\begin{align}
 \<\psi^\text{ac}_{m}|\psi_\one\> &= \frac 1{\sqrt{2} \sqrt{2^{N_p-1}} }
= 2^{-\frac{N_p}{2}}  ,
\nonumber\\
 \<\psi^\text{ac}_{m}|\psi_e\> &= 0 ,
\nonumber\\
 \<\psi^\text{ac}_{m}|\psi_m\> &= \frac 1{\sqrt{2} \sqrt{2^{N_p-1}} }
= 2^{-\frac{N_p}{2}}  ,
\nonumber\\
 \<\psi^\text{ac}_{m}|\psi_f\> &= 0.
\end{align}
After removing the area term $2^{-\frac{N_p}{2}}$, we see that the
wave function overlaps $W^{\one a}_{\cB\cA,1}$ are given by integers
$W^{\one a}_{\cB\cA,1}=(1,0,1,0)$, $a=\one,e,m,f$, when the $\cB$ phase is given by
$m$-particle condensation.

Also $|\psi^\text{ac}_{e}\>$ have the same overlap $2^{-N_l/2}$ with any string
configuration.  We find
\begin{align}
  \<\psi^\text{ac}_{e}|\psi_\one\> &= \frac{2\times 2^{N_p-1}\times 2^{-N_l/2}}
{\sqrt 2\sqrt{2^{N_p-1}} }
= 2^{\frac{N_p-N_l}{2}}=2^{-\frac{N_v}{2}}  , \nonumber\\
 \<\psi^\text{ac}_{e}|\psi_e\> &=  \frac {2\times 2^{N_p-1}\times
2^{-N_l/2}}{\sqrt 2\sqrt{2^{N_p-1}} }
= 2^{\frac{N_p-N_l}{2}}=2^{-\frac{N_v}{2}}  ,
\nonumber\\
 \<\psi^\text{ac}_{e}|\psi_m\> &=  \frac {2^{N_p-1}\times
2^{-N_l/2}-2^{N_p-1}\times 2^{-N_l/2}}{\sqrt 2\sqrt{2^{N_p-1}} }
= 0  ,
\nonumber\\
 \<\psi^\text{ac}_{e}|\psi_f\> &=  \frac {2^{N_p-1}\times
2^{-N_l/2}-2^{N_p-1}\times 2^{-N_l/2}}{\sqrt 2\sqrt{2^{N_p-1}} }
= 0  .
\end{align}
After removing the area term $2^{-\frac{N_v}{2}}$, we see that the
wave function overlaps $W^{\one a}_{\cB\cA,1}$ are given by integers
$W^{\one a}_{\cB\cA,1}=(1,1,0,0)$, $a=\one,e,m,f$, when the $\cB$ phase is given by
$e$-particle condensation.  

The above example supports our previous results that the area independent part
of wave function overlap on torus are universal and are given by integers.

From a more experimental point of view, suppose there is a small physical
system (say realized by a quantum simulator) where we can tune several
paremeters to force phase transitions, and the wave function overlaps can be
measured (by interference for example). If \emph{quantization} is observed in
the wave function overlaps, it may relate to the universal integer part as
above, and such quantization is a sign of topological order and anyon
condensation.


\section{Gapped boundaries of a 2+1D topological order}
\label{GapBound_Z2S3}

We can use the results developed in this paper to study the gapped boundaries
of 2+1D topological order, for example try to find out how many different
gapped boundaries a topological order can have.  In this section, we study some
simple examples.

\subsection{Boundary of $Z_2$ topological order}

Let us choose phase $\cA$ to be $Z_2$ topological order (denoted by $\cZ_2$)
and phase $\cB$ to be trivial.  A domain wall between them is a boundary of
$Z_2$ topological order.  The modular matrices for the $Z_2$ topological order
are given by
\begin{align}
  T_{\cZ_2}&=
\begin{pmatrix}
    1&0&0&0\\
    0&1&0&0\\
    0&0&1&0\\
    0&0&0&-1
  \end{pmatrix}, 
&
  S_{\cZ_2}&=\frac12 \begin{pmatrix}
    1&1&1&1\\
    1&1&-1&-1\\
    1&-1&1&-1\\
    1&-1&-1&1
  \end{pmatrix}.
\end{align}
in the basis $(\one,e,m,f)$.  The modular  matrices for the trivial topological
order are $S_{\cB}=1,\ T_{\cB}=1$.  From modular transformation for normalized
wave function overlap
\begin{align}
 \t{W}_{\cZ_2,g=1} = \t{W}_{\cZ_2,g=1} S_{\cZ_2},  \quad
 \t{W}_{\cZ_2,g=1} = \t{W}_{\cZ_2,g=1} T_{\cZ_2}.
\end{align}
we find two types of gapped boundaries, characterized by two integer vector
solutions of the above equation:
\begin{align}
\label{Toric_MeMm}
  \t{W}_{\cZ_2,g=1}^e&= (1,1,0,0),
\nonumber \\
  \t{W}_{\cZ_2,g=1}^m&= (1,0,1,0) .
\end{align}

Next, let us consider the case of a genus-2 manifold $\Sigma_{2}$: 
\begin{eqnarray}\label{DenhTwist5}
\small
\begin{tikzpicture}[baseline={(current bounding box.center)}]

\draw  [gray](-60*0.8pt,0pt) arc (-120:-60:40*0.8pt);
\draw  [gray](-52.5*0.8pt,-3*0.8pt) arc (130:50:20*0.8pt);
\draw  [gray](60*0.8pt,0pt) arc (-60:-120:40*0.8pt);
\draw  [gray](52.5*0.8pt,-3*0.8pt) arc (50:130:20*0.8pt);

\draw [thick][gray]
(50*0.8pt,25*0.8pt)..controls (65*0.8pt,25*0.8pt) and (82*0.8pt,15*0.8pt)..
(82.5*0.8pt,0pt).. controls (82*0.8pt,-15*0.8pt) and (65*0.8pt,-25*0.8pt).. 
(50*0.8pt,-25*0.8pt)..controls (25*0.8pt,-25*0.8pt) and (10*0.8pt,-15*0.8pt)..
(0pt,-15*0.8pt)..controls (-10*0.8pt,-15*0.8pt) and (-25*0.8pt,-25*0.8pt)..
(-50*0.8pt,-25*0.8pt)..controls (-65*0.8pt,-25*0.8pt) and (-82*0.8pt,-15*0.8pt)..(-82.5*0.8pt,0pt)..controls (-82*0.8pt,15*0.8pt) 
and (-65*0.8pt,25*0.8pt)..(-50*0.8pt,25*0.8pt)..controls (-25*0.8pt,25*0.8pt) 
and (-10*0.8pt,15*0.8pt)..(0pt,15*0.8pt).. controls (10*0.8pt,15*0.8pt) and (25*0.8pt,25*0.8pt)..(50*0.8pt,25*0.8pt);

\draw (30pt,-4.5pt) arc (120:240:8.6pt);
\draw[densely dashed] (30pt,-4.5pt) arc (60:-60:8.6pt);

\draw [densely dashed] (-30pt,-4.5pt) arc (120:240:8.6pt);
\draw (-30pt,-4.5pt) arc (60:-60:8.6pt);

\draw (40*0.8pt,15*0.8pt)..controls (75*0.8pt,13*0.8pt) and (75*0.8pt,-13*0.8pt)..(40*0.8pt,-15*0.8pt);
\draw (40*0.8pt,15*0.8pt)..controls (5*0.8pt,13*0.8pt) and (5*0.8pt,-13*0.8pt)..(40*0.8pt,-15*0.8pt);

\draw (-40*0.8pt,15*0.8pt)..controls (-75*0.8pt,13*0.8pt) and (-75*0.8pt,-13*0.8pt)..(-40*0.8pt,-15*0.8pt);
\draw (-40*0.8pt,15*0.8pt)..controls (-5*0.8pt,13*0.8pt) and (-5*0.8pt,-13*0.8pt)..(-40*0.8pt,-15*0.8pt);

\draw (-25*0.8pt,-2*0.8pt)..controls (-25*0.8pt,14*0.8pt) and (25*0.8pt,14*0.8pt)..(25*0.8pt,-2*0.8pt);
\draw [densely dashed] (-25*0.8pt,-2*0.8pt)..controls (-25*0.8pt,-14*0.8pt) and (25*0.8pt,-14*0.8pt)..(25*0.8pt,-2*0.8pt);

\node at (0pt,3pt){$\gamma$};

\node at (-36*0.8pt, -33*0.8+2pt){$a_2$};
\node at (36*0.8pt, -33*0.8+2pt){$a_1$};

\node at (-73*0.8pt, 0*0.8-1pt){$b_2$};
\node at (73*0.8pt, 0*0.8-1pt){$b_1$};

\draw[>=stealth,->] (0.5pt,8pt)--(0.52pt,8pt);
\draw[>=stealth,->] (39.52*0.8pt,15*0.8pt)--(39.5*0.8pt,15*0.8pt);
\draw[>=stealth,->] (-40.5*0.8pt,15*0.8pt)--(-40.52*0.8pt,15*0.8pt);

\draw[>=stealth,->] (-25.8pt,-11.2pt)--(-25.8pt,-11pt);
\draw[>=stealth,->] ( 25.8pt,-11.2pt)--( 25.8pt,-11pt);

\end{tikzpicture}
\end{eqnarray}
For MCG$(\Sigma_{2})$, 
there are five generators which are Dehn twists along the closed curves 
$a_1$, $b_1$, $a_2$, $b_2$,
and $\gamma$, as shown in \eqref{DenhTwist5}.
Denoting the projective representations of these five Dehn twists as 
$T_{a_1}, \, T_{b_1}, \, T_{a_2}, \, T_{b_2}$, and $T_{\gamma}$ respectively, 
we can use $T_{a_i}$ and $T_{b_i}$ to construct $S_i$ matrix that acts on the 
left (right) half of $\Sigma_{2}$, with $S_i=T_{b_i}\cdot T^{-1}_{a_i}\cdot T_{b_i}
=T_{a_i}^{-1}\cdot T_{b_i}\cdot T_{a_i}^{-1}$.
Then we have the following projective representations of
the five generators of MCG$(\Sigma_2)$:
\begin{equation}
T_1, \, S_1, \, T_2, \, S_2, \, T_{5}.
\end{equation}
where we have denoted $T_1:=T_{a_1}$, $T_2:=T_{a_2}$, and
$T_5:=T_{\gamma}$.
Then based on Eq.\eqref{MNNM2}, we have
\begin{equation}\label{Genus2_gapping_condition}
\t{W}_{\cZ_2,g=2}=\t{W}_{\cZ_2,g=2} \,\cdot R_{\cZ_2},
\end{equation}
where $R_{\cZ_2}=S_1, T_1, S_2, T_2, T_5$.
It is noted that for an Abelian theory, the basis in Fig.\ref{O-O} can be represented as 
\begin{eqnarray}\label{Genus_2_basis_abelian}
\begin{tikzpicture}[baseline={(current bounding box.center)}]
\draw[>=stealth,<-] (0pt,0pt) arc (180:-180:20pt) ;
\draw[>=stealth,<-] (50pt,0pt) arc (180:-180:20pt) ;
\node at (20pt, 25pt){$i$};
\node at (70pt, 26pt){$j$};
\end{tikzpicture}
\end{eqnarray}
That is, the anyon $z$ in Fig.\ref{O-O} corresponds to the identity anyon
$\mathbf{1}$ now.  Then the two anyon loops in \eqref{Genus_2_basis_abelian} are
decoupled.  With this basis, it is straightforward to check that for
$R_{\cZ_2}=S_1, T_1, S_2$ and $T_2$, the solutions to
Eq.\eqref{Genus2_gapping_condition} are simply tensor product of genus-1
solutions (see Eq.\eqref{Genus2_Genus1}):
\begin{equation}\label{4solution_Z2}
\begin{split}
\t{W}^{(1)}_{\cZ_2,g=2}=\t{W}_{\cZ_2,g=1}^e\otimes \t{W}_{\cZ_2,g=1}^e,\\
\t{W}^{(2)}_{\cZ_2,g=2}=\t{W}_{\cZ_2,g=1}^e\otimes \t{W}_{\cZ_2,g=1}^m,\\
\t{W}^{(3)}_{\cZ_2,g=2}=\t{W}_{\cZ_2,g=1}^m\otimes \t{W}_{\cZ_2,g=1}^e,\\
\t{W}^{(4)}_{\cZ_2,g=2}=\t{W}_{\cZ_2,g=1}^m\otimes \t{W}_{\cZ_2,g=1}^m,\\
\end{split}
\end{equation}
where $\t{W}_{\cZ_2,g=1}^e$ and $\t{W}_{\cZ_2,g=1}^m$ are the genus-1 results as expressed in Eq.\eqref{Toric_MeMm}.
However, the solutions $\t{W}^{(2)}_{\cZ_2,g=2}$ and $\t{W}^{(3)}_{\cZ_2,g=2}$ in \eqref{4solution_Z2} are illegal, 
since both phase $\cZ_2$ and phase $\cI$ are homogeneous here and we do not consider the case that  $e$-condensation
and $m$-condensation boundaries coexist.
In the following, we show that $\t{W}^{(2)}_{\cZ_2,g=2}$ and $\t{W}^{(3)}_{\cZ_2,g=2}$ are ruled out
by considering $R_{\cZ_2}=T_5$ in Eq.\eqref{Genus2_gapping_condition}.

For an Abelian theory, since the fusion result of $i\otimes \bar{j}$ is unique, 
the basis in \eqref{Genus_2_basis_abelian} can be rewritten as
\begin{eqnarray}\label{Genus_2_basis_abelian_fuse}
\begin{tikzpicture}[baseline={(current bounding box.center)}]
\draw[>=stealth,->] (10*0.7pt,17*0.7pt) arc (40:180:30*0.7pt);
\draw  (-43*0.7pt,-2*0.7pt) arc (180:320:30*0.7pt);
\draw [>=stealth,->] (10*0.7pt,17*0.7pt) arc (140:0:30*0.7pt);
\draw  (63*0.7pt,-2*0.7pt) arc (0:-140:30*0.7pt);
\draw (10*0.7pt,17*0.7pt) -- (10*0.7pt, -2*0.7pt);
\draw [>=stealth,<-] (10*0.7pt, -2*0.7pt) -- (10*0.7pt,-22*0.7pt);
\node at (-36*0.7+1pt,0pt){$i$};
\node at (55*0.7pt,0pt){$\bar{j}$};
\node at (16*0.7pt,0pt){{ $z'$}};
\end{tikzpicture}
\end{eqnarray}
with $i\otimes \bar{j}=z'$.
Then acting $T_5$ on the basis in \eqref{Genus_2_basis_abelian} or equivalently
\eqref{Genus_2_basis_abelian_fuse}
results in a phase $\theta_{\bar{z}'}=\theta_{z'}=e^{2\pi i s_{z'}}$ (see Fig.\ref{twist}).
In other words, the matrix $T_5$ in the basis \eqref{Genus_2_basis_abelian}
is a diagonal matrix with the diagonal elements being $\theta_{z'}$.
For $Z_2$ topological order, we have $\theta_{\mathbf{1}}=\theta_m=\theta_e=1$, 
and $\theta_f=-1$.
Given a vector $M_{\cZ_2, g=2}$, one can find that if the vector element corresponding to
$\theta_f=-1$ is non-zero, then it cannot be the solution of $\t{W}_{\cZ_2, g=2}=M_{\cZ_2, g=2}\cdot T_5$
in Eq.\eqref{Genus2_gapping_condition}.

More explicitly, let us take the vector $M^{(2)}_{\cZ_2,g=2}$ in Eq.\eqref{4solution_Z2} for example.
Denoting the basis in \eqref{Genus_2_basis_abelian} as 
\begin{equation}
\begin{split}
\{ij\}=&(\mathbf{1}, e, m, f)\otimes (\mathbf{1}, e, m, f)\\
=&
(\mathbf{11}, \mathbf{1}e, \mathbf{1}m, \mathbf{1}f;\,\, e\mathbf{1}, ee, em, ef;\\
&m\mathbf{1}, me, mm, mf;\,\, f\mathbf{1}, fe, fm, ff),
\end{split}
\end{equation}
then $T_5$ is a diagonal matrix of the form
\begin{equation}
\begin{split}
\text{Diag}(T_5)
=&
(1, 1, 1, -1;\,\, 1, 1, -1, 1;\,\,
1, -1, 1, 1;\,\, -1, 1, 1, 1),
\end{split}
\end{equation}
and
\begin{equation}
\begin{split}
\t{W}^{(2)}_{\cZ_2,g=2}=\big(1, 1, 0, 0;\,\, 0,0, 0, 0;\,\,
1, 1, 0, 0;\,\, 0, 0, 0, 0).
\end{split}
\end{equation}
One can check explicitly that $\t{W}^{(2)}_{\cZ_2,g=2}\neq \t{W}^{(2)}_{\cZ_2,g=2}\cdot T_5$.
Similarly, one can find that $\t{W}^{(3)}_{\cZ_2,g=2}\neq \t{W}^{(3)}_{\cZ_2,g=2}\cdot T_5$,
$\t{W}^{(1)}_{\cZ_2,g=2}= \t{W}^{(1)}_{\cZ_2,g=2}\cdot T_5$, and 
$\t{W}^{(4)}_{\cZ_2,g=2}= \t{W}^{(4)}_{\cZ_2,g=2}\cdot T_5$. 
That is, only $\t{W}^{(1)}_{\cZ_2,g=2}$ and $\t{W}^{(4)}_{\cZ_2,g=2}$
in Eq.\eqref{4solution_Z2} are the true solutions of Eq.\eqref{Genus2_gapping_condition}
by considering all the five generators $R_{\cZ_2}=T_1, S_1,T_2, S_2$ and $T_5$.

\begin{table*}[!ht]
\centering
\begin{tabular}{c||c|c|c|c|c|c|c|c}
$\otimes$ & $\bm 1$ &  $a^1$  & $a^2$ &  $b$  &  $b^1$  &  $b^2$  & $c$  & $c^1$    \\
\hline
\hline
$ \bm 1 $ & $\bm 1$ & $a^1$ & $a^2$   & $b$  & $b^1$  &  $b^2$          & $c$  & $c^1$    \\
$a^1$ & $a^1$ & $\bm 1$ & $a^2$	  & $b$  &$b^1$  & $b^2$      &  $c^1$  & $c$  \\
$a^2$ & $a^2$ & $a^2$   & $\bm 1\oplus a^1\oplus a^2$      & $b^1\oplus b^2$    & $c\oplus b^2$ & $b\oplus b^1$      & $c\oplus c^1$  & $c\oplus c^1$\\
$b$  & $b$  & $b$ & $b^1\oplus b^2$     & $\bm 1\oplus a^1\oplus b$ & $b^2\oplus a^2$  & $b^1\oplus a^2$     & $c\oplus c^1$   & $c\oplus c^1$  \\
$b^1$  & $b^1$   & $b^1$ & $b\oplus b^2$      & $b^2\oplus a^2$ & $\bm 1\oplus a^1\oplus b^1$  & $b\oplus a^2$      & $c\oplus c^1$ & $c\oplus c^1$ \\
$b^2$  & $b^2$  & $b^2$  & $b\oplus b^1$    & $b^1\oplus a^2$ & $b\oplus a^2$  & $\bm 1\oplus a^1\oplus b^2$      & $c\oplus c^1$  & $c\oplus c^1$ \\
$c$ & $c$ & $c^1$ & $c\oplus c^1$     & $c\oplus c^1$   & $c\oplus c^1$  & $c\oplus c^1$     & $\bm 1\oplus a^2\oplus b\oplus b^1\oplus b^2$ & $a^1 \oplus a^2\oplus b\oplus b^1\oplus b^2$  \\
$c^1$ & $c^1$ & $c$  & $c\oplus c^1$     & $c\oplus c^1$ & $c\oplus c^1$  & $c\oplus c^1$     & $a^1 \oplus a^2\oplus b\oplus b^1\oplus b^2$  & $\bm 1\oplus a^2\oplus b\oplus b^1\oplus b^2$ \\
\end{tabular}
\caption{Fusion rules $N^{ab}_c$ of 2+1D $S_3$ topological order. Here $b$
and $c$ correspond to pure flux excitations, $a^1$ and $a^2$ pure charge
excitations, $\bm 1$ the vacuum sector while  $b^1$, $b^2$, and $c^1$ are
charge-flux composites. 
}\label{S3FusionRules}
\end{table*}

\subsection{Boundary of $\cS_3$ topological order}

$\cS_3$ topological order (described by quantum double of finite group $S_3$
with fusion rule given by Table \ref{S3FusionRules}) is more interesting, since
it is a non-abelian theory.  Let us choose phase $\cA$ to be $\cS_3$ topological
order (denoted as $\cS_3$) and phase $\cB$ to be trivial  (denoted as $\cI$).
The modular matrices for the $\cS_3$ topological order are given by (with basis
$(\one, a^1, a^2,b,b^1,b^2,c,c^1)$)
\begin{align}\label{ST_S3}
  \text{Diag}(T)&=(1,1,1,1,\ee^{\frac{2\pi\ii}{3}},\ee^{-\frac{2\pi\ii}{3}},1,-1),\\
  \small
  S&=\frac{1}{6}
  \begin{pmatrix}
    1&1&2&2&2&2&3&3\\
    1&1&2&2&2&2&-3&-3\\
    2&2&4&-2&-2&-2&0&0\\
    2&2&-2&4&-2&-2&0&0\\
    2&2&-2&-2&-2&4&0&0\\
    2&2&-2&-2&4&-2&0&0\\
    3&-3&0&0&0&0&3&-3\\
    3&-3&0&0&0&0&-3&3
  \end{pmatrix}.
\end{align}
\Eqn{commute} has five solutions with $M_{\cS_3}^{\mathbf{11}}=1$:
\begin{equation}\label{Genus1_solution}
\begin{split}
  \t{W}_{\cS_3,g=1}^{(1)}&= \begin{pmatrix} 1&0&0&1&0&0&1&0 \end{pmatrix},\\
  \t{W}_{\cS_3,g=1}^{(2)}&= \begin{pmatrix} 1&0&1&0&0&0&1&0 \end{pmatrix},\\
  \t{W}_{\cS_3,g=1}^{(3)}&= \begin{pmatrix} 1&1&0&2&0&0&0&0 \end{pmatrix},\\
  \t{W}_{\cS_3,g=1}^{(4)}&= \begin{pmatrix} 1&1&2&0&0&0&0&0 \end{pmatrix},\\
  \t{W}_{\cS_3,g=1}^{(5)}&= \begin{pmatrix} 1&1&1&1&0&0&0&0 \end{pmatrix}.\\
  \end{split}
\end{equation}
It is found that the last solution does not satisfy the stable condition in \eqn{stable}, \ie
\begin{align}\label{Stable_S3}
 (\t{W}_{\cS_3,g=1}^{(5)})^{a^2}
 (\t{W}_{\cS_3,g=1}^{(5)})^{b}  \not \leq \sum_i N^{a^2,b}_i (\t{W}_{\cS_3,g=1}^{(5)})^{i}.
\end{align}
So the $S_3$ topological order has only 4 types of gapped boundaries.

In fact, as discussed in the following, without resorting to the stable
condition in \eqref{stable}, we show that the fake solution $\t{W}_{\cS_3,g=1}^{(5)}$
can be ruled out by the genus-2 gapping boundary condition in Eq.\eqref{MNNM2},
\textit{i.e.},
\begin{equation}\label{Genus2_gapping_condition_S3}
\t{W}_{\cS_3,g=2}=\t{W}_{\cS_3,g=2} \,\cdot R_{\cS_3},
\end{equation}
which is a linear condition.
Since the $S_3$ topological order is multiplicity-free (\textit{i.e.}, 
$N^{ij}_k\le 1$), we denote the component of $\t{W}_{\cS_3,g=2}$ as
$\t{W}^{i,j,z}_{\cS_3,g=2}$ (See also Eq.\eqref{WFoverlap_genus2}).
Here the anyon types $i,\, j,\, z$ are used to label the basis vectors in the Hilbert space
of degenerate ground states on a genus-2 manifold $\Sigma_2$ (see Fig.\ref{O-O}).
It is noted that both $\t{W}^{i,j,z}_{\cS_3,g=2}$ and the projective representations $R_{\cS_3}^U$ of
MCG($\Sigma_2$) depend on the choice of basis vectors (see Appendix \ref{Appendix: S3}).
In the following discussion, we use the basis vectors in Fig.\ref{O-O}.

Similar to the previous subsection on $Z_2$ topological order, 
we check whether the genus-1 solutions in Eq.\eqref{Genus1_solution} can be embedded in the 
genus-2 solutions of Eq.\eqref{Genus2_gapping_condition_S3}. 
Our logic is as follows:
First, it is straightforward to find that $\t{W}_{\cS_3,g=1}^{(I)}\otimes
\t{W}_{\cS_3,g=1}^{(J)}$
(where $I,\, J=1,\,2,\, 3,\, 4,\, 5$) exhaust all the solutions $\t{W}^{i,j,z}_{\cS_3,g=2}$ of Eq.\eqref{Genus2_gapping_condition_S3}
 with $z=\mathbf{1}$  if we simply consider $R_{\cS_3}=T_1,\, S_1,\, T_2,\,S_2$.
Second, we need to check if $\t{W}_{\cS_3,g=1}^{(I)}\otimes   \t{W}_{\cS_3,g=1}^{(J)}$ are solutions of 
Eq.\eqref{Genus2_gapping_condition_S3} for $R_{\cS_3}=T_5$. In general, the operation
$R_{\cS_3}=T_5$ will mix the components $\t{W}_{\cS_3,g=2}^{i,j,z= \mathbf{1}}$ with 
$\t{W}_{\cS_3,g=2}^{i',j',z\neq \mathbf{1}}$, and it is not apparent that 
$\t{W}_{\cS_3,g=1}^{(I)}\otimes   \t{W}_{\cS_3,g=1}^{(J)}$ are solutions of 
Eq.\eqref{Genus2_gapping_condition_S3} for $R_{\cS_3}=T_5$.
In this case, a careful study of Eq.\eqref{Genus2_gapping_condition_S3} is needed. 
Third, we need to solve for the components $\t{W}^{i,j,z}_{\cS_3,g=2}$ with $z\neq \mathbf{1}$ in 
Eq.\eqref{Genus2_gapping_condition_S3}.

In the following, we solve all the components $\t{W}^{i,j,z}_{\cS_3,g=2}$ that are solutions 
of Eq.\eqref{Genus2_gapping_condition_S3}.
The results can be mainly summarized as follows: 

\begin{enumerate}

\item

It is found there are in total $4$ sets of independent solutions of the genus-2 condition in Eq.\eqref{Genus2_gapping_condition_S3},
which we denote as $\t{W}_{\cS_3,g=2}^{(i)}$, with $i=1,\,2,\, 3,\, 4$.
They embed the genus-1 solutions of the form $ \t{W}_{\cS_3,g=1}^{(i)}\otimes
\t{W}_{\cS_3,g=1}^{(i)}$ with $i=1,\, 2, \, 3,\,4$.
Other pairings of $\t{W}_{\cS_3,g=1}^{(i)}\otimes   \t{W}_{\cS_3,g=1}^{(j)}$ (including 
$\t{W}_{\cS_3,g=1}^{(5)}\otimes   \t{W}_{\cS_3,g=1}^{(5)}$) are ruled out by 
Eq.\eqref{Genus2_gapping_condition_S3}, as summarized in Table.\ref{S3_Genus2_Genus1}.

 \begin{table}[!ht]
 \small
\centering
\begin{tabular}{c||c|c|c|c|c|c|c|c}
$\otimes$ & $\t{W}_{\cS_3,g=1}^{(1)}$ &  $\t{W}_{\cS_3,g=1}^{(2)}$  &
$\t{W}_{\cS_3,g=1}^{(3)}$ &  $\t{W}_{\cS_3,g=1}^{(4)}$  &  $\t{W}_{\cS_3,g=1}^{(5)}$    \\
\hline
\hline
$ \t{W}_{\cS_3,g=1}^{(1)} $ & $\checkmark$ & $\times$  &$\times$   & $\times$   &  $\times$       \\
\hline
$\t{W}_{\cS_3,g=1}^{(2)}$ &  & $\checkmark$ & $\times$  &$\times$ &$\times$   \\
\hline
$\t{W}_{\cS_3,g=1}^{(3)}$ &   &    & $\checkmark$      &$\times$    &$\times$\\
\hline
$\t{W}_{\cS_3,g=1}^{(4)}$  &  &   &     & $\checkmark$ &$\times$ \\
\hline
$\t{W}_{\cS_3,g=1}^{(5)}$  &  &   &        &   &$\times$ \\
\hline
\end{tabular}
\caption{ Only four pairings of genus-1 solutions are allowed by the genus-2 condition. }
\label{S3_Genus2_Genus1}
\end{table}

\item

For the $4$ sets of independent solutions $\t{W}_{\cS_3,g=2}^{(i)}$ with $i=1,\,2,\, 3,\,4$,
all the components can be \textit{uniquely} determined as follows (see Appendix \ref{Appendix: S3}
for details):

-- For $\t{W}_{\cS_3,g=2}^{(1)}$, which embeds the genus-1 solution of the
form $\t{W}_{\cS_3,g=1}^{(1)}\otimes   \t{W}_{\cS_3,g=1}^{(1)}$,
 the condensed anyons are pure flux $\mathbf{1}$, $b$, and $c$. With the genus-2 condition in
Eq.\eqref{Genus2_gapping_condition_S3}, one can obtain the nonzero components of
$\t{W}^{i,j,z}_{\cS_3,g=2}$ with $z\neq \mathbf{1}$ as
\begin{equation}
\small
\left\{
\begin{split}
&\t{W}_{\cS_3, g=2}^{b,c,z=b}=\t{W}_{\cS_3, g=2}^{c,b,z=b}=1, \\
&\t{W}_{\cS_3, g=2}^{c,c,z=b}=\sqrt{2},\\
&\t{W}_{\cS_3, g=2}^{b,b,z=b}=\frac{1}{\sqrt{2}}.
\end{split}
\right.\nonumber
\end{equation}
The nonzero components $\t{W}_{\cS_3, g=2}^{i,j,z}$  with $z= \mathbf{1}$ can be expressed
as the product of genus-1 results as
$\t{W}_{\cS_3, g=2}^{i,j,z=\mathbf{1}}=(\t{W}_{\cS_3,g=1}^{(1)})^i \cdot
(\t{W}_{\cS_3,g=1}^{(1)})^j$, where 
$i,j\in\{\mathbf{1}, b, c\}$.
For all the non-zero components $\t{W}_{\cS_3,g=2}^{i,j,z}$, one can find that $i, j, z\in\{\mathbf{1}, b, c\}$.

-- For $\t{W}_{\cS_3,g=2}^{(2)}$, which embeds the genus-1 solution of the
form $\t{W}_{\cS_3,g=1}^{(2)}\otimes   \t{W}_{\cS_3,g=1}^{(2)}$,
the condensed anyons are $\mathbf{1}$, $a^2$, and $c$. 
One can obtain the nonzero component of $\t{W}^{i,j,z}_{\cS_3,g=2}$ with $z\neq \mathbf{1}$ as
\begin{equation}
\small
\left\{
\begin{split}
&\t{W}_{\cS_3, g=2}^{a^2,c,z=a^2}=\t{W}_{\cS_3,g=2}^{c,a^2,z=a^2}=-1, \\
&\t{W}_{\cS_3,g=2}^{c,c,z=a^2}=\sqrt{2},\\
&\t{W}_{\cS_3,g=2}^{a^2,a^2,z=a^2}=\frac{1}{\sqrt{2}}.
\end{split}
\right.\nonumber
\end{equation}
The nonzero components $\t{W}_{\cS_3, g=2}^{i,j,z}$  with $z= \mathbf{1}$ can be expressed
as the product of genus-1 results as
$\t{W}_{\cS_3, g=2}^{i,j,z=\mathbf{1}}=(\t{W}_{\cS_3,g=1}^{(2)})^i \cdot
(\t{W}_{\cS_3,g=1}^{(2)})^j$, where 
$i,j\in\{\mathbf{1}, a^2, c\}$.
Again, for all the non-zero components $\t{W}_{\cS_3,g=2}^{i,j,z}$, one has $i, j, z\in\{\mathbf{1}, a^2, c\}$.

-- For $\t{W}_{\cS_3,g=2}^{(3)}$, which embeds the genus-1 solution of the
form $\t{W}_{\cS_3,g=1}^{(3)}\otimes   \t{W}_{\cS_3,g=1}^{(3)}$, 
the condensed anyons are $\mathbf{1},\, a^1$, and $b$.
As studied in Appendix \ref{Appendix: S3}, all the components $\t{W}_{\cS_3, g=2}^{i,j,z}$  with $z\neq \mathbf{1}$ 
are zero. The only non-zero components can be considered as the product of genus-1 results as
$\t{W}_{\cS_3, g=2}^{i,j,z=\mathbf{1}}=(\t{W}_{\cS_3,g=1}^{(3)})^i \cdot
(\t{W}_{\cS_3,g=1}^{(3)})^j$, where 
$i,j\in\{\mathbf{1}, a^1, b\}$.

-- For $\t{W}_{\cS_3,g=2}^{(4)}$, which embeds the genus-1 solution of the
form $\t{W}_{\cS_3,g=1}^{(4)}\otimes   \t{W}_{\cS_3,g=1}^{(4)}$,
the condensed anyons are pure charges with $\mathbf{1},\, a^1$, and $a^2$.
Similar to the case of $\t{W}_{\cS_3,g=2}^{(3)}$, all the components $\t{W}_{\cS_3, g=2}^{i,j,z}$ with $z\neq \mathbf{1}$ are zero, 
and the only non-zero components correspond to the product of genus-1 results as
$\t{W}_{\cS_3, g=2}^{i,j,z=\mathbf{1}}=(\t{W}_{\cS_3,g=1}^{(4)})^i \cdot
(\t{W}_{\cS_3,g=1}^{(4)})^j$,
where $i,j\in\{\mathbf{1}, a^1, a^2\}$.

\end{enumerate}

\subsubsection{Rule out the fake solution with genus-2 condition}
\label{Sec: RuleOutFakeSolution}

As an illustration of how to determine the solutions of the genus-2 condition in 
Eq.\eqref{Genus2_gapping_condition_S3},
here we give an example on how $\t{W}_{\cS_3,g=1}^{(5)}\otimes   \t{W}_{\cS_3,g=1}^{(5)}$ is ruled out.
The details of finding \textit{all} the solutions of Eq.\eqref{Genus2_gapping_condition_S3}
can be found in Appendix \ref{Appendix: S3}.

It is convenient to consider the following two choices of basis vectors:
\begin{eqnarray}\label{BasisI}
\text{basis I:}\quad
\begin{tikzpicture}[baseline={(current bounding box.center)}]
\draw[>=stealth,<-] (0pt,0pt) arc (180:-180:20pt) ;
\draw[>=stealth,<-] (60pt,0pt) arc (180:-180:20pt) ;
\draw [>=stealth,->] (60pt,0pt)--(50pt,0pt);
\draw (50pt,0pt)--(40pt,0pt);
\node at (20pt, 25pt){$i$};
\node at (80pt, 25pt){$j$};
\node at (50pt, 5pt){$z$};
\node at (35pt, 0pt){\small $\nu$};
\node at (65pt, 0pt){\small $\mu$};
\end{tikzpicture}
\end{eqnarray}
and
\begin{eqnarray}\label{BasisII}
\text{basis II:}\quad
\begin{tikzpicture}[baseline={(current bounding box.center)}]
\draw[>=stealth,->] (10*0.7pt,17*0.7pt) arc (40:180:30*0.7pt);
\draw  (-43*0.7pt,-2*0.7pt) arc (180:320:30*0.7pt);
\draw (10*0.7pt,17*0.7pt) arc (140:0:30*0.7pt);
\draw [>=stealth,<-] (63*0.7pt,-2*0.7pt) arc (0:-140:30*0.7pt);
\draw (10*0.7pt,17*0.7pt) -- (10*0.7pt, -2*0.7pt);
\draw [>=stealth,<-] (10*0.7pt, -2*0.7pt) -- (10*0.7pt,-22*0.7pt);
\node at (-35*0.7pt,0pt){$i$};
\node at (55*0.7pt,0pt){$j$};
\node at (15*0.7pt,0pt){{ $z$}};
\node at (7.5pt, 18pt){\small $\mu$};
\node at (7.5pt, -20pt){\small $\nu$};
\end{tikzpicture}
\end{eqnarray}
The concrete expression of Eq.\eqref{Genus2_gapping_condition_S3} will depend on the choices of basis vectors.
For example, choosing $R=T_5$ in Eq.\eqref{Genus2_gapping_condition_S3}, $T_5$ is a diagonal matrix in the 
basis II, but in general not diagonal in basis I.
Denoting the wave function on a genus-2 manifold after anyon condensation as $|\Psi^{ac}_{g=2}\rangle$, then
the wavefunciton overlap $\t{W}^{i,j,z}_{\cS_3,g=2}$ in Eq.\eqref{Genus2_gapping_condition_S3} depends on the choice of bases as follows
\begin{equation}\label{Def_W}
\t{W}^{\text{I (II)}; i, j, z}_{\cS_3,g=2}:=\langle \Psi^{ac}_{g=2}|\psi^{\text{I (II)}; i, j, z}\rangle,
\end{equation}

 Suppose $\t{W}_{\cS_3,g=1}^{(5)}\otimes   \t{W}_{\cS_3,g=1}^{(5)}$ is the solution of Eq.\eqref{Genus2_gapping_condition_S3}, then 
 based on the expression in \eqref{Genus1_solution}, we have
 $\t{W}^{\text{I}; a_2, b, z=\mathbf{1}}_{\cS_3,g=2}=1$. In the following, we will show that the genus-2
condition imposes that  $\t{W}^{\text{I}; a_2, b, z=\mathbf{1}}_{\cS_3,g=2}=0$,
 and therefore $\t{W}_{\cS_3,g=1}^{(5)}\otimes   \t{W}_{\cS_3,g=1}^{(5)}$ cannot be the solution of Eq.\eqref{Genus2_gapping_condition_S3}.

To study the effect of $T_5$, it is convenient to consider the basis vectors II in \eqref{BasisII}.
Let us focus on the components with $i=a^2$ and $j=b$. 
(Recall that $T_5$ does not change $i$ and $j$.)
Considering the fusion rule
$a^2\otimes b=b_1\oplus b_2$, then in the basis vectors $|\psi^{\text{II}; a^2, b, z}\rangle$, 
the anyon type $z$ can only be chosen as
$b_1$ or $b_2$, which have non-trivial topological spins [see \eqref{ST_S3}].
Then, considering $R_{\cS_3}=T_5$ in Eq.\eqref{Genus2_gapping_condition_S3}, we have 
\begin{equation}
\t{W}_{\cS_3,g=2}^{\text{II}; a_2, b, z=b^1}=\t{W}_{\cS_3,g=2}^{\text{II}; a_2, b, z=b^2}=0.
\end{equation}
By inserting a complete set of basis vectors I in the expression \eqref{Def_W},
$\t{W}^{\text{II}; a_2, b, z}$ can be expressed as
$
\t{W}_{\cS_3,g=2}^{\text{II}; a_2, b, z=b^1}=\t{W}_{\cS_3,g=2}^{\text{I}; a^2, b, z=\mathbf{1}}\cdot \langle \psi^{\text{I}; a^2,b,\mathbf{1}}|\psi^{\text{II}; a^2,b,b^1}\rangle
+\t{W}_{\cS_3,g=2}^{\text{I}; a^2, b, z=a^1}\cdot \langle \psi^{\text{I}; a^2,b,a^1}|\psi^{\text{II}; a^2,b,b^1}\rangle$, and 
$\t{W}_{\cS_3,g=2}^{\text{II}; a_2, b, z=b^2}=\t{W}_{\cS_3,g=2}^{\text{I}; a^2, b, z=\mathbf{1}}\cdot \langle \psi^{\text{I}; a^2,b,\mathbf{1}}|\psi^{\text{II}; a^2,b,b^2}\rangle
+\t{W}_{\cS_3,g=2}^{\text{I}; a^2, b, z=a^1}\cdot \langle \psi^{\text{I}; a^2,b,a^1}|\psi^{\text{II}; a^2,b,b^2}\rangle$,
where we have considered the fusion rules 
$a^2\otimes a^2=\mathbf{1}\oplus a^1\oplus a^2$, and $b\otimes b=\bm 1\oplus a^1\oplus b$.
In the appendix (see Eq.\ref{Wza1_appendix}), 
one can explicitly show that $\t{W}_{\cS_3,g=2}^{\text{I}; a^2, b, z=a^1}=0$ by considering the genus-2 condition in 
Eq.\eqref{Genus2_gapping_condition_S3} with $R_{\cS_3}=S_1$ or $S_2$, which correspond to the punctured $S$ matrix.
Then one arrives at
\begin{equation}\label{W1_zero}
\left\{
\begin{split}
\t{W}_{\cS_3,g=2}^{\text{I}; a^2, b, z=\mathbf{1}}\cdot \langle \psi^{\text{I}; a^2,b,\mathbf{1}}|\psi^{\text{II}; a^2,b,b^1}\rangle=0,\\
\t{W}_{\cS_3,g=2}^{\text{I}; a^2, b, z=\mathbf{1}}\cdot \langle \psi^{\text{I}; a^2,b,\mathbf{1}}|\psi^{\text{II}; a^2,b,b^2}\rangle=0.
\end{split}
\right.
\end{equation}
Furthermore, since $\langle \psi^{\text{I}; a^2,b,\mathbf{1}}|\psi^{\text{II}; a^2,b,b^1}\rangle=\sqrt{d_{a^2} \, d_b \, d_{b^1}}/D$
and $\langle \psi^{\text{I}; a^2,b,\mathbf{1}}|\psi^{\text{II}; a^2,b,b^2}\rangle=\sqrt{d_{a^2} \, d_b \, d_{b^2}}/D$, which are nonzero.
Then based on Eq.\eqref{W1_zero}, one can immediately obtain 
\begin{equation}\label{Zero_solution}
\t{W}_{\cS_3,g=2}^{\text{I}; a^2, b, z=\mathbf{1}}
=0.
\end{equation}
On the other hand, we know that if $\t{W}_{\cS_3,g=1}^{(5)}\otimes   \t{W}_{\cS_3,g=1}^{(5)}$ is the solution of Eq.\eqref{Genus2_gapping_condition_S3}, 
then we have $\t{W}^{\text{I}; a_2, b, z=\mathbf{1}}_{\cS_3,g=2}=1$, which contradicts with \eqref{Zero_solution}. 
Therefore, $\t{W}_{\cS_3,g=1}^{(5)}\otimes   \t{W}_{\cS_3,g=1}^{(5)}$ is ruled out as the solution of 
genus-2 condition in Eq.\eqref{Genus2_gapping_condition_S3}.

One can refer to Appendix \ref{Appendix: S3} for
 \textit{all} the solutions of Eq.\eqref{Genus2_gapping_condition_S3}
 for $\cS_3$ topological order.

\section{Summary}

In this paper, we develop a systematic approach to the gapped domain walls
between two topological orders $\cA$ and $\cB$.  (If $\cB$ is the trivial topological
order, the domain wall becomes the boundary of topological order $\cA$.) Our
systematic approach is based on the topological partition function $W^{I_{\cB}
I_{\cA}}_{\cB\cA,g}$ of the domain wall $\Si_g$, which is a Riemann surface of
arbitrary genus $g$, which is a multi-component partition function labeled by
$I_{\cA},I_{\cB}$.  The multi-component partition function $W^{I_{\cB} I_{\cA}}_{\cB\cA,g}$ is
expected since the domain wall has a non-invertible gravitational
anomaly\cite{JW190513279} as characterized by topological orders $\cA$ and $\cB$.
The topological partition function $W^{I_{\cB} I_{\cA}}_{\cB\cA,g}$ can also be viewed as
the overlap of the degenerate ground states of $\cA$ and $\cB$ on $\Si_g$, where
$I_{\cA}$ ($I_{\cB}$) labels the ground states of topological order $\cA$ on $\Si_g$.
This allows us to derive the following linear conditions on the data $W^{I_{\cB}
I_{\cA}}_{\cB\cA,g}$ that characterized the domain walls:
\begin{align}
\label{RWWR}
 R_{\cB}^{I_{\cB},J_{\cB}} W^{J_{\cB} I_{\cA}}_{\cB\cA,g} &= W^{I_{\cB} J_{\cA}}_{\cB\cA,g} R_{\cA}^{J_{\cA},I_{\cA}}, 
\nonumber\\
W^{J_{\cB} I_{\cA}}_{\cB\cA,1} & \in \N, \ \ \ \ (\text{in quasi-particle basis}),
\end{align}
where $R_{\cA}$ ($R_{\cB}$) is the mapping-class-group representation for topological
order $\cA$ (topological order $\cB$) for genus-$g$ Riemann surface.  Eqn.
(\ref{RWWR}) is a special case of a more general condition proposed in
\Ref{JW190513279}, for the partition function with non-invertible gravitational
anomaly.

In this paper, through some simple examples, we try to demonstrate the validity
of the condition \eq{RWWR} (and the condition in \Ref{JW190513279}), by showing
that the condition gives rise to a classification of gapped domain walls
between two topological orders.  In particular, we show that the topological
partition function $W^{I_{\cB} I_{\cA}}_{\cB\cA,1}$ for genus-1 surface, plus the linear
condition \eq{RWWR}, is not enough.\cite{LWW1414} We need to, at least, use the
topological partition function $W^{I_{\cB} I_{\cA}}_{\cB\cA,2}$ for genus-2 surface and its
condition \eq{RWWR}, to obtain a correct classification of gapped domain walls.

At moment, we do not known if the topological partition functions $W^{I_{\cB}
I_{\cA}}_{\cB\cA,g}$ for arbitrary genus-$g$ surface can fully characterize the gapped
domain wall or not (although we think that, very likely, they do).  In Appendix
\ref{structure}, using the connection between anyon condensation and domain
walls, we develop a classifying theory of gapped domain wall based on the
structure coefficients $M^{kc,u}_{ia,jb}$ that describe the condensable algebra
in a topological order.  We also give the conditions \eqn{eq.strucoeasso},
\eqn{eq.strucoeiso}, and \eqn{eq.strucoeacom} on the structure coefficients
$M^{kc,u}_{ia,jb}$, so that they can described a gapped domain wall.  However,
those conditions are non-linear and is very hard to solve.

We see that we have two systematic ways to describe the gapped domain walls
between two topological orders.  The first approach is based on  topological
partition functions $W^{I_{\cB} I_{\cA}}_{\cB\cA,g}$, which is easier to solve, but not
known to fully classify the domain walls.  The second approach is based on the
structure coefficients $M^{kc,u}_{ia,jb}$, which classify the gapped domain
walls, but is hard to solve.  Try to gain a deeper understanding of the two
approaches may help us to find an easier way to fully classify gapped domain
walls.

XW is supported by  the Gordon and Betty Moore Foundations EPiQS initiative
through Grant No.GBMF4303 at MIT.  LK is supported by the Science, Technology
and Innovation Commission of Shenzhen Municipality (Grant Nos.
ZDSYS20170303165926217 and JCYJ20170412152620376) and Guangdong Innovative and
Entrepreneurial Research Team Program (Grant No. 2016ZT06D348), and by NSFC
under Grant No. 11071134 and 11971219.  XGW is partially supported by NSF
Grant No.  DMR-1506475 and DMS-1664412 and by the Simons Collaboration on
Ultra-Quantum Matter, which is a grant from the Simons Foundation (651440).

\appendix

\allowdisplaybreaks

\section{Topological path integral on a space time with domain walls and world
lines } \label{toppath}

\begin{figure}[tb]
\begin{center}
\includegraphics[scale=0.6]{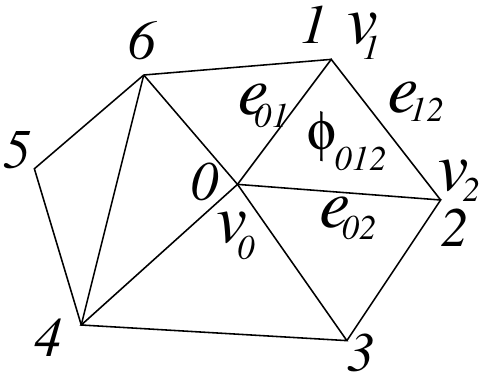} \end{center}
\caption{ 
A 2-dimensional complex.  The vertices (0-simplices) are labeled by $i$.  The
edges (1-simplices) are labeled by $\<ij\>$.  The faces (2-simplices) are
labeled by $\<ijk\>$.  The degrees of freedoms may live on the vertices
(labeled by $v_i$), on the edges (labeled by $e_{ij}$) and on the faces
(labeled by $\phi_{ijk}$).
}
\label{comp}
\end{figure}
\begin{figure}[tb]
\begin{center}
\includegraphics[scale=0.6]{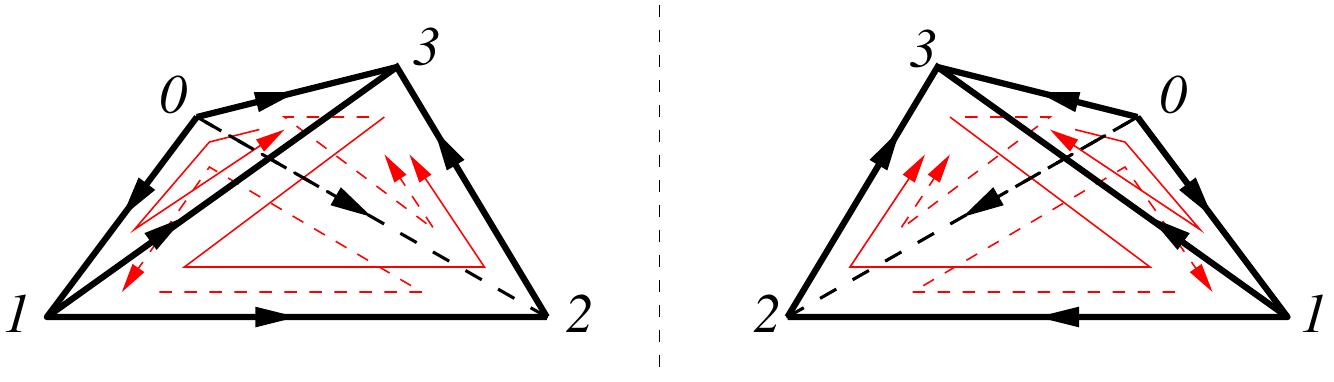} \end{center}
\caption{ (Color online) Two branched simplices with opposite orientations.
(a) A branched simplex with positive orientation and (b) a branched simplex
with negative orientation.  }
\label{mir}
\end{figure}

\subsection{Space-time lattice and branching structure}

To find the conditions on the domain-wall data, we need to use extensively the
space-time path integral.  So we will first describe how to define a space-time
path integral. We first triangulate the $3$-dimensional space-time to obtain a
simplicial complex $M^3$ (see Fig. \ref{comp}).  Here we assume that all
simplicial complexes are of bounded geometry in the sense that the number of
edges that connect to one vertex is bounded by a fixed value.  Also the number
of triangles that connect to one edge is bounded by a fixed value, \etc.

In order to define a generic lattice theory on the space-time complex $M^3$,
it is important to give the vertices of each simplex a local order.  A nice
local scheme to order  the vertices is given by a branching
structure.\cite{C0527,CGL1172,CGL1204} A branching structure is a choice of
orientation of each edge in the $n$-dimensional complex so that there is no
oriented loop on any triangle (see Fig. \ref{mir}).

The branching structure induces a \emph{local order} of the vertices on each
simplex.  The first vertex of a simplex is the vertex with no incoming edges,
and the second vertex is the vertex with only one incoming edge, \etc.  So the
simplex in  Fig. \ref{mir}a has the following vertex ordering: $0<1<2<3$.

The branching structure also gives the simplex (and its sub simplexes) an
orientation denoted by $s_{ij \cdots k}=1,*$.  Fig. \ref{mir} illustrates two
$3$-simplices with opposite orientations $s_{0123}=1$ and $s_{0123}=*$.  The
red arrows indicate the orientations of the $2$-simplices which are the
subsimplices of the $3$-simplices.  The black arrows on the edges indicate the
orientations of the $1$-simplices.

The degrees of freedom of our lattice model live on the vertices  (denoted by
$v_i$ where $i$ labels the vertices), on the edges (denoted by $e_{ij}$ where
$\<ij\>$ labels the edges), and on other high dimensional simplicies of the
space-time complex (see Fig. \ref{comp}).

\begin{figure}[t]
\begin{center}
\includegraphics[scale=0.6]{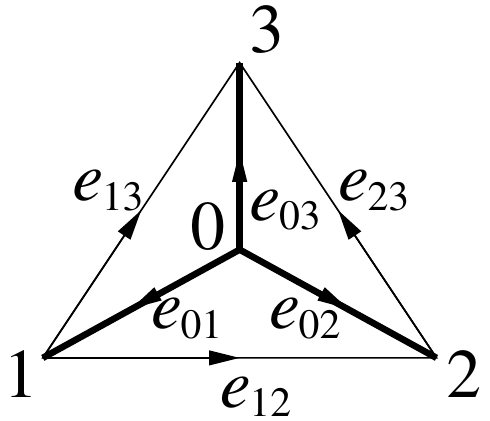}
\end{center}
\caption{
The tensor $\tC{C}0123$ is associated with a tetrahedron, which has a branching
structure.  If the vertex-0 is above the triangle-123, then the tetrahedron
will have an orientation $s_{0123}=*$.  If the vertex-0 is below the
triangle-123, the tetrahedron will have an orientation $s_{0123}=1$. The
branching structure gives the vertices a local order: the $i^{th}$ vertex has
$i$ incoming edges.  
}
\label{tetr}
\end{figure}

\subsection{Discrete path integral}

In this paper, we will only consider a type of 2+1D path integral that can be
constructed from a tensor set $T$ of two real and one complex tensors:
$T=(w_{v_0}, \tAw{d}01,\tC{C}0123)$.  The complex tensor $\tC{C}0123$ can be
associated with a tetrahedron, which has a branching structure (see Fig.
\ref{tetr}).  A branching structure is a choice of orientation of each edge in
the complex so that there is no oriented loop on any triangle (see Fig.
\ref{tetr}).  Here the $v_0$ index is associated with the vertex-0, the
$e_{01}$ index is associated with the edge-$01$, and the $\phi_{012}$ index is
associated with the triangle-$012$.  They represents the degrees of freedom on
the vertices, edges, and the triangles.

Using the tensors, we can define the path integral on any 3-complex
that has no boundary:
\begin{align}
\label{Z3d}
 Z(M^3)&=\sum_{ v_0,\cdots; e_{01},\cdots; \phi_{012},\cdots}
\prod_\text{vertex} w_{v_{0}} 
\prod_\text{edge} \tAw{d}01\times
\\
&\ \ \ \ \ \ \ \ \ \ 
\prod_\text{tetra} [\tC{C}0123 ]^{s_{0123}}
\nonumber 
\end{align}
where $\sum_{v_0,\cdots; e_{01},\cdots; \phi_{012},\cdots}$ sums over all the
vertex indices, the edge indices, and the face indices, $s_{0123}=1$ or $*$
depending on the orientation of tetrahedron (see Fig.  \ref{tetr}).  We believe
such type of path integral can realize any 2+1D topological order.

\subsection{Path integral on space-time with natural boundary}

On the complex $M^3$ with boundary: $B^2= \prt M^3$, the partition
function is defined differently:
\begin{align}
\label{Z3dB}
 Z(M^3) & =\sum_{ \{ v_i; e_{ij}; \phi_{ijk} \} }
\prod_{\text{vertex}\notin B^2} w_{v_{0}} 
\prod_{\text{edge}\notin B^2} \tAw{d}01\times
\\
&\ \ \ \ \ \ \ \ \ \ 
\prod_\text{tetra} [\tC{C}0123 ]^{s_{0123}}
\nonumber 
\end{align}
where $\sum_{v_i; e_{ij}; \phi_{ijk}}$ only sums over the vertex indices, the
edge indices, and the face indices that are not on the boundary.  The resulting
$Z(M^3)$ is actually a complex function of $v_{i}$'s, $e_{ij}$'s, and
$\phi_{ijk}$'s on the boundary $B^2$: $Z(M^3;\{v_{i};e_{ij};\phi_{ijk}\})$.
Such a function is a vector in the vector space $\cV_{B^2}$.  (The vector
space $\cV_{B^2}$ is the space of all complex function of the boundary
indices on the boundary complex $B^2$: $\Psi(\{v_{i};e_{ij};\phi_{ijk}\})$.)
We will denote such a vector as $|\Psi(M^3)\>$.
boundary) are attached with the tensors $w_{v_{i}}$ and $\tAw{d}01$.  
The boundary \eq{Z3dB} defined above is called a natural boundary
of the path integral.

We also note that only the vertices and the edges in the bulk (\ie not on the
But when we glue two boundaries together, those tensors $w_{v_{i}}$ and
$\tAw{d}ij$ are added back.  For example, let $M^3$ and $N^3$ to have the
same boundary (with opposite orientations)
\begin{align}
 \prt M^3 = -  \prt N^3 = B^2
\end{align}
which give rise to wave function on the boundary $|\Psi(M^3)\>$ and
$\<\Psi(N^3)|$ after the path integral in the bulk.  Gluing two boundaries
together is like doing the inner product $\<\Psi(N^3)|\Psi(M^3)\>$.  So the
tensors $w_{v_{i}}$ and $\tAw{d}ij$ defines the inner product in the boundary
Hilbert space $\cV_{B^2}$.  Therefore, we require $w_{v_{i}}$ and $\tAw{d}ij$
to satisfy the following unitary condition
\begin{align}
 w_{v_{i}} > 0, \ \ \ \tAw{d}ij >0.
\end{align}

\subsection{Topological path integral}
\label{pathtop}

We notice that the above path integral is defined for any space-time lattice.
The partition function $Z(M^3)$ depends on the choices of of the space-time
lattice.  For example, $Z(M^3)$ depends on the number of the cells in
space-time, which give rise to the leading volume dependent term,
in the large space-time limit (\ie the thermodynamic limit)
\begin{align}
 Z(M^3) = \ee^{-\eps V} Z^\text{top}(M^3)
\end{align}
where $V$ is the space-time volume, $\eps$ is the energy density of the
ground state, and $Z^\text{top}(M^3)$ is the volume independent partition
function. It was conjectured that the volume independent partition function
$Z^\text{top}(M^3)$ in the thermodynamic limit, as a function of closed
space-time $M^3$, is a topological invariant that can fully characterize
topological order.\cite{KW1458,WW180109938} So it is very desirable to fine
tune the path integral to make the energy density $\eps=0$.  This can be
achieved by fine tuning the tensors $w_{v_{i}}$ and $\tAw{d}ij$.  But we can be
better.  We can also choose the tensor $(w_{v_0}$, $\tAw{d}01$, $\tC{C}0123)$
to be the fixed-point tensor-set under the renormalization group flow of the
tensor network.\cite{LN0701,GW0931} In this case, not only the volume factor
$\ee^{-\eps V}$ disappears, the volume independent partition function
$Z^\text{top}(M^3)$ is also  re-triangulation invariant, for any size of
space-time lattice. In this case, we refer the path integral as a topological
path integral, and denote the resulting partition function as
$Z^\text{top}(M^3)$.  $Z^\text{top}$ is also referred as the volume
independent the partition function, which is a very important concept, since
only volume independent the partition functions correspond to topological
invariants.  In particular, it was conjectured that such kind of  topological
path integrals describe all the topological order with gappable boundary.  For
details, see \Ref{KW1458,WW180109938}.  

\begin{figure}[t]
\begin{center}
\includegraphics[scale=0.5]{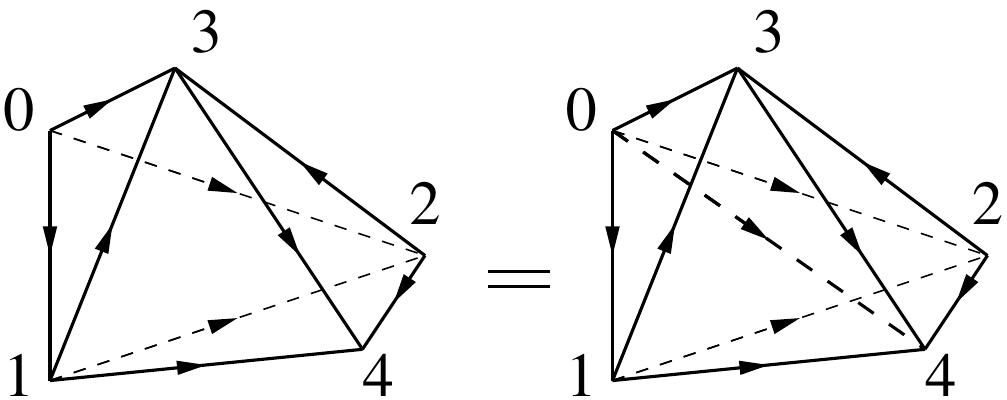}
\end{center}
\caption{
A re-triangulation of a 3D complex.
}
\label{2to3}
\end{figure}
\begin{figure}[t]
\begin{center}
\includegraphics[scale=0.5]{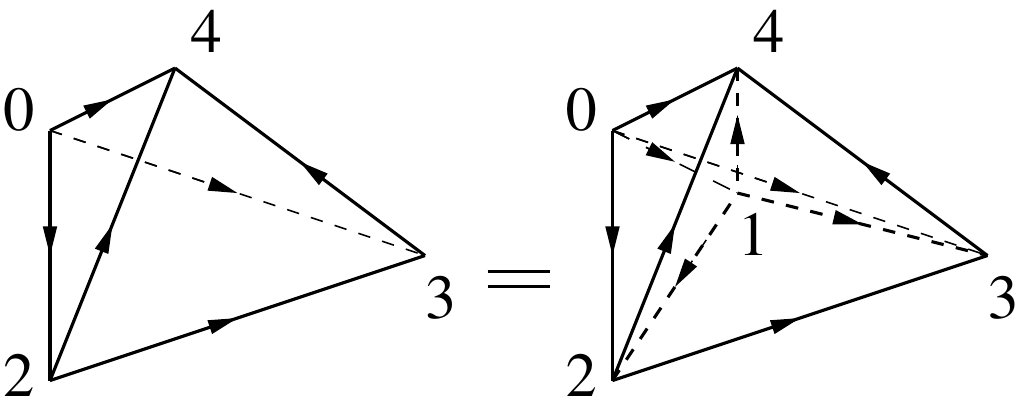}
\end{center}
\caption{
A re-triangulation of another 3D complex.
}
\label{1to4}
\end{figure}

The invariance of partition function $Z$ under the re-triangulation in Fig.
\ref{2to3} and \ref{1to4} requires that
\begin{align}
\label{CC23}
&\ \ \
\sum_{\phi_{123}} \tC{C}0123 \tC{C}1234
\nonumber\\
&=
\sum_{e_{04}} \tAw{d}04
\sum_{ \phi_{014} \phi_{024} \phi_{034} }
\tC{C}0124 
\\
&\ \ \ \ \ \ \
\tC{C^*}0134
\tC{C}0234 .
\nonumber 
\end{align}
\begin{align}
\label{CC14}
& \ \ \ \
\tC{C}0234
\\
&=
\sum_{e_{01}e_{12}e_{13}e_{14},v_1} w_{v_1} 
\tAw{d}01 \tAw{d}12 \tAw{d}13 \tAw{d}14 
\hskip -10mm
\sum_{ 
\phi_{012} 
\phi_{013} 
\phi_{014} 
\phi_{123} 
\phi_{124} 
\phi_{134} 
}
\nonumber\\
&\ \ \ \ \ 
\tC{C}0123
\tC{C^*}0124 
\nonumber\\
&\ \ \ \ \ 
\tC{C}0134
\tC{C}1234
\nonumber 
\end{align}
We would like to mention that there are other similar conditions for different
choices of the branching structures.  The branching structure of a tetrahedron
affects the labeling of the vertices. For more details, see \Ref{ZW180809394}.

\subsection{Topological path integral with domain walls and world lines}
\label{pathtopB}

When the spacetime $M^3$ has a domain wall in it, we can have different
tensor sets, $(w_{A,v_0}$, $\tAw{d_{\cA}}01$, $\tC{C_{\cA}}0123)$ and $(w_{B,v_0}$,
$\tAw{d_{B}}01$, $\tC{C_{B}}0123)$ , on the two sides of the domain wall.  Here we
will assume that the two tensor sets define topological path integrals in the
bulk.  The domain wall is defined via a different tensor set for the simplexes
that touch the domain wall.  Again we can choose the domain wall tensors to
make the partition function with domain wall (after summing over the bulk and
domain wall degrees of freedom) to be re-triangulation invariant (even for the
re-triangulations that involve the domain wall).  Therefore, the topological
path integral can also be defined for spacetime with domain walls.  Different
choices of domain wall tensors give rise to different domain walls.  Those
domain walls can be characterized by the data introduced in Section
\ref{walldata}.

To find the conditions on those domain-wall data, we also need to use the
space-time path integral with world-lines of topological excitations.  We
denote the resulting partition function as
\begin{align}
\label{Zlines}
Z \bpm \includegraphics[scale=.40]{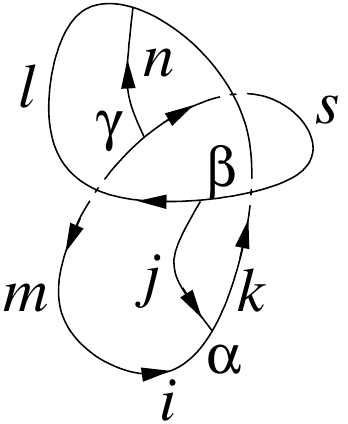} \epm ,
\end{align}
where $i,j,k,\cdots \in \{1,2,\cdots,N\}$ label the type of topological
excitations, and $\al,\bt,\ga$ label the fusion channels (\ie
different choice of actions at the junction of three world-lines).  Again, the
world line is defined via a different tensor set for the simplexes that touch
the world line.  We can choose the world line tensors to make the partition
function with world line to be re-triangulation invariant (even for the
re-triangulations that involve the world line).  Therefore, the topological
path integral can also be defined for spacetime with world lines.  Different
choices of world line tensors give rise to different world lines, which are
labeled by the types of topological excitations.  In this paper, we will only
consider those topological path integrals with re-triangulation invariance.

\section{Categorical approach to evaluate topological path integral with
world lines}
\label{catcal}

\subsection{Planar world-lines and unitary m-fusion category}

The topological path integrals with world lines (\ie the re-triangulation
invariant path integrals with world lines) $Z^\text{top}$ can be computed via
an algebraic approach (or more precisely an categorical approach).  In the
following, we will give a brief introduction of such an approach. More details
can be found in \Ref{W150605768}.

First, let us consider the partition functions with only planar world-line
configurations.  In this case, we may pretend the space-time to be
2-dimensional (or the  space-time is actually 2-dimensional).

The partition functions with different world-line configurations can be related
by some linear relations. For example
\begin{align}
\label{IHwave} 
Z^\text{top} \bpm \includegraphics[scale=.40]{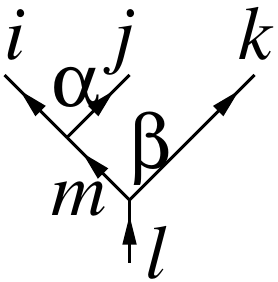} \epm = \sum_{n\chi\del}
F^{ijm,\alpha \bt}_{kln,\chi\del} Z^\text{top} \bpm \includegraphics[scale=.40]{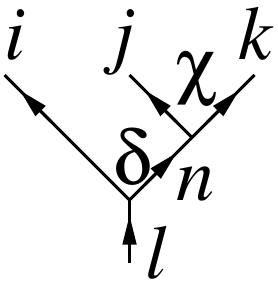} \epm .
\end{align} 
This is because the above partition functions describe the amplitude of fusion
type-$i,j,k$ topological excitations into degenerate type-$l$ topological
excitations.  The two sides of the equation just correspond to different order
of fusion which give rise to the same end product. Thus those amplitudes are
related.

Let us consider the fusion of two  topological excitations of type-$i,j$.  From
far away, the  two  topological excitations may be viewed as single topological
excitation.  But such a single topological excitation may correspond to several
different topological excitations which happen to have the same energy.  For
example, two spin-1/2 excitations can be viewed as spin-0 and spin-1
excitations that happen to have the same energy.  So to describe the fusion of
$i$ and $j$, we need to introduce $N^{ij}_k \in \N$, which count the number of
the type-$k$ topological excitation which happen to have the same energy that
appear in the fusion of type-$i,j$ topological excitations.  Note that
$N^{ij}_k$ may not equal to $N^{ji}_k$.  

Therefore, we have
\begin{align}
 \label{Feq0} &
F^{ijm,\al\bt}_{kln,\chi\del} = 0 \text{ when} \\
 & N^{ij}_{m}=0 \text{ or }
N^{mk}_{l}=0 \text{ or } N^{jk}_{n}=0 \text{ or } N^{in}_{l}=0 . 
\nonumber
\end{align} When $N^{ij}_{m}=0$ or $ N^{mk}_{l}=0$, the left-hand-side of
\eqn{IHwave} is always zero.  Thus $F^{ijm,\al\bt}_{kln,\chi\del} = 0$ when
$N^{ij}_{m}=0$ or $ N^{mk}_{l}=0$.  When $N^{jk}_{n}=0$ or  $N^{in}_{l}=0$,
amplitude on the right-hand-side of \eqn{IHwave} is always zero.  So we can
choose $F^{ijm,\al\bt}_{kln,\chi\del} = 0$ when $N^{jk}_{n}=0$ or
$N^{in}_{l}=0$.

For fixed $i$, $j$, $k$, and $l$, the fusion type-$i,j,k$ topological
excitations produce $N^{ijk}_l$ type-$l$ topological excitations that happen
to have the same energy.
We have
\begin{align}
 N^{ijk}_l = \sum_m N^{ij}_m N^{mk}_l = \sum_n N^{in}_l N^{jk}_n .
\end{align}
Thus the matrix $F^{ij}_{kl}$ with matrix elements
$(F^{ij}_{kl})^{m,\al\bt}_{n,\chi\del} = F^{ijm,\al\bt}_{kln,\chi\del} $ is a
matrix of dimension $N^{ijk}_l \times N^{ijk}_l $.  The  matrix describe the
relation between two different orders of fusions same set of degenerate $l$
particles.  Thus the matrix $F^{ij}_{kl}$ must be unitary:  
\begin{align} 
\label{2FFstar}
\sum_{n\chi\del} F^{ijm',\al'\bt'}_{kln,\chi\del}
(F^{ijm,\al\bt}_{kln,\chi\del})^* =\del_{m,m'}\del_{\al,\al'}\del_{\bt,\bt'}.
\end{align}

The second type of linear relations re-express the amplitude for $\bmm
\includegraphics[scale=.35]{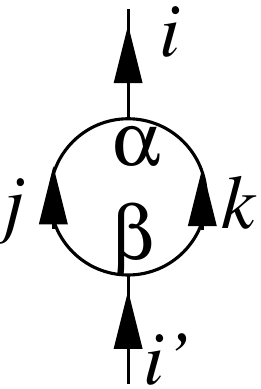} \emm$ in terms of the amplitude for $\bmm
\includegraphics[scale=.35]{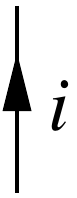} \emm$:
\begin{align}
 \label{PhiO} Z^\text{top} \bpm
\includegraphics[scale=.40]{iOip} \epm =  O^{jk,\al\bt}_i \del_{ii'} Z^\text{top} \bpm
\includegraphics[scale=.40]{iline} \epm  . 
\end{align}
 We call such local change of graph an O-move.

For fixed $i,j$,  $\bmm \includegraphics[scale=.35]{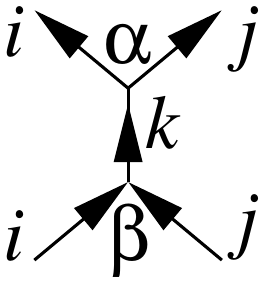} \emm$ for different
$k\al\bt$ describe all the possible  processes.  So the amplitude for
$\bmm \includegraphics[scale=.35]{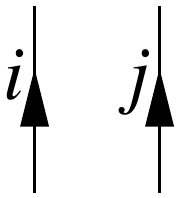} \emm$ can be expressed in terms of the
amplitudes of  $\bmm \includegraphics[scale=.35]{ijkY} \emm$:
\begin{align}
\label{PhiY} \sum_{k,\al\bt} Y^{ij}_{k,\al\bt} Z^\text{top} \bpm
\includegraphics[scale=.40]{ijkY} \epm = Z^\text{top} \bpm \includegraphics[scale=.40]{ij}
\epm
\end{align}
 We will call such a local change as a Y-move.  We can choose
\begin{align} 
\label{Yz} 
Y^{ij}_{k,\al\bt}=0, \text{ if }& N^{ij}_k<1.
\end{align}

We can adjust the action at the triple world-line junction
to simplify 
$O^{jk,\al\bt}_i$ and
$Y^{ij}_{k,\al\bt}$. 
After the simplification, 
\begin{align}
O^{ij,\al\bt}_{k} &= \sqrt{\frac{d_id_j}{d_k}}\del^{ij}_k \del_{\al\bt}, 
\nonumber \\
Y^{ij}_{k,\al\bt} &= \sqrt{ \frac {d_k} {d_id_j}
}\del^{ij}_k \del_{\al\bt}, \ \ \ \ \
d_i>0, 
\label{OY}
\end{align}
 where $\delta^{jk}_i=1$ for
$N^{jk}_i>0$ and $\delta^{jk}_i=0$ for $N^{jk}_i=0$. 
Also $d_i$ are the real and positive solutions from
\begin{align}
 \label{Nddd} \sum_{ij} d_id_j N^{ij}_k =
d_kD^2,\ \ \  D\equiv \sqrt{\sum_l d_l^2}, 
\end{align}
which are called the quantum dimensions of type-$k$ topological excitation.

We see that the partition function $A(X)$ for any world-line configurations can
be characterized by tensor data $(N,N^{ij}_{k}, F^{ijm,\al\bt}_{kln,\ga\la})$.
However, only  certain tensor data $(N,N^{ij}_{k},
F^{ijm,\al\bt}_{kln,\ga\la})$, that satisfy some conditions can
self-consistently describe partition function $A(X)$.  Those conditions form a
set of non-linear equations whose variables are $N^{ij}_{k}$,
$F^{ijm,\al\bt}_{kln,\ga\la}$, $d_i$ (where $d_i$ can be determined by
$N^{ij}_k$ alone):
\begin{align}
 \label{Neq} &\bullet\ \sum_{m=0}^N N^{ij}_{m} N^{mk}_{l}
=\sum_{n=0}^N N^{jk}_{n} N^{in}_l
\nonumber\\
 &\bullet\ \sum_{jk} (N^{jk}_i)^2
\geq 1 ;
\nonumber\\
\end{align}
\begin{align} \label{Feq} & \bullet\ \sum_{n\chi\del}
F^{ijm',\al'\bt'}_{kln,\chi\del} (F^{ijm,\al\bt}_{kln,\chi\del})^*
=\del_{m,m'}\del_{\al,\al'}\del_{\bt,\bt'},
\nonumber\\
 &\bullet\
F^{ijm,\al\bt}_{kln,\chi\del} = 0 \text{ when}
\nonumber \\
 & \ \ \ \
N^{ij}_{m}<1 \text{ or } N^{mk}_{l}<1 \text{ or } N^{jk}_{n}<1 \text{ or }
N^{in}_{l}<1 ,
\nonumber\\
 &\bullet\ \sum_{t} \sum_{\eta=1}^{N^{jk}_{t}}
\sum_{\vphi=1}^{N^{it}_{n}} \sum_{\ka=1}^{N^{tl}_{s}}
F^{ijm,\al\bt}_{knt,\eta\vphi} F^{itn,\vphi\chi}_{lps,\ka\ga}
F^{jkt,\eta\ka}_{lsq,\del\phi}
\nonumber\\
 &= \sum_{\eps=1}^{N^{mq}_{p}}
F^{mkn,\bt\chi}_{lpq,\del\eps} F^{ijm,\al\eps}_{qps,\phi\ga} . 
\end{align}
\begin{align} \label{Oeq} \bullet\  \sum_{i,j} d_id_j N^{ij}_k =d_kD^2,\ \ \
D=\sqrt{\sum_l d_l^2}. 
\end{align}
\begin{align} \label{dFeq} \bullet\  \sum_{n\chi\del} d_n F^{km^\prime i,\alpha
\chi}_{jnl,\bt^\prime\del^\prime}{(F^{kmi,\alpha \chi}_{jnl,\bt\del})}^*
=\frac{d_id_l}{d_m}
\delta_{mm^\prime}\delta_{\al\al^\prime}\delta_{\bt\bt^\prime},
\end{align}

We like to mention that, in the above we did not assume the existence of
trivial particle, which fuse with other particles as an identity.  We will call
such kind of fusion as unitary m-fusion category

\begin{figure}[tb] 
\centerline{ \includegraphics[scale=0.42]{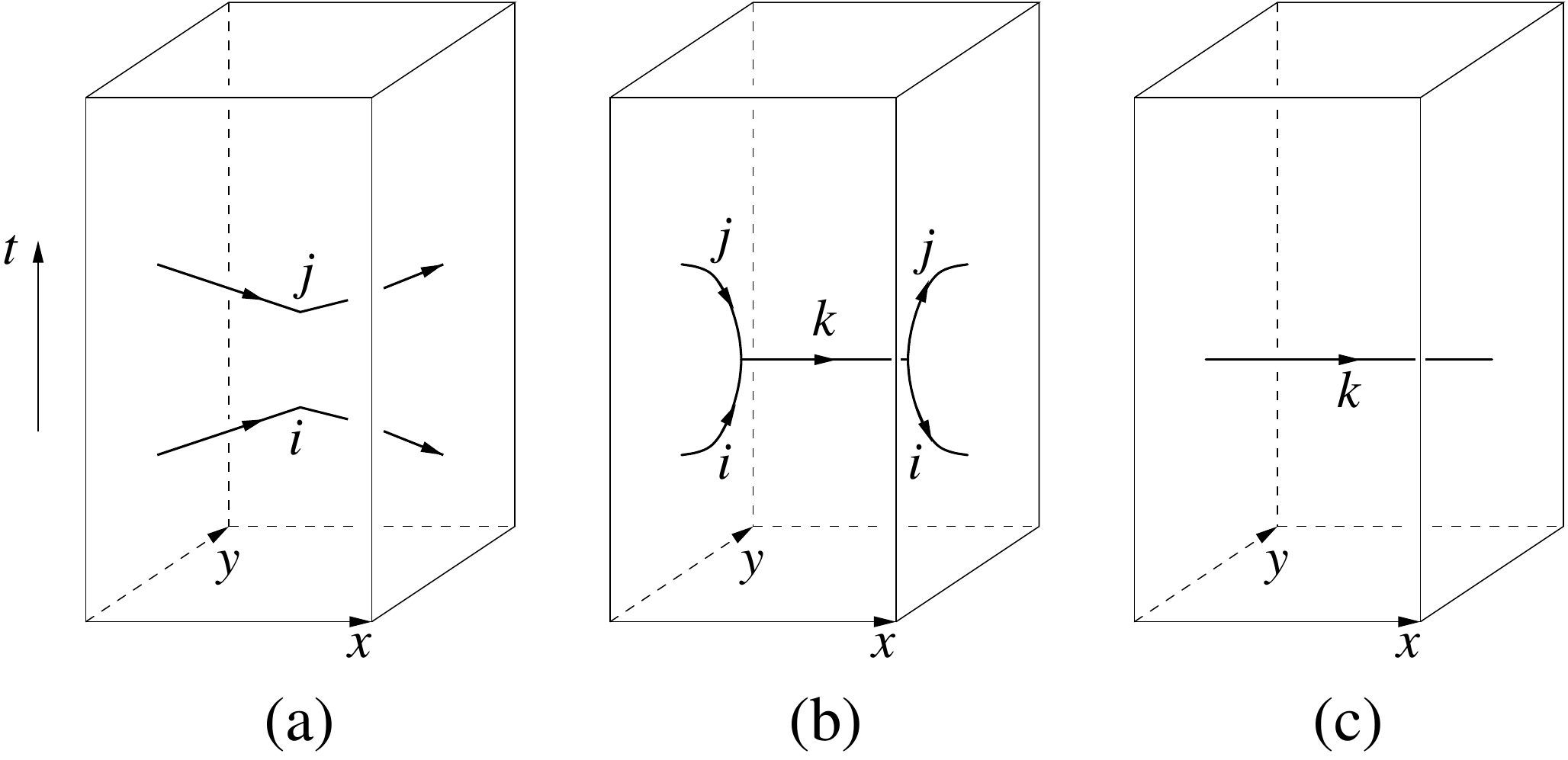} } 
\caption{
(a): Two world-lines created by operators $W^x_i$ and $W^x_j$. (b): The world-lines can be deformed according to the Y-move (c):
The O-move can reduce (b) to (c).  } 
\label{falgT} 
\end{figure}

As an application of the above algebraic structure
-- the unitary m-fusion category, let us consider 
two world-lines of type-$i$ and type-$j$, wrapping around a torus $S^1_x\times S^1_y$ in
the $x$-direction.
Let $W^x_i$ and $W^x_j$ 
be the string operators that creates the world-lines.
Applying the Y-move and then the O-move, and using \eqn{OY} (see
Fig.  \ref{falgT}), we find that 
\begin{align}
\label{WWNW} 
W^x_i
W^x_j&=\sum_{k=1}^N \sum_{\al=1}^{N^{ij}_k} Y^{ij}_{k,\al} O^{ij,\al}_k
W^x_k
\nonumber\\
 &= \sum_{k}
N^{ij}_{k} W^x_k . 
\end{align}
 We see that the algebra of the loop operator
$W_i$ forms a representation of fusion algebra $i\otimes j=\sum_{k}
N^{ij}_{k} k$.

\subsection{Presence of trivial particle and  unitary fusion category}

Now let us assume such a trivial particle type to exist, and
denoted it by $\one$, which satisfies the following fusion rule 
\begin{align}
\one\otimes i=i\otimes \one=i .  
\end{align} 
Thus $N^{ij}_k$ satisfies
\begin{align}
N^{\one i}_j=N^{i \one}_j=\del_{ij}. 
\end{align}
 We also requires that for every $i$
there exists a unique $\bar i$ such that
\begin{align}
 \bar{\bar i}=i,\ \ \
\bar \one =\one ,\ \ \ N^{ij}_\one =\del_{i\bar j}
\end{align}

We can represent a type-$\one$ string by a dash line.  By examine the O-move
with $k=\one$:
\begin{align}
 Z^\text{top} \bpm \includegraphics[scale=.40]{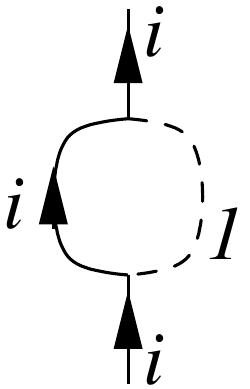} \epm =  Z^\text{top}
\bpm \includegraphics[scale=.40]{iline} \epm  . 
\end{align}
we see that we can remove or add any vertex with dash line 
without changing $Z^\text{top}$.

With the presence of trivial particle type, we can determine the amplitude for
a loop of $i$-string.  Using the rule of adding dash lines (the trivial
strings) and $O$-move \eqn{PhiO}, we find
\begin{align}
Z^\text{top} \bpm \includegraphics[scale=.40]{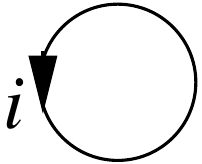} \epm &= Z^\text{top} \bpm
\includegraphics[scale=.40]{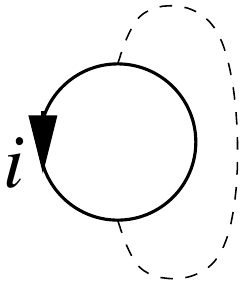} \epm = O^{i\bar i}_1 Z^\text{top} \bpm
\includegraphics[scale=.40]{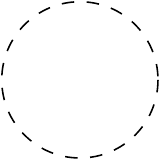} \epm
\nonumber\\
 &= d_i Z^\text{top} \bpm
\includegraphics[scale=.40]{1lp} \epm
\end{align}
Thus a loop of type-$i$ world-line has an amplitude $d_i$.

\subsection{Non-planar diagram and braided fusion category}

We have being considering planar graphs and the related fusion category theory.
In this section we will consider non-planar graphs.  Since the particles now
live in 2-dimensional space (or higher), the fusion of the particles satisfies
\begin{align}
 i\otimes
j =j\otimes i,
\end{align}
 and thus
\begin{align}
 N^{ij}_k=N^{ji}_k.
\end{align} 
So the fusion of 2D particles are commutative (while the fusion of 1D particles
may not be commutative).  

\begin{figure}[tb] \centerline{ \includegraphics[scale=0.5]{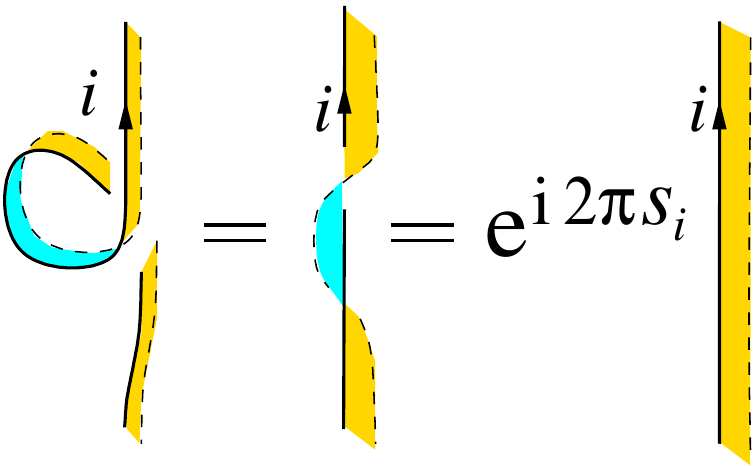} } 
\caption{ (Color online) 
A ``self-loop" with canonical framing corresponds to a twist by
$2\pi$.  A twist by $2\pi$ induces a phase $\ee^{\ii 2\pi s_i}$ that defines
the spin $s_i$ of the particle.}
\label{twist} 
\end{figure}

Here, we also like point out that a world-line of a particle are always framed
(\ie having a shadow world-line running parallel to it).  When we draw a graph
on a plane, there is canonical framing, obtain by shifting the graphs
perpendicular to the plane (see Fig. \ref{twist}). We have been using such a
canonical framing in our previous discussion, and we have omitted drawing the
framing.  But if we do not use this canonical framing, then we need to draw the
framing explicitly, as in Fig. \ref{twist}.

\begin{figure}[tb] \centerline{ \includegraphics[scale=0.5]{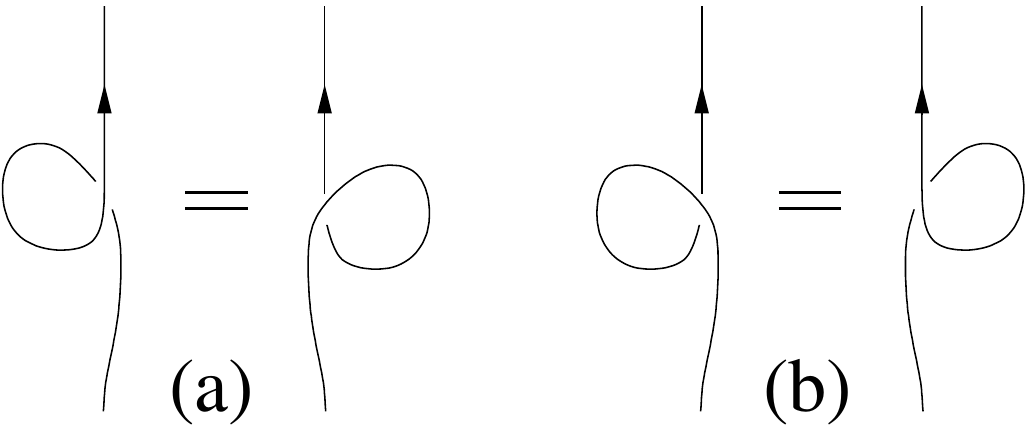} } 
\caption{
The two ``self-loops'' in (a) are ``right-handed'' and correspond to the same
twist.  The two ``self-loops'' in (b) are ``left-handed'' and also correspond
to the same twist that is opposite to that in (a).  } 
\label{twist2}
\end{figure}

\begin{figure}[tb] 
\centerline{ \includegraphics[scale=0.6]{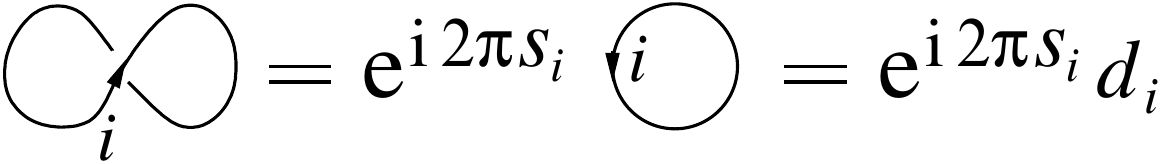} } 
\caption{A figure ``8'' of type-$i$ string has an amplitude $\ee^{2\pi \ii s_i} d_i$.} 
\label{twsidi} 
\end{figure}

Let us consider simple string configuration with crossing: a ``self-loop'' with
the canonical framing (see Fig. \ref{twist}).  Such a ``self-loop'' corresponds
to a straight line with a $2\pi$ twist, which is equal to a untwisted straight
line with a phase $\ee^{\ii 2\pi s_i}$.  Here $s_i$ is the \emph{spin} of the
type-$i$ topological excitation, which is defined mod 1.  We also note that the
handness of the ``self-loop'' determines the direction of the twist (see Fig.
\ref{twist2}).  As a result, a figure ``8'' of type-$i$ string has an amplitude
$\ee^{2\pi \ii s_i} d_i$ (see Fig. \ref{twsidi}).

\section{The gapped domain walls as 1+1D anomalous topological order}

There is another ways to fully characterize the domain walls. We note that the
gapped 1+1D domain walls can be viewed as  1+1D anomalous topological
orders.\cite{KW1458} The 1+1D anomalous topological orders are characterized by
unitary fusion categories (UFC) which are described by the following data (1)
$ N\in \N$: the number of types (including the trivial type) of topological
excitations on the domain wall.  We will use $i,j,k,$ \etc, to label the types
of topological exitations and use $1$ to label the trivial type.  (2) $
N^{ij}_{k}\in \N$: the fusion coefficients of the topological excitations.  (3)
$ F^{ijm,\al\bt}_{kln,\ga\la}$: the unitary relation between different fusion
spaces obtained via different fusion paths.

Those data $( N,  N^{ij}_k,  F^{ijm,\al\bt}_{kln,\ga\la})$ satisfy
\begin{align}
\label{Neq}
&
\sum_{m=0}^{ N}  N^{ij}_{m}  N^{mk}_{l} =\sum_{n=0}^{ N}  N^{jk}_{n}  N^{in}_l
\nonumber\\
&
 N^{i \one}_j = N^{1i}_j =\del_{ij} ;\ \ \
\sum_{k=1}^{ N}  N_1^{i k} N_1^{ kj}=\del_{ij}
\end{align}
\begin{align}
\label{Feq}
&
  F^{ijm,\al\bt}_{kln,\chi\del} = 0 \text{ when}
\nonumber \\
&
\ \ \ \
 N^{ij}_{m}<1 \text{ or }
 N^{mk}_{l}<1 \text{ or }
 N^{jk}_{n}<1 \text{ or }
 N^{in}_{l}<1
,
\nonumber\\
&
 \sum_{n\chi\del}
 F^{ijm',\al'\bt'}_{kln,\chi\del}
( F^{ijm,\al\bt}_{kln,\chi\del})^*
=\del_{m,m'}\del_{\al,\al'}\del_{\bt,\bt'},
\nonumber\\
&  \sum_{n\chi\del}  d_n  F^{km^\prime i,\alpha \chi}_{jnl,\bt^\prime\del^\prime}{( F^{kmi,\alpha \chi}_{jnl,\bt\del})}^* =\frac{ d_i d_l}{ d_m}
\delta_{mm^\prime}\delta_{\al\al^\prime}\delta_{\bt\bt^\prime},
\nonumber\\
&
\sum_{t}
\sum_{\eta=1}^{ N^{jk}_{t}}
\sum_{\vphi=1}^{ N^{it}_{n}}
\sum_{\ka=1}^{ N^{tl}_{s}}
 F^{ijm,\al\bt}_{knt,\eta\vphi}
 F^{itn,\vphi\chi}_{lps,\ka\ga}
 F^{jkt,\eta\ka}_{lsq,\del\phi}
\nonumber\\
&=
\sum_{\eps=1}^{ N^{mq}_{p}}
 F^{mkn,\bt\chi}_{lpq,\del\eps}
 F^{ijm,\al\eps}_{qps,\phi\ga}
.
\end{align}
Here $ d_i$ is the largest left eigenvalue of the matrix
$ N_i$ (defined as $( N_i)_{kj}= N^{ij}_k$).

Fusion category $( N,  N^{ij}_k,  F^{ijm,\al\bt}_{kln,\ga\la})$ and weighted
wave function overlap $W_{\cB\cA,g}^{I_{\cB}I_{\cA}}$ provide 
two very different ways to characterize the
same domain wall.  It is amazing that to two sets of data should have a
one-to-one correspondence.

\section{More details on the gapped boundary of $S_3$ topological order}
\label{Appendix: S3}

\subsection{Some preliminaries}

In this section, we give the details of solving the gapped boundaries, using the
genus-2 condition in \eqref{Genus2_gapping_condition_S3}, for normalized
wave function overlap $\t{W}_{\cS_3,g=2}$.  For $g=2$, it is convenient to
consider the following two choices of basis vectors:
\begin{eqnarray}\label{BasisI_appendix}
\text{basis I:}\quad
\begin{tikzpicture}[baseline={(current bounding box.center)}]
\draw[>=stealth,<-] (0pt,0pt) arc (180:-180:20pt) ;
\draw[>=stealth,<-] (60pt,0pt) arc (180:-180:20pt) ;
\draw [>=stealth,->] (60pt,0pt)--(50pt,0pt);
\draw (50pt,0pt)--(40pt,0pt);
\node at (20pt, 25pt){$i$};
\node at (80pt, 25pt){$j$};
\node at (50pt, 5pt){$z$};
\node at (35pt, 0pt){\small $\nu$};
\node at (65pt, 0pt){\small $\mu$};
\end{tikzpicture}
\end{eqnarray}
and
\begin{eqnarray}\label{BasisII_appendix}
\text{basis II:}\quad
\begin{tikzpicture}[baseline={(current bounding box.center)}]
\draw[>=stealth,->] (10*0.7pt,17*0.7pt) arc (40:180:30*0.7pt);
\draw  (-43*0.7pt,-2*0.7pt) arc (180:320:30*0.7pt);
\draw (10*0.7pt,17*0.7pt) arc (140:0:30*0.7pt);
\draw [>=stealth,<-] (63*0.7pt,-2*0.7pt) arc (0:-140:30*0.7pt);
\draw (10*0.7pt,17*0.7pt) -- (10*0.7pt, -2*0.7pt);
\draw [>=stealth,<-] (10*0.7pt, -2*0.7pt) -- (10*0.7pt,-22*0.7pt);
\node at (-35*0.7pt,0pt){$i$};
\node at (55*0.7pt,0pt){$j$};
\node at (15*0.7pt,0pt){{ $z$}};
\node at (7.5pt, 18pt){\small $\mu$};
\node at (7.5pt, -20pt){\small $\nu$};
\end{tikzpicture}
\end{eqnarray}
In particular, `basis I' will be useful in making a connection to the genus-1 solution by choosing $z=\mathbf{1}$, 
and `basis II' will be useful in studying the effect of Dehn twist operator $T_5$ (or $T_{\gamma}$) 
in Fig.\ref{DenhTwist5}. Within basis II, $T_5$ will be a diagonal matrix with the diagonal elements corresponding
to the topological spin $\theta_z$ of anyon $z$.

Before any concrete calculation, it is noted that for the solutions of gapped boundaries, if 
$\t{W}^{\text{I (II)}; i,j,z}_{\cS_3,g=2}\neq 0$,
then the topological spins of $i$, $j$, and $z$ in \ref{BasisI_appendix} and \ref{BasisII_appendix} must be trivial.
This can be understood by considering $R_{\cS_3}=T_1,\, T_2,\, (S_1)^{-4}, \, (S_2)^{-4},$ and $T_5$ 
in Eq.\eqref{Genus2_gapping_condition},

We denote basis I and basis II in \eqref{BasisI_appendix} and \eqref{BasisII_appendix} as 
\begin{equation}
|\psi^{\text{I};i,j,z; \mu, \nu}\rangle,\quad \text{and}\quad |\psi^{\text{II}; i,j,z; \mu, \nu}\rangle,
\end{equation}
respectively. Since the $S_3$ topological order is multiplicity free, we can write the above basis vectors as 
\begin{equation}\label{TwoBases}
|\psi^{\text{I}; i,j,z}\rangle,\quad \text{and}\quad |\psi^{\text{II}; i,j,z}\rangle.
\end{equation}
These two basis vectors are normalized as follows:
\begin{equation}
\begin{split}
&\langle \psi^{\text{I}; i,j,z} | \psi^{\text{I}; i',j',z'}\rangle=\delta_{i,i'} \delta_{j,j'} \delta_{z,z'},\\
&\langle \psi^{\text{II}; i,j,z} | \psi^{\text{II}; i',j',z'}\rangle=\delta_{i,i'} \delta_{j,j'} \delta_{z,z'}.
\end{split}
\end{equation}
In addition, they are related to each other as,
\begin{equation}\label{basisTransform}
|\psi^{\text{I}; i,j,z}\rangle=\sum_{z'}\, [F^{i\bar{j}}_{i\bar{j}}]_{(z,z')}\,\, | \psi^{\text{II}; i,j,z'}\rangle,
\end{equation}
where $[F^{i\bar{j}}_{i\bar{j}}]_{(z,z')}$ is defined by
\begin{equation}\label{F_move_II}
\small
\begin{tikzpicture}[baseline={(current bounding box.center)}]
\draw [>=stealth,->](-8pt,-16pt)--(-8pt,-6pt);\draw (-8pt,-6pt)--(-8pt,0pt);
\draw [>=stealth,->](-8pt,0pt)--(-8pt,10pt); \draw(-8pt,10pt)--(-8pt,16pt);

\draw [>=stealth,->](8pt,-16pt)--(8pt,-6pt);\draw (8pt,-6pt)--(8pt,0pt);
\draw [>=stealth,->](8pt,0pt)--(8pt,10pt); \draw(8pt,10pt)--(8pt,16pt);

\draw [>=stealth,->](8pt,-4pt)--(0pt,0pt); \draw(0pt,0pt)--(-8pt,4pt);

\node at (-8pt,-20pt){$c$};
\node at (-8pt,20pt){$a$};
\node at (8pt,-20pt){$d$};
\node at (8pt,21pt){$b$};
\node at (0pt,4pt){$e$};
\node at (-12pt,5pt){$\alpha$};
\node at ( 12pt,-4pt){$\beta$};
\end{tikzpicture}
=
\sum_{f,\mu,\nu}
\left[
F^{ab}_{cd}
\right]_{(e,\alpha,\beta),(f,\mu,\nu)}
\begin{tikzpicture}[baseline={(current bounding box.center)}]
\draw [>=stealth,->](-10pt,-18pt)--(-5pt,-13pt);\draw(-5pt,-13pt)--(0pt,-8pt);
\draw [>=stealth,->](10pt,-18pt)--(5pt,-13pt);\draw(5pt,-13pt)--(0pt,-8pt);
\draw [>=stealth,->](0pt,-8pt)--(0pt,2pt); \draw(0pt,2pt)--(0pt,8pt);
\draw [>=stealth,->](0pt,8pt)--(-7pt,15pt);\draw(-7pt,15pt)--(-10pt,18pt);
\draw [>=stealth,->](0pt,8pt)--(7pt,15pt);\draw(7pt,15pt)--(10pt,18pt);
\node at (-13pt,-18pt){$c$};
\node at (-13pt,18pt){$a$};
\node at (13pt,-18pt){$d$};
\node at (13pt,18pt){$b$};
\node[right] at (0pt,6pt){$\mu$};
\node[right] at (-10pt,0pt){$f$};
\node[right] at (0pt,-6pt){$\nu$};
\end{tikzpicture},
\end{equation}
Here $[F^{i\bar{j}}_{i\bar{j}}]_{(z,z')}$ is related to the conventional $F$-matrix as defined in \eqref{IHwave}
through the following relation:
\begin{equation}\label{F_ab_cd}
\left[
F^{ab}_{cd}
\right]_{(e,\alpha,\beta),(f,\mu,\nu)}
=
\sqrt{\frac{d_e d_f}{d_a d_d}} \left[F^{{c}{e}{a},\alpha\beta}_{{b}{f}{d},\mu\nu}
\right]^{\ast}.
\end{equation}
It is convenient to study the effect of Dehn twist $T_5$ in basis II, which simply results in a phase factor, 
\textit{i.e.}, 
\begin{equation}
T_5\, |\psi^{\text{II}; i,j,z}\rangle=\theta_z\,|\psi^{\text{II}; i,j,z}\rangle.
\end{equation}
Within basis I, one has
\begin{equation}\label{T5_act_basisI}
T_5\, |\psi^{\text{I};i,j,z}\rangle=\sum_{z',z''} [F^{i\bar{j}}_{i \bar{j}}]_{(z,z')} \cdot \theta_{z'}
\cdot [F^{i\bar{j}}_{i\bar{j}}]^{-1}_{(z',z'')} 
 |\psi^{\text{I}; i,j, z''}\rangle.
\end{equation}
Apparently, $T_5$ is in general not diagonal in basis I. However, if 
if the theory is abelian, then one has  $z=\mathbf{1}$ and $[F^{ij}_{ij}]_{\mathbf{1}z'}=N_{ij}^{z'}$.
Then $T_5$ is a diagonal matrix with the matrix elements
\begin{equation}
\langle \psi^{\text{I}; i,j,z''}|T_5 |\psi^{\text{I}; i,j,z}\rangle=
\delta_{\mathbf{1}, z} \delta_{\mathbf{1},z''} \theta_{z'} N_{i\bar{j}}^{z'}.
\end{equation}

Another interesting case is that if $\theta_{z'}=1$ for all $[F^{i\bar{j}}_{i\bar{j}}]_{(z,z')}\ne 0$, then
based on Eq.\eqref{T5_act_basisI}, one has
\begin{equation}\label{T5_trivialSpin}
T_5|\psi^{\text{I}; i,j,z}\rangle=|\psi^{\text{I}; i, j,z}\rangle, \quad \text{if }\theta_{z'}=1 \text{ for all } [F^{i\bar{j}}_{i\bar{j}}]_{(z,z')}\ne 0.
\end{equation}

For $S_3$ topological order, all the $F$ matrices have been obtained in Ref.
 \onlinecite{cui2015universal}
The so-called punctured $S$ matrix
\begin{eqnarray}\label{PuncturedSmatrix}
S_{a,\mu;b,\nu}^{(z)}&=
\frac{1}{D}
\cdot 
\frac{1}{\sqrt{d_z}}
\cdot 
\begin{tikzpicture}[baseline={(current bounding box.center)}]
\draw (20pt,0pt) circle (20pt);
\draw[line width=6pt, draw=white] (0pt,0pt) circle (20pt);
\draw (0pt,0pt) circle (20pt);
\draw [>=stealth,->] (20pt,0.1pt)--(20pt,0.11pt);
\draw [>=stealth,->] (40pt,0.1pt)--(40pt,0.11pt);
\draw[line width=6pt, draw=white]  (0pt,0pt) arc (-180:-270:20pt);
\draw (0pt,0pt) arc (-180:-270:20pt);
\node at (-15+40pt,0pt){$a$};
\node at (5+40pt,0pt){$b$};
\draw[>=stealth,->]  (20pt,-20pt)..controls (15pt,-30pt) and (5pt,-30pt)..(0pt,-20pt);
\node at (10pt,-32pt){$z$};
\node at (-5pt,-25pt){\small$\mu$};
\node at (22pt,-25pt){\small$\nu$};
\end{tikzpicture}
\label{eq.puncturedS}
\end{eqnarray}
can be expressed in terms of $F$-matrix as (in the multiplicity-free case)
\begin{eqnarray}\label{PuncturedS}
\small
S_{a,b}^{(z)}
=\frac{1}{D}
\cdot 
\frac{1}{\sqrt{d_z}}
\cdot\frac{d_ad_b}{\theta_a\theta_b}
\sum_{f} 
F^{{a}{a}{z}}_{{b}{b}{f}}\cdot \theta_f \cdot
[F^{{a}{a}{f}}_{{b}{b}\mathbf{1}}]^\ast,
\end{eqnarray}
based on which we can obtain the punctured $S^{(z)}$ matrix.
For our motivation of studying the gapped boundaries, we only need to consider
$S^{(z)}$ with $\theta_z=1$ in \eqref{BasisI_appendix}.
Based on Eq.\eqref{ST_S3} and the fusion rules of $i\otimes \bar{i}$ in Table \ref{S3FusionRules}, 
we only need to check the cases of $z=a^1,\, a^2,\, b$. 

The results of punctures $S^{(z)}$ and $T^{(z)}$ matrices with $z=a^1,\, a^2,\, b$
are summarized as follows:

$\bullet$ For $z=a^1$, one has (with the basis $a^2$, $b$, $b^1$, and $b^2$):
\begin{equation}
S^{(z=a^1)}=
\frac{1}{3}(\omega^2-\omega)\cdot 
\left(
\begin{array}{cccc}
0 &1  &1 &-1\\
1 &0 &1 &1\\
1 &1 &-1 &0\\
-1 &1 &0 &1
\end{array}
\right)
\end{equation}
and $T^{(z=a^1)}=\text{diag}(1, 1,\omega, \omega^2)$,
where $\omega=e^{\frac{2\pi i}{3}}$.

$\bullet$ For $z=a^2$, one has (with the basis $c^1$, $c$, and $a^2$):
\begin{equation}\label{Sa_2}
S^{(z=a^2)}=
\left(
\begin{array}{cccc}
\frac{1}{2} &\frac{1}{2}  &\frac{1}{\sqrt{2}}\\
\frac{1}{2}  &\frac{1}{2}  &-\frac{1}{\sqrt{2}}\\
\frac{1}{\sqrt{2}} &-\frac{1}{\sqrt{2}} &0
\end{array}
\right),
\end{equation}
and 
$T^{(z=a^2)}=\text{diag}(-1, 1, 1)$.

$\bullet$ For $z=b$, one can obtain (in the basis $c^1$, $c$, and $b$):
\begin{equation}
S^{(z=b)}=
\left(
\begin{array}{cccc}
\frac{1}{2} &-\frac{1}{2}  &\frac{1}{\sqrt{2}}\\
-\frac{1}{2}  &\frac{1}{2}  &\frac{1}{\sqrt{2}}\\
\frac{1}{\sqrt{2}} &\frac{1}{\sqrt{2}} &0
\end{array}
\right),
\end{equation}
and 
$T^{(z=b)}=\text{diag}(-1, 1, 1)$.
For the punctured $S^{(z)}$ and $T^{(z)}$ presented above, one can check explicitly 
that they satisfied the so-called modular relation
\begin{equation}
\left(S^{(z)}T^{(z)}\right)^3=\left(S^{(z)}\right)^2,\quad \left(S^{(z)}\right)^4=\theta_z^*.
\end{equation}

Now we are ready to study the solutions of 
\begin{equation}\label{WWR}
W=W\cdot R,
\end{equation}
with $R=S^{(z)}$ and $T^{(z)}$.

$\bullet$ For $z=a^1$, since $T^{(z=a^1)}$ is diagonal, one can find that $W$ has the form
$W=(x, y, 0, 0)$.  Then choosing $R=S^{(z=a^1)}$ in \eqref{WWR}, one can find that
$x=y=0$. That is, there is no non-zero solution of \eqref{WWR} for
$R=S^{(z=a^1)}$ and $T^{(z=a^1)}$.
More explicitly, with the basis in \eqref{BasisI_appendix}, we have
\begin{equation}\label{Wza1_appendix}
\t{W}^{\text{I}; i,j,z=a^1}_{\cS_3,g=2}=0,
\end{equation}
where $i,j\in\{a^2,\, b,\, b^1,\, b^2\}$.

$\bullet$ For $z=a^2$, since $T^{(z=a^2)}=\text{diag}(-1, 1, 1)$, we have $W=(0, x, y)$. Then 
by choosing $R=S^{(z=a^2)}$ in \eqref{WWR}, one can find that
$x=-\sqrt{2}y$. That is,
\begin{equation}\label{Wza2}
W=(0, \, -\sqrt{2}y, \, y),
\end{equation}
where $y$ is to be determined by other conditions.
With the basis vectors chosen in \eqref{BasisI_appendix},
Eq.\eqref{Wza2} indicates that
\begin{equation}
\begin{split}\label{Puncture_z_a2}
\t{W}^{\text{I}; i,c,z=a^2}_{\cS_3,g=2}=&-\sqrt{2}\cdot  \t{W}^{\text{I}; i,a^2,z=a^2}_{\cS_3,g=2},\\
\t{W}^{\text{I}; c,i,z=a^2}_{\cS_3,g=2}=&-\sqrt{2}\cdot  \t{W}^{\text{I}; a^2,i,z=a^2}_{\cS_3,g=2},
\end{split}
\end{equation}
where $i\in\{a^2,\, c,\, c^1\}$.

$\bullet$ For $z=b$, since $T^{(z=b)}=\text{diag}(-1, 1, 1)$, we have $W=(0, x, y)$.
Then by choosing $R=S^{(z=b)}$ in \eqref{WWR}, one can find that
$x=\sqrt{2}y$. That is,
\begin{equation}\label{Wzb}
W=(0, \, \sqrt{2}y, \, y),
\end{equation}
where again $y$ is to be determined by other conditions.
With the basis vectors chosen in \eqref{BasisI_appendix},
Eq.\eqref{Wzb} indicates that
\begin{equation}\label{Puncture_z_b}
\begin{split}
\t{W}^{\text{I}; i,c,z=b}_{\cS_3,g=2}=&\sqrt{2}\cdot  \t{W}^{\text{I}; i,b,z=b}_{\cS_3,g=2},\\
\t{W}^{\text{I}; c,i,z=b}_{\cS_3,g=2}=&\sqrt{2}\cdot  \t{W}^{\text{I}; b,i,z=b}_{\cS_3,g=2},
\end{split}
\end{equation}
where $i\in\{b,\, c,\, c^1\}$.

\subsection{Solutions of genus-2 condition}

In this subsection, we solve the genus-2 condition in Eq.\eqref{Genus2_gapping_condition_S3}, and find there are 
in total $4$ sets of independent solutions, which embed the genus-1 solutions
$\t{W}^{(i)}_{\cS_3,g=1}\otimes \t{W}^{(i)}_{\cS_3,g=1}$
with $i=1,\, 2,\, 3,\, 4$, respectively.

 \begin{table}[!h]
\centering
\begin{tabular}{c||c|c|c|c|c|c|c|c}
$\otimes$ & $\t{W}_{\cS_3, g=1}^{(1)}$ &  $\t{W}_{\cS_3, g=1}^{(2)}$  &
$\t{W}_{\cS_3, g=1}^{(3)}$ &  $\t{W}_{\cS_3, g=1}^{(4)}$  &  $\t{W}_{\cS_3, g=1}^{(5)}$    \\
\hline
\hline
$ \t{W}_{\cS_3, g=1}^{(1)} $ & $\textcolor{blue}{\checkmark}$ & \shortstack{$\textcolor{blue}{\times}$\\ $a^2\otimes b$}  & \shortstack{$\textcolor{blue}{\times}$\\ $a^1\otimes c$}   & \shortstack{$\textcolor{blue}{\times}$\\$a^1\otimes c$\\ $a^2\otimes b$ }   &  \shortstack{$\textcolor{blue}{\times}$\\$a^1\otimes c$\\ $a^2\otimes b$ }       \\
\hline
$\t{W}_{\cS_3, g=1}^{(2)}$ &  & $\textcolor{blue}{\checkmark}$ & \shortstack{$\textcolor{blue}{\times}$\\ $a^1\otimes c$\\ $a^2\otimes b$}	  &\shortstack{$\textcolor{blue}{\times}$\\ $a^1\otimes c$} &\shortstack{$\textcolor{blue}{\times}$\\$a^1\otimes c$\\ $a^2\otimes b$}   \\
\hline
$\t{W}_{\cS_3, g=1}^{(3)}$ &   &    & $\textcolor{blue}{\checkmark}$      &\shortstack{$\textcolor{blue}{\times}$\\ $a^2\otimes b$}    &\shortstack{$\textcolor{blue}{\times}$\\ $a^2\otimes b$ }\\
\hline
$\t{W}_{\cS_3, g=1}^{(4)}$  &  &   &     & $\textcolor{blue}{\checkmark}$ &\shortstack{$\textcolor{blue}{\times}$\\ $a^2\otimes b$ }  \\
\hline
$\t{W}_{\cS_3, g=1}^{(5)}$  &  &   &        &   &\shortstack{$\textcolor{blue}{\times}$\\ $a^2\otimes b$ }  \\
\hline
\end{tabular}
\caption{ Only four sets of solutions are allowed by the genus-2 condition. 
Related fusion rules are $a^1\otimes c=c^1$, and $a^2\otimes b=b^1\oplus b^2$.
}
\label{S3_solution_appendix}
\end{table}

Before solving the genus-2 condition in Eq.\eqref{Genus2_gapping_condition_S3}, 
it is first noted that the ground state degeneracy of a topological order on genus-2 closed manifold is
\begin{equation}
\text{GSD}(\Sigma_2)=\sum_i\Big(\frac{1}{S_{0i}}\Big)^2,
\end{equation}
where we sum over all the anyon types $i$.
One can check that for the $S_3$ topological order, we have $\text{GSD}(\Sigma_2)=116$.
To obtain the solutions of genus-2 condition in Eq.\eqref{Genus2_gapping_condition} means 
we need to obtain the 116 components of $\t{W}_{\cS_3, g=2}^{\text{I (II)}; i, j, z}$ if we choose the 
basis vectors I (II).

Let us start by considering different pairings of genus-1 solutions in Table \ref{S3_solution_appendix}.
First, it is noted that one common feature of the pairings
$\t{W}^{(p)}_{\cS_3,g=1}\otimes \t{W}^{(q)}_{\cS_3,g=1}$
with $(p,q)=(1,2),\, (1,4),\, (1,5), \, (2,3),\,(2,5),\,(3,4),\,(3,5),\,(4,5)$, and $(5,5)$ is that the component
$\t{W}_{\cS_3,g=2}^{\text{I}; a^2,b,z=\mathbf{1}}$ is non-zero.
As illustrated in Sec.\ref{Sec: RuleOutFakeSolution} in the main text, the genus-2 condition imposes that 
$\t{W}_{\cS_3,g=2}^{\text{I}; a^2,b,z=\mathbf{1}}=0$, and therefore the above pairings cannot be the 
solutions of Eq.\eqref{Genus2_gapping_condition_S3}.
Second, the common feature of pairings $\t{W}^{(p)}_{\cS_3,g=1}\otimes \t{W}^{(q)}_{\cS_3,g=1}$ with 
$(p,q)=(1,3),\, (1,4),\,(1,5),\,(2,3),\,(2,4)$, and $(2,5)$ is that 
the component $\t{W}_{\cS_3,g=2}^{\text{I}; a^1,c,z=\mathbf{1}}$ is non-zero.
In the following, we will show that the genus-2 condition imposes that $\t{W}_{\cS_3,g=2}^{\text{I}; a^1,c,z=\mathbf{1}}=0$,
and therefore the above pairings cannot be the solutions of Eq.\eqref{Genus2_gapping_condition_S3}.

Considering the fusion rule $a^2\otimes c=c^1$, the only allowed component $|\psi^{\text{II}; a^2, c, z}\rangle$
in \eqref{BasisII_appendix} is for $z=c^1$. Since $\theta_{c^1}=-1$, we have 
\begin{equation}\label{Overlap_a1c_0}
\langle \Psi^{ac}_{g=2}|\psi^{\text{II}; a^2, c, z=c^1}\rangle=0,
\end{equation}
which results from Eq.\eqref{Genus2_gapping_condition_S3} with $R_{\cS_3}=T_5$.
Now we insert a complete set of basis vectors I into the above equation, and obtain
\begin{equation}\label{Overlap_a1c}
\small
\begin{split}
\langle \Psi_{g=2}^{ac}|\psi^{\text{II}; a^1,c,c_1}\rangle
=&\sum_{i,j,z}\langle \Psi_{g=2}^{ac}| \psi^{\text{I}; i,j,z}\rangle
\langle  \psi^{\text{I}; i,j,z}|\psi^{\text{II}; a^1,c,c_1}\rangle\\
=&
\langle \Psi_{g=2}^{ac}| \psi^{\text{I}; a^1,c,z=\textbf{1}}\rangle
\langle  \psi^{\text{I}; a^1,c,z=\textbf{1}}|\psi^{\text{II}; a^1,c,c_1}\rangle\\
\end{split}
\end{equation}
where we have considered the fusion rules 
$a^1\otimes a^1=\mathbf{1}$, 
and $c\otimes c=\mathbf{1}\oplus a^2\oplus b\oplus b^1\oplus b^2$, and the only allowed component in
$| \psi^{\text{I}; a^1,c,z}\rangle$ is for $z=\textbf{1}$.
Since $\langle  \psi^{\text{I}; a^1,c,z=\textbf{1}}|\psi^{\text{II}; a^1,c,c_1}\rangle=\sqrt{d_c\, d_{a^1} d_{c^1}}/D$
which is nonzero in Eq.\eqref{Overlap_a1c}, then based on Eqs.\eqref{Overlap_a1c_0} and \eqref{Overlap_a1c}, we obtain
\begin{equation}
\t{W}_{\cS_3,g=2}^{\text{I}; a^1,c,z=\textbf{1}}:=\langle \Psi_{g=2}^{ac}| \psi^{\text{I}; a^1,c,z=\textbf{1}}\rangle=0.
\end{equation}
This means the pairings $\t{W}^{(p)}_{\cS_3,g=1}\otimes \t{W}^{(q)}_{\cS_3,g=1}$ with 
$(p,q)=(1,3),\, (1,4),\,(1,5),\,(2,3),\,(2,4)$, and $(2,5)$ cannot be the solutions of genus-2 condition in
Eq.\eqref{Genus2_gapping_condition_S3}.
\\
\\
\underline{\textit{Solutions of genus-2 condition:}}
\\
\\
Now let us check the pairings of genus-1 solutions
$\t{W}^{(p)}_{\cS_3,g=1}\otimes \t{W}^{(p)}_{\cS_3,g=1}$ 
with $p=1,\, 2,\, 3,\,4$ respectively in Table \ref{S3_solution_appendix}.
\\
\\
\underline{$\t{W}^{(1)}_{\cS_3,g=1}\otimes \t{W}^{(1)}_{\cS_3,g=1}$:}

The genus-1 solution $\t{W}^{(1)}_{\cS_3,g=1}$ in \eqref{Genus1_solution} corresponds to flux condensations
of anyons $\mathbf{1}$, $b$, and $c$. (The relevant fusion rules are 
$b\otimes b=\mathbf{1}\oplus a^1\oplus b$,
$c\otimes c=\mathbf{1}\oplus a^2\oplus b\oplus b^1\oplus b^2$,
and $b\otimes c=c\oplus c^1$.)
Based on the genus-1 solution, we have
\begin{equation}
\t{W}_{\cS_3,g=2}^{\text{I}; i, j, z=\mathbf{1}}=1, \quad \forall \, i,j\in\{\mathbf{1},\, b,\, c\},
\end{equation}
and 
$\t{W}_{\cS_3,g=2}^{\text{I}; i, j, z=\mathbf{1}}=0$, if there $\exists \, i,j\notin\{\mathbf{1},\, b,\, c\}$.

In the following, we will show that $\t{W}_{\cS_3,g=2}^{\text{I}; i, j, z=\mathbf{1}}$ are coupled to certain 
$\t{W}_{\cS_3,g=2}^{\text{I}; i, j, z}$ with $z\neq\mathbf{1}$ through the operation $R_{\cS_3}=T_5$. 
We will frequently use the basis transformation in Eq.\eqref{basisTransform}.
Let us start from $i=b$ and $j=c$.
Then for $|\psi^{\text{I}; b, c, z}\rangle=[F^{bc}_{bc}]_{zz'} \, |\psi^{\text{II}; b, c, z'}\rangle$, 
where $z\in\{\mathbf{1}, b\}$ and $z'\in\{c, \, c_1\}$, one can find that
\begin{equation}\label{F_bc_bc}
\small
[F^{bc}_{bc}]_{zz'}=\frac{1}{\sqrt{2}}
\left(
\begin{array}{cccc}
1 &1\\
1 &-1
\end{array}
\right).
\end{equation}
Since the action of $T_5$ on $|\psi^{\text{II}; b, c, z'}\rangle$ is simply 
$T_5 |\psi^{\text{II}; b, c, z'}\rangle=\theta_{z'}|\psi^{\text{II}; b, c, z'}\rangle$, 
then based on the basis transformation in Eqs.\eqref{basisTransform} and \eqref{F_bc_bc}
one can find that $T_5$ in the basis $\{ |\psi^{\text{I}; b, c, \mathbf{1}}\rangle, |\psi^{\text{I}; b, c, b}\rangle\}$
has the expression:
\begin{equation}\label{T5_bc_bc}
\small
T_5=
\left(
\begin{array}{cccc}
0 &1\\
1 &0
\end{array}
\right).
\end{equation}
By solving Eq.\eqref{Genus2_gapping_condition_S3}, one can obtain
\begin{equation}\label{W_bc_z}
\t{W}_{\cS_3,g=2}^{\text{I}; b, c, b}=\t{W}_{\cS_3,g=2}^{\text{I}; b, c, \mathbf{1}}=1.
\end{equation}
For the case of $i=c$ and $j=b$, the discussion is almost the same, and one can find that
\begin{equation}\label{W_cbb}
\t{W}_{\cS_3,g=2}^{\text{I}; c, b, b}=\t{W}_{\cS_3,g=2}^{\text{I}; c, b, \mathbf{1}}=1.
\end{equation}

Next, let us consider the case $i=j=c$.
For $|\psi^{\text{I}; c, c, z}\rangle=[F^{cc}_{cc}]_{zz'} \, |\psi^{\text{II}; c, c, z'}\rangle$, 
where $z, \, z'\in\{\mathbf{1}, a^2, \, b, \, b^1, \, b^2\}$, one can find that
\begin{equation}\label{F_cc_cc}
\small
[F^{cc}_{cc}]_{zz'}=\frac{1}{3}
\left(
\begin{array}{ccccc}
1 &\sqrt{2}  &\sqrt{2} &\sqrt{2} &\sqrt{2}\\
\sqrt{2} &2  &-1 &-1 &-1 \\
\sqrt{2} &-1  &2 &-1 &-1 \\
\sqrt{2} &-1  &-1 &-1 &2 \\
\sqrt{2} &-1  &-1 &2 &-1 \\
\end{array}
\right).
\end{equation}
Combining with $T_5 |\psi^{\text{II}; c, c, z'}\rangle=\theta_{z'}|\psi^{\text{II}; c, c, z'}\rangle$, 
one can find the expression of $T_5$ in the basis
$\{ |\psi^{\text{I}; c, c, \mathbf{1}}\rangle,\, |\psi^{\text{I}; c, c, a^2}\rangle, \,
|\psi^{\text{I}; c, c, b}\rangle,\, |\psi^{\text{I}; c, c, b^1}\rangle,\, |\psi^{\text{I}; c, c, b^2}\rangle\}$ 
as follows
\begin{equation}\label{T5_z_a1}
\small
T_5=\frac{1}{3}\left(
\begin{array}{cccccc}
1 &\sqrt{2} &\sqrt{2} &\sqrt{2}e^{-\frac{i2\pi}{3}} &\sqrt{2}e^{\frac{i 2 \pi}{3}}\\
\sqrt{2} &2 &-1 &e^{\frac{i\pi}{3}} & e^{-\frac{i\pi}{3}}\\
\sqrt{2} &-1 &2 &e^{\frac{i\pi}{3}} &e^{-\frac{i\pi}{3}}\\
\sqrt{2}e^{-\frac{i2\pi}{3}} &e^{\frac{i\pi}{3}} &e^{\frac{i\pi}{3}} &e^{-\frac{i\pi}{3}} &2\\
\sqrt{2}e^{\frac{i2\pi}{3}} &e^{-\frac{i\pi}{3}} &e^{-\frac{i\pi}{3}} &2 &e^{\frac{i\pi}{3}}
\end{array}
\right),
\end{equation}
By solving Eq.\eqref{Genus2_gapping_condition_S3} with
$\t{W}_{\cS_3,g=2}^{\text{I}; c, c, z=\mathbf{1}}=1$ and
$\t{W}_{\cS_3,g=2}^{\text{I}; c, c, b^1}=\t{W}_{\cS_3,g=2}^{\text{I}; c, c, b^2}=0$, one can obtain
\begin{equation}\label{W_cca2_ccb}
\t{W}_{\cS_3,g=2}^{\text{I}; c, c, a^2}+\t{W}_{\cS_3,g=2}^{\text{I}; c, c, b}=\sqrt{2}.
\end{equation}
To further determine the concrete value of $\t{W}_{\cS_3,g=2}^{\text{I}; c, c, a^2}$ and
$\t{W}_{\cS_3,g=2}^{\text{I}; c, c, b}$, we resort to Eqs.\eqref{Puncture_z_a2} and \eqref{Puncture_z_b},
based on which one can find that
\begin{equation}\label{W_cca2_ca2a2}
\small
\t{W}_{\cS_3,g=2}^{\text{I}; c, c,
a^2}=-\sqrt{2}\,\t{W}_{\cS_3,g=2}^{\text{I}; c, a^2, a^2}= -\sqrt{2}\,\t{W}_{\cS_3,g=2}^{\text{I};  a^2,c, a^2}
=2\, \t{W}_{\cS_3,g=2}^{\text{I}; a^2, a^2, a^2},
\end{equation}
and
\begin{equation}\label{Wccb_Wcbb}
\small
\t{W}_{\cS_3,g=2}^{\text{I}; c, c, b}=\sqrt{2}\,\t{W}_{\cS_3,g=2}^{\text{I};
c, b, b}=\sqrt{2} \, \t{W}_{\cS_3,g=2}^{\text{I}; b, c, b}=2 \, \t{W}_{\cS_3,g=2}^{\text{I}; b, b, b}.
\end{equation}

Since we have obtained $\t{W}_{\cS_3,g=2}^{\text{I}; c, b, b}=1$ in
Eq.\eqref{W_cbb}, then based on Eq.\eqref{Wccb_Wcbb} we have
\begin{equation}
\small
\t{W}_{\cS_3,g=2}^{\text{I}; c, c, b}=\sqrt{2}, \,
\t{W}_{\cS_3,g=2}^{\text{I}; c, b, b}=\t{W}_{\cS_3,g=2}^{\text{I}; b, c,
b}=1,\, \t{W}_{\cS_3,g=2}^{\text{I}; b, b, b}=\frac{1}{\sqrt{2}}.
\end{equation}
Then from Eq.\eqref{W_cca2_ccb}, one has $\t{W}_{\cS_3,g=2}^{\text{I}; c, c, a^2}=0$. 
Based on Eq.\eqref{W_cca2_ca2a2}, one has 
$\t{W}_{\cS_3,g=2}^{\text{I}; c, c, a^2}=\t{W}_{\cS_3,g=2}^{\text{I}; c,
a^2, a^2}= \t{W}_{\cS_3,g=2}^{\text{I};  a^2,c, a^2}
=\t{W}_{\cS_3,g=2}^{\text{I}; a^2, a^2, a^2}=0$.

Till now, we have obtained certain $\t{W}_{\cS_3,g=2}^{\text{I}; i, j, z}$ that
embeds $\t{W}^{(1)}_{\cS_3,g=1}\otimes \t{W}^{(1)}_{\cS_3,g=1}$. 
In particular, the nonzero components as follows:
\begin{equation}\label{W_zb_nonzero}
\small
\left\{
\begin{split}
&\t{W}_{\cS_3, g=2}^{\text{I};b,c,z=b}=\t{W}_{\cS_3, g=2}^{\text{I};c,b,z=b}=1, \\
&\t{W}_{\cS_3, g=2}^{\text{I};c,c,z=b}=\sqrt{2},\\
&\t{W}_{\cS_3, g=2}^{\text{I};b,b,z=b}=\frac{1}{\sqrt{2}},
\end{split}
\right.
\end{equation}
and
\begin{equation}\label{W(1)_nonzero}
\t{W}_{\cS_3, g=2}^{\text{I};i,j,z=\mathbf{1}}=1, \quad \forall \, i,j\in\{\mathbf{1}, b, c\}.
\end{equation}

Furthermore, we claim that these are the only non-zero components of 
$\t{W}_{\cS_3,g=2}^{\text{I}; i, j,z}$ that embeds $\t{W}^{(1)}_{\cS_3,g=1}\otimes
\t{W}^{(1)}_{\cS_3,g=1}$.
That is, for all the non-zero components $\t{W}_{\text{I}; S_3,g=2}^{i,j,z}$, 
one can find that $i, j, z\in\{\mathbf{1}, b, c\}$.
All the other ($116-13=103$) components are zero. 
This can be checked explicitly as follows.

First, for $\t{W}_{\cS_3,g=2}^{\text{I}; i, j,z=\mathbf{1}}$ which is the tensor product of genus-1 
solutions, there are $55$ components that are zero.
They are $\t{W}_{\cS_3,g=2}^{\text{I}; i, j,z=\mathbf{1}}$ such that there
$\exists\,i,\, j \notin\{\mathbf{1}, b,c\}$.
Second, there are $18$ components of $\t{W}_{\cS_3,g=2}^{\text{I}; i, j,z}$
which are zero, with $\theta_z\neq 1$.
They correspond to $\t{W}_{\cS_3,g=2}^{\text{I}; i, j,z}$ with $z=b^1,\, b^2$.

The rest components with $z\neq \mathbf{1}$ and $\theta_z=1$ need more careful study.
Let us check them case by case. 

-- For $z=a^1$, we have shown in Eq.\eqref{Wza1_appendix} that
$\t{W}^{\text{I}; i,j,z=a^1}_{\cS_3,g=2}=0$, $\forall \, i,j\in\{a^2,\, b,\, b^1,\, b^2\}$. There are in total $16$ components.

-- For $z=a^2$, as we have analyzed above, all the 
$9$ components $\t{W}_{\cS_3,g=2}^{\text{I}; i, j,z}$ with $z=a^2$ are zero.

-- For $z=b$, we have $\t{W}_{\cS_3,g=2}^{\text{I}; i, j,z=b}=0$, if there $\exists\, i,j=c^1$.
There are in total $5$ components.

-- For $z=c$, this is not allowed by the fusion rules.

Altogether, for all the $116$ genus-2 components $\t{W}_{\cS_3,g=2}^{\text{I}; i, j,z}$ that embed
$\t{W}^{(1)}_{\cS_3,g=1}\otimes \t{W}^{(1)}_{\cS_3,g=1}$,
one can find there are indeed $103$ components 
that are zero, and all the $13$ non-zero components are listed in 
Eqs.\eqref{W_zb_nonzero} and \eqref{W(1)_nonzero}.
\\
\\
\underline{$\t{W}^{(2)}_{\cS_3,g=1}\otimes \t{W}^{(2)}_{\cS_3,g=1}$:}

The genus-1 solution $\t{W}^{(2)}_{\cS_3,g=1}$ in \eqref{Genus1_solution} corresponds to the condensation
of anyons $\mathbf{1}$, $a^2$, and $c$. (The relevant fusion rules are $a^2\otimes a^2=\mathbf{1}\oplus a^1\oplus a^2$,
$c\otimes c=\mathbf{1}\oplus a^2\oplus b\oplus b^1\oplus b^2$,
and $a^2\otimes c=c\oplus c^1$.)
Based on the genus-1 solution, we have
\begin{equation}
\t{W}_{\cS_3,g=2}^{\text{I}; i, j, z=\mathbf{1}}=1, \quad \forall \, i,j\in\{\mathbf{1},\, a^2,\, c\},
\end{equation}
and 
$\t{W}_{\cS_3,g=2}^{\text{I}; i, j, z=\mathbf{1}}=0$, if there $\exists \, i,j\notin\{\mathbf{1},\, a^2,\, c\}$.

Next, let us consider the components with $z\neq \mathbf{1}$.
Let us start from $i=a^2$ and $j=c$.
Then for $|\psi^{\text{I}; a^2, c, z}\rangle=[F^{a^2c}_{a^2c}]_{zz'} \, |\psi^{\text{II}; a^2, c, z'}\rangle$, 
where $z\in\{\mathbf{1}, a^2\}$ and $z'\in\{c, \, c_1\}$, one can find that
\begin{equation}\label{F_a2c_a2c}
\small
[F^{a^2c}_{a^2c}]_{zz'}=\frac{1}{\sqrt{2}}
\left(
\begin{array}{cccc}
1 &1\\
-1 &1
\end{array}
\right).
\end{equation}
Recalling that the action of $T_5$ on $|\psi^{\text{II}; b, c, z'}\rangle$ is simply 
$T_5 |\psi^{\text{II}; a^2, c, z'}\rangle=\theta_{z'}|\psi^{\text{II}; a^2, c, z'}\rangle$, 
based on the basis transformation in \eqref{F_a2c_a2c}, one can obtain the expression of $T_5$
in the basis $\{|\psi^{\text{I}; a^2, c, \mathbf{1}}\rangle,\, |\psi^{\text{I}; a^2, c, a^2}\rangle\}$ as follows
\begin{equation}\label{T5_a2c_a2c}
\small
T_5=
\left(
\begin{array}{cccc}
0 &-1\\
-1 &0
\end{array}
\right).
\end{equation}
By solving Eq.\eqref{Genus2_gapping_condition_S3}, one can obtain
\begin{equation}\label{W_a2c_z}
\t{W}_{\cS_3,g=2}^{\text{I}; a^2, c, a^2}=-\t{W}_{\cS_3,g=2}^{\text{I}; a^2, c, \mathbf{1}}=-1.
\end{equation}
With the same procedure, one can find that
\begin{equation}\label{W_ca2_z}
\t{W}_{\cS_3,g=2}^{\text{I}; c, a^2, a^2}=-\t{W}_{\cS_3,g=2}^{\text{I}; c,a^2, \mathbf{1}}=-1.
\end{equation}
Next, we consider the effect of punctured $S^{(z)}$ matrices. Based on Eq.\eqref{Puncture_z_a2}, 
we have
\begin{equation}\label{W_cca2_ca2a2_02}
\small
\t{W}_{\cS_3,g=2}^{\text{I}; c, c,
a^2}=-\sqrt{2}\,\t{W}_{\cS_3,g=2}^{\text{I}; c, a^2, a^2}= -\sqrt{2}\,\t{W}_{\cS_3,g=2}^{\text{I};  a^2,c, a^2}
=2\, \t{W}_{\cS_3,g=2}^{\text{I}; a^2, a^2, a^2}.
\end{equation}
Then we can obtain
\begin{equation}
\small
\t{W}_{\cS_3,g=2}^{\text{I}; c, c, a^2}=\sqrt{2}, \,
\t{W}_{\cS_3,g=2}^{\text{I}; c, a^2, a^2}=\t{W}_{\cS_3,g=2}^{\text{I}; a^2,
c, a^2}=-1,\, \t{W}_{\cS_3,g=2}^{\text{I}; a^2, a^2, a^2}=\frac{1}{\sqrt{2}}.
\end{equation}
Then we need to solve Eq.\eqref{Genus2_gapping_condition_S3} with $T_5$ in \eqref{T5_z_a1}.
One can find that $\t{W}_{\cS_3,g=2}^{\text{I}; c, c, z=\mathbf{1}}=1$,
$\t{W}_{\cS_3,g=2}^{\text{I}; c, c, b^1}=\t{W}_{\cS_3,g=2}^{\text{I}; c, c, b^2}=0$, and
$\t{W}_{\cS_3,g=2}^{\text{I}; c, c, a^2}+\t{W}_{\cS_3,g=2}^{\text{I}; c, c, b}=\sqrt{2}$.
Since $\t{W}_{\cS_3,g=2}^{\text{I}; c, c, a^2}=\sqrt{2}$, one can obtain
$\t{W}_{\cS_3,g=2}^{\text{I}; c, c, b}=0$. Together with Eq.\eqref{Wccb_Wcbb}, we have
\begin{equation}
\t{W}_{\cS_3,g=2}^{\text{I}; c, c, b}=\t{W}_{\cS_3,g=2}^{\text{I}; c, b,
b}=\t{W}_{\cS_3,g=2}^{\text{I}; b, c, b}=\t{W}_{\cS_3,g=2}^{\text{I}; b, b, b}=0.
\end{equation}
Till now, we have obtained certain $\t{W}_{\cS_3,g=2}^{\text{I}; i, j, z}$ that
embeds $\t{W}^{(2)}_{\cS_3,g=1}\otimes \t{W}^{(2)}_{\cS_3,g=1}$, with the 
non-zero components as follows:
\begin{equation}\label{W_za2_nonzero}
\small
\left\{
\begin{split}
&\t{W}_{\cS_3, g=2}^{\text{I};a^2,c,z=a^2}=\t{W}_{\cS_3, g=2}^{\text{I};c,a^2,z=a^2}=-1, \\
&\t{W}_{\cS_3, g=2}^{\text{I};c,c,z=a^2}=\sqrt{2},\\
&\t{W}_{\cS_3, g=2}^{\text{I};a^2,a^2,z=a^2}=\frac{1}{\sqrt{2}},
\end{split}
\right.
\end{equation}
and
\begin{equation}\label{W(2)_nonzero}
\t{W}_{\cS_3, g=2}^{\text{I};i,j,z=\mathbf{1}}=1, \quad \forall \, i,j\in\{\mathbf{1}, a^2, c\}.
\end{equation}
It is emphasized that for $\t{W}_{\cS_3,g=2}^{\text{I}; i, j,z}$ that embeds 
$\t{W}^{(2)}_{\cS_3,g=1}\otimes \t{W}^{(2)}_{\cS_3,g=1}$, the components in
Eqs.\eqref{W_za2_nonzero} 
and \eqref{W(2)_nonzero} exhaust all the non-zero components.
The discussion is almost the same as that for the case of
$\t{W}^{(1)}_{\cS_3,g=1}\otimes \t{W}^{(1)}_{\cS_3,g=1}$, 
except for the following difference:

-- For $\t{W}_{\cS_3,g=2}^{\text{I}; i, j,z=\mathbf{1}}$ which is the tensor product of genus-1 
solutions, there are $55$ components that are zero.
They are $\t{W}_{\cS_3,g=2}^{\text{I}; i, j,z=\mathbf{1}}$ such that there
$\exists\,i,\, j \notin\{\mathbf{1}, a^2,c\}$.

-- For $z=b$, all the $9$ components $\t{W}_{\cS_3,g=2}^{\text{I}; i, j,z}$ with $z=b$ are zero.

-- For $z=a^2$, we have $\t{W}_{\cS_3,g=2}^{\text{I}; i, j,z=a^2}=0$, if there $\exists\, i,j=c^1$.
There are in total $5$ components of this kind.

Altogether, for all the $116$ genus-2 components $\t{W}_{\cS_3,g=2}^{\text{I}; i, j,z}$ that embed
$\t{W}^{(2)}_{\cS_3,g=1}\otimes \t{W}^{(2)}_{\cS_3,g=1}$,
one can find there are $103$ components that are zero, and all the $13$ non-zero components are listed in 
Eqs.\eqref{W_za2_nonzero} and \eqref{W(2)_nonzero}.
\\
\\
\underline{$\t{W}^{(3)}_{\cS_3,g=1}\otimes \t{W}^{(3)}_{\cS_3,g=1}$:}

The genus-1 solution $\t{W}^{(3)}_{\cS_3,g=1}$ in \eqref{Genus1_solution} corresponds to the condensation
of anyons $\mathbf{1}$, $a^1$, and $b$. Relevant fusion rules are $a^1\otimes a^1=\mathbf{1}$, $a^1\otimes b=b$,
and $b\otimes b=\mathbf{1}\oplus a^1\oplus b$..
Based on the genus-1 solution, we have
\begin{equation}\label{sol_1}
\small
\t{W}_{\cS_3,g=2}^{\text{I}; \mathbf{1}, \mathbf{1}, z=\mathbf{1}}=\t{W}_{\cS_3,g=2}^{\text{I}; a^1, \mathbf{1}, z=\mathbf{1}}=
\t{W}_{\cS_3,g=2}^{\text{I}; \mathbf{1}, a^1, z=\mathbf{1}}=\t{W}_{\cS_3,g=2}^{\text{I}; a^1, a^1, z=\mathbf{1}}=1,
\end{equation}
\begin{equation}\label{sol_2}
\small
\t{W}_{\cS_3,g=2}^{\text{I}; \mathbf{1}, b, z=\mathbf{1}}=\t{W}_{\cS_3,g=2}^{\text{I}; b, \mathbf{1}, z=\mathbf{1}}=
\t{W}_{\cS_3,g=2}^{\text{I}; a^1, b, z=\mathbf{1}}=\t{W}_{\cS_3,g=2}^{\text{I}; b, a^1, z=\mathbf{1}}=2,
\end{equation}
and
\begin{equation}\label{sol_3}
\t{W}_{\cS_3,g=2}^{\text{I}; b, b, z=\mathbf{1}}=4.
\end{equation}
One remarkable feature of
$|\psi^{\text{I}; i,j,z=\mathbf{1}}\rangle$ with $i,j\in \{\mathbf{1},a^1,b\}$
is that for the fusion results of $i\times \bar{j}=\sum_z N_{i \bar{j}}^z\, z$ with $N_{i\bar{j}}^z>0$, 
all the topological spins are trivial, \textit{i.e.}, $\theta_z=1$. In this case, based on Eq.\eqref{T5_trivialSpin}, one has 
\begin{equation}
T_5\, |\psi^{\text{I};i,j,z=\mathbf{1}}\rangle=|\psi^{\text{I}; i, j, z=\mathbf{1}}\rangle, \quad i,j\in \{\mathbf{1},a^1, b\}.
\end{equation}
Then one can find that the solutions $\t{W}_{\cS_3,g=2}^{\text{I}; i,j, z=\mathbf{1}}$ in Eqs.\eqref{sol_1}, \eqref{sol_2}, and \eqref{sol_3}
are decoupled from other components $\t{W}_{\cS_3,g=2}^{\text{I}; i,j, z}$ with $z\neq \mathbf{1}$.
In the following, we will show that Eqs.\eqref{sol_1}, \eqref{sol_2}, and \eqref{sol_3} exhaust all the non-zero 
components of $\t{W}_{\cS_3,g=2}^{\text{I}; i,j, z}$ that embed
$\t{W}^{(3)}_{\cS_3,g=1}\otimes \t{W}^{(3)}_{\cS_3,g=1}$.

First, similar to previous discussions, for $\t{W}_{\cS_3,g=2}^{\text{I}; i, j,z=\mathbf{1}}$ which is the tensor product of genus-1 
solutions, there are $55$ components that are zero.
They are $\t{W}_{\cS_3,g=2}^{\text{I}; i, j,z=\mathbf{1}}$ such that there
$\exists\,i,\, j \notin\{\mathbf{1}, a^1,b\}$.
Second, there are $18$ components of $\t{W}_{\cS_3,g=2}^{\text{I}; i, j,z}$
which are zero, with $\theta_z\neq 1$.
They correspond to $\t{W}_{\cS_3,g=2}^{\text{I}; i, j,z}$ with $z=b^1,\, b^2$.

The rest components with $z\neq \mathbf{1}$ and $\theta_z=1$ can be checked 
case by case as follows. 

-- For $z=a^1$, we have shown in Eq.\eqref{Wza1_appendix} that
$\t{W}^{\text{I}; i,j,z=a^1}_{\cS_3,g=2}=0$, $\forall \, i,j\in\{a^2,\, b,\, b^1,\, b^2\}$. There are in total $16$ components.

-- For $z=a^2$ and $b$, 
we need to solve Eq.\eqref{Genus2_gapping_condition_S3} with $T_5$ in \eqref{T5_z_a1}.
Different from the case of $\t{W}^{(1)}_{\cS_3,g=1}\otimes
\t{W}^{(1)}_{\cS_3,g=1}$ and $\t{W}^{(2)}_{\cS_3,g=1}\otimes \t{W}^{(2)}_{\cS_3,g=1}$,
now we have $\t{W}_{\cS_3,g=2}^{\text{I}; c, c, z=\mathbf{1}}=0$, 
which is set by the genus-1 solutions $\big(\t{W}^{(3)}_{\cS_3,g=1}\big)^c
\cdot \big(\t{W}^{(3)}_{\cS_3,g=1}\big)^{c}=0$.
Note also that $\t{W}_{\cS_3,g=2}^{\text{I}; c, c, b^1}=\t{W}_{\cS_3,g=2}^{\text{I}; c, c, b^2}=0$ because of the nontrivial 
topological spins of $b^1$ and $b^2$.
Then by solving Eq.\eqref{Genus2_gapping_condition_S3} with $T_5$ in \eqref{T5_z_a1} we can get
\begin{equation}
\t{W}_{\cS_3,g=2}^{\text{I}; c, c, a^2}=\t{W}_{\cS_3,g=2}^{\text{I}; c, c, b}=0.
\end{equation}
Then based on Eqs.\eqref{Puncture_z_a2} and \eqref{Puncture_z_b}, we can find that all the 
$18$ components of $\t{W}_{\cS_3,g=2}^{\text{I}; i, j, a^2}$ and $\t{W}_{\cS_3,g=2}^{\text{I}; i, j, b}$ are zero. 

In short, for $\t{W}_{\cS_3,g=2}^{\text{I}; i,j,z}$ that embed
$\t{W}^{(3)}_{\cS_3,g=1}\otimes \t{W}^{(3)}_{\cS_3,g=1}$, one can find 
there are in total $107$ components that are zero, and $9$ non-zero components which are listed in 
Eqs.\eqref{sol_1}, \eqref{sol_2}, and \eqref{sol_3}.
\\
\\
\underline{$\t{W}^{(4)}_{\cS_3,g=1}\otimes \t{W}^{(4)}_{\cS_3,g=1}$:}

The genus-1 solution $\t{W}^{(4)}_{\cS_3,g=1}$ in \eqref{Genus1_solution} corresponds to the condensation
of anyons $\mathbf{1}$, $a^1$, and $a^2$. Relevant fusion rules are $a^1\otimes a^1=\mathbf{1}$, $a^1\otimes a^2=a^2$,
and $a^2\otimes a^2=\mathbf{1}\oplus a^1\oplus a^2$.
Based on the genus-1 solution, we have
\begin{equation}\label{sol_1b}
\small
\t{W}_{\cS_3,g=2}^{\text{I}; \mathbf{1}, \mathbf{1}, z=\mathbf{1}}=\t{W}_{\cS_3,g=2}^{\text{I}; a^1, \mathbf{1}, z=\mathbf{1}}=
\t{W}_{\cS_3,g=2}^{\text{I}; \mathbf{1}, a^1, z=\mathbf{1}}=\t{W}_{\cS_3,g=2}^{\text{I}; a^1, a^1, z=\mathbf{1}}=1,
\end{equation}
\begin{equation}\label{sol_2b}
\small
\t{W}_{\cS_3,g=2}^{\text{I}; \mathbf{1}, a^2, z=\mathbf{1}}=\t{W}_{\cS_3,g=2}^{\text{I}; a^2, \mathbf{1}, z=\mathbf{1}}=
\t{W}_{\cS_3,g=2}^{\text{I}; a^1, a^2, z=\mathbf{1}}=\t{W}_{\cS_3,g=2}^{\text{I}; a^2, a^1, z=\mathbf{1}}=2,
\end{equation}
and
\begin{equation}\label{sol_3b}
\t{W}_{\cS_3,g=2}^{\text{I}; a^2, a^2, z=\mathbf{1}}=4.
\end{equation}
Similar to the case of $\t{W}^{(3)}_{\cS_3,g=1}\otimes \t{W}^{(3)}_{\cS_3,g=1}$, in the fusion results of 
$i\times \bar{j}=\sum_z N_{i \bar{j}}^z\, z$ with $N_{i\bar{j}}^z>0$
and $i,j\in \{\mathbf{1},a^1,a^2\}$,
all the topological spins of anyon $z$ are trivial, \textit{i.e.}, $\theta_z=1$. In this case, based on Eq.\eqref{T5_trivialSpin}, one has 
\begin{equation}
T_5\, |\psi^{\text{I};i,j,z=\mathbf{1}}\rangle=|\psi^{\text{I}; i, j, z=\mathbf{1}}\rangle, \quad i,j\in \{\mathbf{1},a^1,a^2\}.
\end{equation}
Then, again, one can find that the solutions $\t{W}_{\cS_3,g=2}^{\text{I}; i,j, z=\mathbf{1}}$ in Eqs.\eqref{sol_1b}, \eqref{sol_2b}, and \eqref{sol_3b}
are decoupled from other components $\t{W}_{\cS_3,g=2}^{\text{I}; i,j, z}$ with $z\neq \mathbf{1}$.
We will show that Eqs.\eqref{sol_1b}, \eqref{sol_2b}, and \eqref{sol_3b} exhaust all the non-zero 
components of $\t{W}_{\cS_3,g=2}^{\text{I}; i,j, z}$ that embed
$\t{W}^{(4)}_{S_,g=1}\otimes \t{W}^{(4)}_{\cS_3,g=1}$.

The discussion is the same as that for $\t{W}^{(3)}_{\cS_3,g=1}\otimes \t{W}^{(3)}_{\cS_3,g=1}$ except for the following difference:
 For $\t{W}_{\cS_3,g=2}^{\text{I}; i, j,z=\mathbf{1}}$ which is the tensor product of genus-1 
solutions, there are $55$ components that are zero.  They are $\t{W}_{\cS_3,g=2}^{\text{I}; i, j,z=\mathbf{1}}$ such that there
$\exists\,i,\, j \notin\{\mathbf{1}, a^1,a^2\}$.

Then one can find that for $\t{W}_{\cS_3,g=2}^{\text{I}; i,j,z}$ that embed 
$\t{W}^{(4)}_{\cS_3,g=1}\otimes \t{W}^{(4)}_{\cS_3,g=1}$, 
there are in total $107$ components that are zero, and $9$ non-zero components which are listed in 
Eqs.\eqref{sol_1b}, \eqref{sol_2b}, and \eqref{sol_3b}.

\underline{Summary:}

As a summary of this section, we solve all the components of $\t{W}_{\cS_3,g=2}^{\text{I}; i, j,z}$ with genus-2 condition
 in Eq.\eqref{Genus2_gapping_condition_S3} with the basis vectors chosen in \eqref{BasisI_appendix}.
 There are in total $4$ sets of independent solutions, which embed the genus-1 solutions of the form
 $\t{W}^{(i)}_{\cS_3,g=1}\otimes \t{W}^{(i)}_{\cS_3,g=1}$ with $i=1,\, 2,\, 3,\, 4$, respectively.
 The genus-1 solution $\t{W}^{(5)}_{\cS_3,g=1}$ in Eq.\eqref{Genus1_solution} is ruled out by the genus-2 condition.
 
 In the genus-2 solutions, the only non-zero components are listed in 
 Eqs.\eqref{W_zb_nonzero} and \eqref{W(1)_nonzero} for the solutions that embed 
  $\t{W}^{(1)}_{\cS_3,g=1}\otimes \t{W}^{(1)}_{\cS_3,g=1}$, Eqs.\eqref{W_za2_nonzero} 
and \eqref{W(2)_nonzero} for the solutions that embed
$\t{W}^{(2)}_{\cS_3,g=1}\otimes \t{W}^{(2)}_{\cS_3,g=1}$, 
Eqs.\eqref{sol_1}, \eqref{sol_2}, and \eqref{sol_3} for the solutions that embed
$\t{W}^{(3)}_{\cS_3,g=1}\otimes \t{W}^{(3)}_{\cS_3,g=1}$, 
and Eqs.\eqref{sol_1b}, \eqref{sol_2b}, and \eqref{sol_3b} for the solutions that 
embed $\t{W}^{(4)}_{\cS_3,g=1}\otimes \t{W}^{(4)}_{\cS_3,g=1}$.

\section{Structure coefficients of condensable algebra and ground-state overlap}
\label{structure}

It is known that the gapped boundary is closely related to anyon condensation,
whose data is fully encoded in the so-called condensable
algebra~\cite{K13078244} in the unitary modular tensor category (UMTC) that
describes the anyons.

Let $(\CC,c)$ and $(\mathcal{D},c)$ be two 2d topological orders with the same
chiral central charge $c$, where
$\CC$ and $\mathcal{D}$ are two UMTC's.
We assume that they are Witt equivalent, in other words, they can be connected by a gapped domain wall. 
There are two equivalent mathematical descriptions of such a gapped domain wall: 
\begin{itemize}

\item A gapped domain wall between $(\CC,c)$ and $(\mathcal{D},c)$ can be described by a unitary fusion category $\CM$, 
which is equipped with a unitary braided monoidal equivalence $$\phi_\CM: \cC
\boxtimes \overline{\DC} \xrightarrow{\simeq} Z(\CM).$$
where
$\CC \boxtimes \overline{\mathcal{D}}$ is the UMTC describing the topological
order obtained by stacking $\cC$ with the time-reversal of $\DC$ and
$Z(\cM)$ denotes the Drinfeld center of $\cM$.

\item A gapped domain wall between $(\CC,c)$ and $(\mathcal{D},c)$ is uniquely
  determined by a Lagrangian condensable algebra $A$ in $\cC\boxtimes
  \overline{\DC}$. Here Lagrangian means that $A$ is ``maximal'' in the sense
  \begin{align}
    (\dim A)^2=\dim \cC\dim \DC.
  \end{align}
  In
  this case, the gapped domain wall is described by the UFC $(\cC\boxtimes
  \overline{\DC})_A$, i.e. the category of right $A$-modules in $\cC\boxtimes
  \overline{\DC}$. A Lagrangian algebra $A$, as an object in $\cC\boxtimes
  \overline{\DC}$, can be decomposed as follows: 
  \begin{align}
    A = \sum_{i\in\Irr(\CC), j\in \Irr(\mathcal{D})} (i\boxtimes j)^{\oplus
    \t{W}^{ji}_{\DC\cC,1}}, 
  \end{align}
where $\Irr(\CC)$ and
$\Irr(\mathcal{D})$ denote the sets of isomorphic classes of simple objects.
\end{itemize}

In particular, when $\DC$ is the trivial topological order, the above
describes a gapped boundary of $\cC$, and $A$ is a Lagrangian algebra in
$\cC$. The existence of a Lagrangian algebra, for example, in terms of the structure
coefficients to be introduced in this section, is a necessary and sufficient condition for
the existence of a gapped boundary. The approached presented in the main text,
using the invariance of wave function overlaps under the modular transformations
(or mapping class group transformations), however, only constitutes necessary
conditions. The advantage of the modular invariance approach is that they are
all linear equations and much easier to solve, as opposed to the non-linear
equations for the structure coefficients such as the associativity condition. For the examples presented in
this paper, each of our solutions does correspond to a Lagrangian algebra. 

Physically, a condensable algebra is just a composite
anyon that becomes the ground state after condensation (\ie it is condensed). For this to be possible,
it is necessary that this composite anyon has special algebraic structures,
which can be thought as a generalization of usual commutative associative
algebra. We know that after picking a basis, the multiplication of a usual
associative algebra can be expressed by the structure coefficients. It is also
the case for the condensable algebra. The structure coefficients encode all the
``special algebraic structures'' of the condensed composite anyon; in
particular, they predict the form of the ground-state overlap.

To explain what a ``basis'' of a composite anyon means in a unitary modular
tensor category, we begin by recalling the (orthonormal) basis of an $n$-dimensional Hilbert
space $H$, which is a
set of state vectors $|\alpha\>$ satisfying
\begin{align}
  \<\alpha|\beta\>=\delta_{\alpha\beta},\quad \sum_{\alpha=1}^n |\alpha\> \<\alpha|=\id_H.
\end{align}
To generalize this notion, note that an $n$-dimensional Hilbert space can be
viewed as a composite anyon, that is composed of $n$ trivial anyons, $H=\oplus
\one^{\oplus n}$. The trivial
anyon is the tensor unit $\one$, namely the ground field $\C$, in the
tensor category. Moreover, a vector $|\alpha\>$ can be identified with the
embedding linear map
$q_\alpha:\C \to H$, by $q_\alpha(\lambda)=\lambda|\alpha\>$. Similarly,
$\<\alpha|$ can be
identified with the projection $p_\alpha: H\to \C$, by
$p_a(|\psi\>)=\<\alpha|\psi\>$.

Now, given a composite anyon 
\begin{align}
A=\oplus i^{\oplus N^A_i}, 
\end{align}
where $i$ labels the
simple anyons, a basis of $A$ is a
set of morphisms (``linear maps'' in generic tensor category)
$p^{i,\alpha}_A: A\to i ,$ $q_{i,\alpha}^A: i\to A,$ $\alpha=1,\dots,N^A_i$
\begin{equation}
  p^{i,\alpha}_A
  q_{j,\beta}^A=\delta_{ij}\delta_{\alpha\beta}\id_i,\quad\sum_{i}\sum_{\alpha=1}^{N^A_i}q_{i,\alpha}^A
  p_A^{i,\alpha}=\id_A. \label{fcdecom}
\end{equation}
We usually require that $p^{i,\alpha}_A,q_{i,\alpha}^A$ are Hermitian conjugates:
\begin{align}
  p^{i,\alpha}_A=(q_{i,\alpha}^A)^\dag.
\end{align}
It is intuitive to use the following graphs for the basis:
\SelectTips{lu}{12}
\begin{equation}
  \g{[u]:|-@{<}^i^>\alpha[d]-|-@{<}^A[d]}=p_A^{i,\alpha},
\ \ \ \ \ \ \ \
\g{[d]:|-@{>}_i_>\alpha[u]-|-@{>}_A[u]}=q_{i,\alpha}^A.
\end{equation}
Our choice of symbol is to remind the reader of the similarity to the basis of
Hilbert spaces (rotate the graph by $90^\circ$ anticlockwise).
They satisfy similar orthonormal and complete conditions:
\begin{gather}
\label{iA}
  \g{:|-@{<}_<i^>{\alpha\ \ }[r]-|-@{<}_>A[r]}\sim \< i,\alpha|,\quad
  \g{:|-@{>}^<i_>{\ \ \alpha}[l]-|-@{>}^>A[l]}\sim |\alpha,i\>,
\\
    \g{:|-@{<}_<i^>{\alpha\ \ }[r]-|-@{<}_A[r],[rrr]:|-@{>}^<j_>\beta[l]}\sim \<
    i,\alpha|\beta,j\>=\delta_{ij}\delta_{\alpha\beta},
\\
    \sum_{i\alpha}\g{:|-@{<}_i^>{\ \ \alpha}[r]-|-@{<}_>A[r],[r]:|-@{>}_>{\alpha\ \ }[l]-|-@{>}^>A[l]}\sim
    \sum_{i\alpha}|\alpha,i\>\<
    i,\alpha|=\id_A.
\end{gather}
The only subtle part is that the tensor product and braiding involving
nontrivial $i$ is different from usual intuitions from vector spaces, which
will be explained below.
First, we need to take a basis of the tensor
product of simple anyons $i\ot j$:
\begin{equation}
 \g{[d]-|-@{>}_i[uuu],[rd]-|-@{>}_j[uuu]}=\sum_{k}\sum_{\alpha=1}^{N_k^{ij}}\g{:|-@{>}_<\alpha_>\alpha_k[u],[u]:[d]
[ld]-|-@{>}^i[ru]-|-@{<}^j[rd],[luu]-|-@{<}_i[rd]-|-@{>}_j[ru]}.
\label{eq.dec}
\end{equation}
Choosing the orthonormal basis is equivalent to
choosing $Y$ and $O$ to be identity matrix. The previous choice of
vertex basis is in fact,
\begin{align}
 \gdir{:_k_>\alpha[u]:_j[ur],[u]:^i[ul]}=\left( \frac{d_i d_j}{d_k}
  \right)^{1/4}
 \g{:|-@{>}_>\alpha_k[u],[luu]-|-@{<}_i[rd]-|-@{>}_j[ru]},
 \nonumber\\
  \gdir{:^i[ur]:_k_<\alpha[u],[rr]:_j[ul]}=\left( \frac{d_i d_j}{d_k}
  \right)^{1/4}\g{[u]:|-@{<}^>\alpha^k[d],
 [ld]-|-@{>}^i[ru]-|-@{<}^j[rd],}
 \label{eq.changebasis}
\end{align}
They may be referred to as the ``rotatable basis'', since after the above
rescaling, rotating a vertex (by bending the legs) leads to at most a phase
factor or a unitary matrix, which cancels each other if we are considering a closed graph. Also, as
explained in Sec.~\ref{sec.wfg2}, in the rotatable basis, a closed graph
representing a wave function is properly normalised to a constant
\eqref{eq.norm} that dose not depend on
the anyon and vertex labels.
However, in this section we prefer to use the
orthonormal basis, indicated by the semicircle in the graph, which is more
convenient for open graphs.

The associativity of tensor product is encoded in the $F$-matrix.
\begin{equation}
\g{[uu]-|-@{<}^j[ld]-|-@{>}^i[lu],[ruu]-|-@{<}^k[d]-|-@{<}@(d,ru)[ldd]:|-@{>}@(lu,d)^r^>\alpha[luu],
[dd]:|-@{>}_l_>\beta[u]}=\sum_{s,\chi,\delta}
F^{{i}{j}{r},\alpha\beta}_{{k}{l}{s},\chi\delta}
\g{[luu]-|-@{<}_i[d]-|-@{<}@(d,lu)[rdd]:|-@{>}@(ru,d)_s_>\chi[ruu],[uu]-|-@{<}_j[rd]-|-@{>}_k[ru],
[dd]:|-@{>}_l_>\delta[u]},
\label{eq.F}
\end{equation}
where we used the definition of $F$ matrix in \eqref{IHwave}.
The braiding can be similarly represented by
the $R$-matrix:
\begin{align}
  \g{[d]:|-@{>}_>\alpha_<k[u],-|-@{>}_>i@(ul,dl)[uur],-|-@{>}^>j|(.65)\hole@(ur,dr)[uul]}
  =\sum_b R^{ij}_{k,\beta\alpha}\g{[d]:|-@{>}_>\beta_<k[u],-|-@{>}^>j[ul],-|-@{>}_>i[ur]}.
  \label{eq.R}
\end{align}
We like to remark that, in general the tensors such as $F,R$ matrices do depend on our
choice of basis. However, it is easily verified that the two choices of basis
\eqref{eq.changebasis} differ by the same overall factor on both sides of
\eqref{eq.F}\eqref{eq.R}. Thus, $F,R$ matrices remain the same under the change
of basis \eqref{eq.changebasis}.

For an algebra $A$ in a tensor category, there is a
multiplication morphism $m:A\ot A\to A$.  First
take a basis of $A$ as in \eqref{fcdecom}
\begin{align}
  \id_A= \g{[u]-|-@{<}[dd]^A}=\sum_{i}\sum_{\alpha=1}^{N^A_i}\
  \g{[dd]:|-@{>}[u],-|-@{<}[d]^A^>\alpha:[d]^i^>\alpha-|-@{<}[d]^A}
 =\sum_{i}\sum_{\alpha=1}^{N^A_i}q_{i,\alpha}^A p^{i,\alpha}_A.
\end{align}
The multiplication morphism $m$ is then
\begin{align}
  m=\g{[ld]-|-@{>}^A[ur]m*\cir{}-|-@{>}[u]_>A,"m"-|-@{<}^A[rd]}=
  \sum_{ijk,\alpha\beta\chi}\hskip-10mm
  \g{[uuu]-|-@{<}[d]^A^>\chi:|-@{<}[d]^k^>\chi-|-@{<}^A[d]m*\cir{},
    "m"-|-@{<}[ld]_A_>\alpha:|-@{<}[ld]_i_>\alpha-|-@{<}_A[ld], 
    "m"-|-@{<}[rd]^A^>\beta:|-@{<}[rd]^j^>\beta-|-@{<}^A[rd],[u]:[u],[lldd]:[ru],[rrdd]:[lu]
}.
\end{align}
The central part can be expressed in terms of basis vertices $p_{ij}^{k,\mu}$
\begin{align}
  \g{[uu]:|-@{<}[d]^k^>\chi-|-@{<}^A[d]m*\cir{},
    [lldd]:|-@{>}^i^>\alpha[ru]-|-@{>}^A"m", 
    [rrdd]:|-@{>}_j_>\beta[lu]-|-@{>}_A "m"}
    &=\sum_u M_{i\alpha,j\beta}^{k\chi,\mu}
\g{[u]:|-@{<}[d]^k^>\mu-|-@{<}[ld]_i,-|-@{<}[rd]^j}\nonumber\\
&=\sum_\mu M_{i\alpha,j\beta}^{k\chi,\mu} p_{ij}^{k,\mu}.
\end{align}
Thus
\begin{align}
  m&=\g{[ld]-|-@{<}^A[ur]m*\cir{}-|-@{>}[u]_>A,"m"-|-@{<}^A[rd]}
  =
  \sum_{ijk,\alpha\beta\chi,\mu}M_{i\alpha,j\beta}^{k\chi,\mu}
  \g{[uu]-|-@{<}[d]^A^>\chi:|-@{<}^k^>\mu[d]:[u],
    :|-@{<}[ld]_i_>\alpha-|-@{<}_A[ld], 
  :|-@{<}[rd]^j^>\beta-|-@{<}^A[rd]}\nonumber\\
  &=\sum_{ijk,\alpha\beta\chi,\mu}M_{i\alpha,j\beta}^{k\chi,\mu}  q_{k,\chi}^A
  p_{ij}^{k,\mu} \left(p_A^{i,\alpha}\ot
  p_A^{j,\beta} \right).
\end{align}
$M_{i\alpha,j\beta}^{k\chi,\mu}$ is the ``structure coefficients'' of the algebra. In the
category of vector spaces $\mathbf{Vec}$, the object labels $i,j,k$ and the vertex label
$\mu$ reduce to trivial, and $M_{i\alpha,j\beta}^{k\chi,\mu}$ reduces to structure
coefficients of usual associative algebra, with $\alpha,\beta,\chi$ the labels of basis
vectors. Again, similar to the usual associative algebra, the structure
coefficients depends on the choice of basis, both the basis of $A$,
$p^{i,\alpha}_A,q^A_{i,\alpha}$
and the vertex basis $p_{ij}^{k,\mu}$. It is easy to write down transformations of
$M_{i\alpha,j\beta}^{k\chi,\mu}$ under a change of basis. For example, using the
rotatable basis \eqref{eq.changebasis}, the corresponding structure
coefficients is
\begin{align}
  \tilde{M}_{i\alpha,j\beta}^{k\chi,\mu}=\left( \frac{d_k}{d_id_j} \right)^{1/4}
  M_{i\alpha,j\beta}^{k\chi,\mu}.
\end{align}
Structure coefficients related by a change of basis are considered equivalent and describe the same algebra.

Now we are ready to define the condensable algebra
$(A,M_{i\alpha,j\beta}^{k\chi,\mu})$, by listing the defining, i.e., sufficient and
necessry conditions of $M_{i\alpha,j\beta}^{k\chi,\mu}$:
\begin{enumerate}
  \item Associative:
  \begin{align}
    \g{m*\cir<2ex>{}="m1",[ur]m*\cir<2ex>{}="m2",
    [ld]-|-@{>}^A"m1"-|-@{>}^A"m2"-|-@{>}^A[ur],"m1"-|-@{<}^A[rd],"m2"-|-@{<}^A[rrdd]}=
    \g{m*\cir<2ex>{}="m1",[ul]m*\cir<2ex>{}="m2",
    [rd]-|-@{>}_A"m1"-|-@{>}_A"m2"-|-@{>}_A[ul],"m1"-|-@{<}_A[ld],"m2"-|-@{<}_A[lldd]},
  \end{align}

    \begin{align}
      \sum_{\omega}M_{i\alpha,j\beta}^{r\omega,\mu}M_{r\omega,k\chi}^{l\delta,\nu}=\sum_{s\xi\psi\zeta}F^{{i}{j}{r},\mu
      \nu}_{{k}{l}{s},\xi\psi}
      M_{j\beta,k\chi}^{s\zeta,\xi}M_{i\alpha,s\zeta}^{l\delta,\psi}.\label{eq.strucoeasso}
    \end{align}
  \item Unital: There exists ``the unit of the multiplication'', the unit morphism
    $\eta:\one\to A$,
  \begin{align}
    \g{m*\cir<2ex>{ }="m",[ld]{\eta}*\cir<2ex>{ }="e",
    "m"-|-@{<}_A"e",[u]-|-@{<}^<A"m"-|-@{<}^A[dd]}
    =\ \g{-|-@{<}^A[dd]}=
    \g{m*\cir<2ex>{ }="m",[rd]{\eta}*\cir<2ex>{ }="e",
    "m"-|-@{<}^A"e",[u]-|-@{<}_<A"m"-|-@{<}_A[dd]}\ ,
  \end{align}
  expressed in terms of embedding basis $\eta=\eta_\alpha
    q_{\one,\alpha}^A$, and
    \begin{align}
      \sum_{\alpha=1}^{N^A_\one} \eta_\alpha M_{\one
      \alpha,j\beta}^{k\chi,1}=\sum_{\alpha=1}^{N^A_\one} \eta_\alpha M_{j\beta,\one
      \alpha}^{k\chi,1}=\delta_{jk}\delta_{\beta\chi}.
    \end{align}
  \item Connected:  The range of index $\alpha$ in the above
    expression is $N^A_\one=1$, thus
    \begin{align}
      M_{\one 1,j\beta}^{k\chi,1}= M_{j\beta,\one
      1}^{k\chi,1}=\eta_1^{-1}\delta_{jk}\delta_{\beta\chi}.\label{eq.unicon}
    \end{align}
  \item Isometric:
  \begin{align}
    \g{m*\cir<2ex>{ }="m",[dd]{m^\dag}*\cir<2ex>{ }="md",
    "m"-|-@{>}_>A[u],"md"-|-@{<}^>A[d],
    "m"-|-@{<}@/l4ex/_A"md", "md"-@/l4ex/"m",
    "m"-|-@{<}@/r4ex/^A"md", "md"-@/r4ex/"m",
  }=\ \g{-|-@{<}^A[dd]}\ .
  \end{align}
    \begin{align}
      \sum_{i\alpha j\beta\mu}
      M_{i\alpha,j\beta}^{k\chi,\mu}\left(M_{i\alpha,j\beta}^{k'\chi',\mu}\right)^*=\delta_{kk'}\delta_{\chi\chi'}.
      \label{eq.strucoeiso}
    \end{align}
  \item  Commutative:
  \begin{align}
    \g{m*\cir<2ex>{ }="m"-|-@{>}_>A[u],"m"-|-@{<}^>A[rd],"m"-|-@{<}_>A[ld]}
    =\g{m*\cir<2ex>{ }="m"-|-@{>}_>A[u], "m"-|-@{<}@(rd,ru)_>A[ldd],
    "m"-|-@{<}@(ld,lu)^>A[rdd]|(0.66)\hole}.
  \end{align}
    \begin{align}
      M_{i\alpha,j\beta}^{k\chi,\mu}=\sum_\nu R^{ij}_{k,\nu\mu}
      M_{j\beta,i\alpha}^{k\chi,\nu}.
      \label{eq.strucoeacom}
    \end{align}
\end{enumerate}
    A unital connected isometric algebra in a unitary tensor category
    automatically satisfies the Frobenius condition~\cite{LR9604008}:
  \begin{align}
    \g{{m^\dag}*\cir<2ex>{}="md"-|-@{>}_A [ur] m*\cir<2ex>{}="m","md"-|-@{>}^>A[uu],"md"-|-@{<}_>A[d],
    "m"-|-@{>}_>A[u],"m"-|-@{<}^>A[dd]}
    =
    \g{m*\cir<2ex>{}="m"-|-@{>}_A [u] {m^\dag}*\cir<2ex>{}="md","m"-|-@{<}_>A[rd],"m"-|-@{<}^>A[ld],
    "md"-|-@{>}^>A[ru],"md"-|-@{>}_>A[lu]}=
    \g{{m^\dag}*\cir<2ex>{}="md"-|-@{>}^A [ul] m*\cir<2ex>{}="m","md"-|-@{>}_>A[uu],"md"-|-@{<}^>A[d],
    "m"-|-@{>}^>A[u],"m"-|-@{<}_>A[dd]}.
    \label{eq.Fro}
  \end{align}
  As a direct corollary, $A$ is self dual and
  \begin{align}
    |\eta_1|^2={\dim A}={\sum_i N^A_i d_i},
  \end{align}
  where $\eta_1$ is the from the unit morphism $\eta=\eta_1 q_{\one,1}^A:\one
  \to A$. 
We choose the phase of $\eta_1$ and let
\begin{align}
 \eta_1=\sqrt{\sum_i N^A_i d_i}.
\end{align}
Also, by attaching counit $\eta^\dag$ to topright $A$
  in \eqref{eq.Fro}, we obtain the following useful formula to exchange the
  upper
  and lower indices of $M_{i\alpha,j\beta}^{k\chi,\mu}$
  \begin{align}
\label{MMFM}
    M_{i\alpha,j\beta}^{k\chi,\mu}=\sum_{\gamma,\nu} \left( M_{k\chi,\bar j\gamma}^{i\alpha,\nu}
  \right)^* F^{k \bar{j}i,\nu\mu}_{jk\one,11} M_{\bar j\gamma,j\beta}^{\one 1,1}\eta_1^*,
  \end{align}
  where $F^{k \bar{j}i,\nu\mu}_{jk\one,11}$ comes from the following 
  \begin{align}
    \g{:^i_>\nu|-@{>}[u]-^k|-@{>}[uu],[u]-_{\bar
    j}|-@{>}[ur]-^j|<@{|}|-@{<}[dd],[uuur]:^\one@{.}^>1[d]}=\sum_\mu
    F^{k \bar{j}i,\nu\mu}_{jk\one,11}\g{:^k^>\mu|-@{<}[d]-^j|-@{<}[rd],[d]-_i|-@{<}[ld]}.
  \end{align}
  Similarly, attaching $\eta^\dag$ to topleft $A$, we obtain
  \begin{align}
    M_{i\alpha,j\beta}^{k\chi,\mu} 
    =\sum_{\gamma,\nu} \left(M_{\bar{i}\gamma,k\chi}^{j\beta,\nu}\right)^*
    \left( F^{i\bar{i} \one, 11}_{kkj,\nu\mu} \right)^*
      M_{i\alpha,\bar{i}\gamma}^{\one 1,1}\eta_1^*.
  \end{align}

Let us compute $M_{\bar i\gamma,i\beta}^{\one 1,1}$.
From \eqn{MMFM}, we obtain
\begin{align}
M_{\one 1,i\beta}^{i\chi,1}
&=
\sum_{\gamma} \left( M_{i\chi,\bar i\gamma}^{\one 1,1}
  \right)^* F^{i \bar{i}\one,11}_{ii\one,11} M_{\bar i\gamma,i\beta}^{\one 1,1}\eta_1^*,
\end{align}
Using $ F^{i\bar{i} \one, 11}_{ii\one,11} =d_i^{-1}$, $M_{i\alpha,\bar{i}\gamma}^{\one 1,1} = M_{\bar{i}\gamma,i\alpha}^{\one
1,1}$ and \eqn{eq.unicon}, the above becomes
\begin{align}
 \del_{\bt\chi}=
\frac{\sum_j N^A_j d_j}{d_i}
\sum_{\gamma} \left(M_{i\chi,\bar{i}\gamma}^{\one 1,1}\right)^*
      M_{i\bt,\bar{i}\gamma}^{\one 1,1} 
\end{align}
We see that
$M_{i\alpha,\bar{i}\bt}^{\one 1,1}$ is proportional to a unitary matrix.
By choosing a proper basis for \eqn{iA}
we can make
\begin{align}
 M_{i\alpha,\bar{i}\bt}^{\one 1,1} = \frac{d_i \del_{\al\bt}}{\sqrt{\sum_j N^A_j d_j}} .
\end{align}
Together with \eqref{eq.strucoeacom}, one can permute any pair of $i\alpha$
indices.

The structure coefficients can be used to compute
  the (topological universal part of) overlap between ground states before and
  after condensation. For simplicity, here we only discuss the case that the
  phase after condensation is topologically trivial and has a unique ground
  state.
  We start with the ground state on a torus. Before condensation, the loops of simple objects form a
  basis of the ground space on a torus:
  \begin{align}
    \gdir{:@(r,r)_i[uu]-@(l,l)[dd]}.
  \end{align}
  After condensation, the ground state is, up to an overall factor that depends
  on the system size, the loop of the corresponding condensable algebra $A$:
  \begin{align}
    \gdir{:@(r,r)_A[uu]-@(l,l)[dd]}
    =\sum_i N^A_i \gdir{:@(r,r)_i[uu]-@(l,l)[dd]}.
  \end{align}
Thus we can extract the multiplicity $N^A_i$ from the ground-state overlap on a
torus, which is exactly the universal part of wave function overlap (the labels
in the trivial phase is omitted, see \eqn{W1} and \eqn{M1}) 
\begin{align}
  \t{W}^i_{\cC;1} =M_{\cC}^i =N^A_i.
\end{align}
On a closed two-dimensional surface with genus $g=2$, one choice of the basis before condensation is given by
\begin{align}
  \gdir{:@(u,u)^j[rr]-|>@{|}|>/-2.5pt/@{(}@(d,d)[ll]_>\nu,:_z[lu]-@(d,d)[ll]-|>@{|}|>/-2.5pt/@{(}|-@{<}@(u,u)^i[rr]^>\mu}\ .
      \label{eq.basis1}
\end{align}
Still, after condensation, the ground state is
\begin{align}
  \gdir{ {m^\dag}*\cir<2ex>{}="md":@(u,u)^A[rr]-@(d,d)"md",[lu]m*\cir<2ex>{}="m",
  "md":_A"m"-@(d,d)[ll]-|-@{<}@(u,u)^A"m"}\ .
\end{align}
Now, the structure coefficients kick in. After straightforward calculation, we
find
\begin{align}
  &\qquad\scalebox{.9}{$\gdir{ {m^\dag}*\cir<2ex>{}="md":@(u,u)^A[rr]-@(d,d)"md",[lu]m*\cir<2ex>{}="m",
  "md":_A"m"-@(d,d)[ll]-|-@{<}@(u,u)^A"m"}$}
  \\
  &=\sum_{ijz\mu\nu}\sum_{\alpha\beta\chi}
  M_{i\alpha,z\chi}^{i\alpha,\mu}
 \left(M_{z\chi,j\beta}^{j\beta,\nu}\right)^*
 \scalebox{.9}{$\gdir{:@(u,u)^j[rr]-|>@{|}|>/-2.5pt/@{(}@(d,d)[ll]_>\nu,:_z[lu]-@(d,d)[ll]-|>@{|}|>/-2.5pt/@{(}|-@{<}@(u,u)^i[rr]^>\mu}$}\nonumber\\
  &=\sum_{ijz\mu\nu} \sum_{\alpha\beta\chi}
  \frac{M_{i\alpha,z\chi}^{i\alpha,\mu}
 \left(M_{z\chi,j\beta}^{j\beta,\nu}\right)^*}{d_z}
 \scalebox{.9}{$\gdir{:@(u,u)^j[rr]-@(d,d)[ll]_>\nu,:_z[lu]-@(d,d)[ll]-|-@{<}@(u,u)^i[rr]^>\mu}$}.
\nonumber 
\end{align}
Therefore, for genus 2, the wave function overlap, up to an overall factor, is
given by (the labels in the trivial phase is omitted) 
\begin{align}
\label{W2MM}
  \t{W}^{(i,\bar j,z,\mu,\nu)}_{\cC;2}\propto \frac 1{d_z}
\sum_{\alpha\beta\chi}
  M_{i\alpha,z\chi}^{i\alpha,\mu}\left(M_{z\chi,j\beta}^{j\beta,\nu}\right)^*
\end{align}
Using (see \eqn{eq.unicon} and \eqn{Genus2_Genus1})
\begin{align}
M_{\one 1,j\beta}^{k\chi,1} = M_{j\beta,\one 1}^{k\chi,1}
&=\frac{1}{\sqrt{\sum_i N^A_i d_i}}\delta_{jk}\delta_{\beta\chi}.
\nonumber\\
 \t{W}^{i,j,\one,1,1}_{\cC;2} 
 &= \t{W}^{i}_{\cC,1} \t{W}^{j}_{\cC,1}
\end{align}
we can fix the overall factor.  Notice that (setting $z=\one$ in \eqn{W2MM})
\begin{align}
\sum_{\alpha\beta}
  M_{i\alpha,\one 1}^{i\alpha,1}\left(M_{\one 1,j\beta}^{j\beta,1}\right)^*
= \frac{N^A_iN^A_j}{\sum_i N^A_i d_i} ,
\end{align}
where we have use the fact that $\al=1,\cdots,N^A_i$ and $\bt=1,\cdots,N^A_j$.
Since $ \t{W}^{1}_{\cC,1}=N^A_{\one}=1$, we find 
\begin{align}
\label{W2MM0}
\t{W}^{(i,\bar j,z,\mu,\nu)}_{\cC;2} 
= 
\frac{\sum_k N^A_k d_k}{d_z}
\sum_{\alpha\beta\chi}
  M_{i\alpha,z\chi}^{i\alpha,\mu}\left(M_{z\chi,j\beta}^{j\beta,\nu}\right)^*
\end{align}

Another choice of basis is
\begin{align}
  \gdir{-|>@{>}@(lu,u)_i[lld]-@(d,ld)[rrd],-|>@{>}@(ru,u)^j[rrd]-@(d,rd)[lld],[dd]:|<@{|}|>@{|}|</2.5pt/@{)}|>/-2.5pt/@{(}^y[uu]|(-0.2){\mu'}|(1.2){\nu'}}\ ,
\end{align}
whose difference from the first choice is given by the $F$ matrix.
Correspondingly, 
\begin{align}
  &\qquad\scalebox{.9}{$\gdir{{m^\dag}*\cir<2ex>{}="md",
    [dd]{m}*\cir<2ex>{}="m",
    "md"-|>@{>}@(lu,u)_A[lld],"m"-@(ld,d)[llu],
    "md"-|>@{>}@(ru,u)^A[rrd],"m"-@(rd,d)[rru],
"m":^A"md"}$}\nonumber\\
&=\sum_{ijy\mu'\nu'}\left(\sum_{\alpha\beta\chi}M_{i\alpha,j\beta}^{y\chi,\mu'}\left({M_{i\alpha,j\beta}^{y\chi,\nu'}}\right)^*\right)
\scalebox{.9}{$\gdir{-|>@{>}@(lu,u)_i[lld]-@(d,ld)[rrd],-|>@{>}@(ru,u)^j[rrd]-@(d,rd)[lld],[dd]:|<@{|}|>@{|}|</2.5pt/@{)}|>/-2.5pt/@{(}^y[uu]|(-0.2){\mu'}|(1.2){\nu'}}$}\ .
\end{align}
The compatibility between the two choices is guaranteed by the Frobenius
condition \eqref{eq.Fro}.
It is straightforward to generalize to surfaces with any genus.

A Lagrangian algebra $A$ is \emph{modular invariant}, which was first proposed
and proved in \cite[Theorem\,3.4]{kr} for a special modular tensor category. But
the proof automatically works for all modular tensor category. Below we explain
the property in detail. Modular invariance means $T$ and $S$ invariance. First, the topological spin of $A$ is trivial
  \begin{align}
    \theta_A= \frac 1{\dim A}
    \gdir{[ld]:@(r,l)^<A[uurr]-@(r,r)[dd]-@(l,r)|-\hole[uull]-@(l,l)[dd]}=1,
  \end{align}
  which is a direct consequence of the commutative and Frobenius conditions. 
  Thus a twist of string $A$ leaves the graph invariant, which generates a
  subset a Dehn twists on the manifold. This property is the $T$
  invariance. Expressing in terms of simple objects,
  \begin{align}
    T_{ii} N^A_i=N^A_i.
  \end{align}

  $A$ also has the following invariance under
  the punctured $S$ transformation, which, together with the twist of $A$ above,
  generates all transformations in the mapping class group,
  \begin{align}
    \frac {d_i}{D}\ \gdir{ [ldl]:_>i|(.78)\hole[uuuu],[lu]m*\cir<2ex>{}="m",
      [rd]:@{|}|>/-2.5pt/@{(}_z_>\beta[lu]:_A"m"-^(0.2)A|(.49)\hole@(d,d)[ll]-|(.3)@{<}@(u,u)^(0.8)A"m"}
    =
  \sum_\alpha \g{[uu]:[d]^i^>\alpha-^A[d]m*\cir{},
    [lldd]:^i^>a[ru]-^A"m", 
  [rrdd]:_z_>\beta[lu]-_A "m"}.\label{eq.pSt}
  \end{align}
Substituting the structure coefficients, the above becomes
\begin{align}
    &\frac {d_i}{D}\ \gdir{ [ldl]:_>i|(.78)\hole[uuuu],[lu]m*\cir<2ex>{}="m",
      [rd]:@{|}|>/-2.5pt/@{(}_z_>\beta[lu]:_A"m"-^(0.2)A|(.49)\hole@(d,d)[ll]-|(.3)@{<}@(u,u)^(0.8)A"m"}=
      \frac {d_i}{D}\sum_{j\alpha,\nu}
      M^{j\alpha,\nu}_{j\alpha,z\beta}\ 
    \gdir{ [ldl]:_>i|(.78)\hole[uuuu],[lu]="m",
      :_z_>\nu[lu]-_(0.2)j|(.49)\hole@(d,d)[ll]-@{|}|>/-2.5pt/@{(}|(.3)@{<}^(0.8)j@(u,u)"m"}
      \nonumber\\
      &= \sum_{j\alpha,\nu,\mu}
      M^{j\alpha,\nu}_{j\alpha,z\beta} S^{(z)}_{j,\nu;i,\mu}
      \g{[u]:|-@{<}[d]^i^>\mu-|-@{<}[ld]_i,-|-@{<}[rd]^z}
      \nonumber\\
      &= \sum_\alpha \g{[uu]:|-@{<}[d]^i^>\alpha-|-@{<}^A[d]m*\cir{},
    [lldd]:|-@{>}^i^>\alpha[ru]-|-@{>}^A"m", 
  [rrdd]:|-@{>}_z_>\beta[lu]-|-@{>}_A "m"}
  =
    \sum_{\alpha,u} M^{i\alpha,\mu}_{i\alpha,z\beta}
    \g{[u]:|-@{<}[d]^i^>\mu-|-@{<}[ld]_i,-|-@{<}[rd]^z},
\end{align}
where we used the punctured $S$ matrix as in \eqref{eq.puncturedS}. 
Therefore, the structure coefficients satisfy the (punctured) $S$ invariance:
\begin{align}
  \sum_{j\nu} S^{(z)}_{j,\nu;i,\mu}\sum_\alpha
  M^{j\alpha,\nu}_{j\alpha,z\beta}=\sum_\alpha
  M^{i\alpha,\mu}_{i\alpha,z\beta}.
\end{align}
For a generic closed surface, as long as an appropriate graph basis
such as \eqref{eq.basis1} is picked, part of the graph will look like \eqref{eq.pSt}, and
thus the wave function overlap is invariant under the punctured $S$
transformation. For example, on genus 2, taking the left half of
\eqref{eq.basis1} and applying punctured $S$ transformation, we find that
\begin{align}
  \sum_{k\chi} S^{(z)}_{k,\chi;i,\mu} \t{W}^{(k,j,z,\chi,\nu)}_{\cC;2}=\t{W}^{(i,j,z,\mu,\nu)}_{\cC;2}. 
\end{align}
In particular, when $z=\one$, $\sum_\alpha M^{i\alpha,1}_{i\alpha,\one 1}=\eta_1^{-1}N^A_i$, we have the $S$
invariance on torus
\begin{align}
  \sum_j S_{ji}N^A_j=N^A_i.
\end{align}

\bibliography{../../bib/wencross,../../bib/all,../../bib/publst,./local}

\end{document}